\documentclass[12pt, 
]{article}

\usepackage{graphicx}
\usepackage{amsfonts}
\usepackage{amsmath}
\usepackage[usenames]{color}
\usepackage{amssymb}
\usepackage[mathscr]{eucal} 
\usepackage{longtable}

\usepackage{color}
\input{colordvi.tex}

\def\kesu#1{}

\def\N{\mathbb{N}}
\def\Z{\mathbb{Z}}

\def\R{\mathbb{R}}
\def\C{\mathbb{C}}
\def\P{\mathbb{P}}

\def\Re{\mathrm{Re}}

\newcommand{\sfrac}[2]{\left(\frac{#1}{#2}\right)}

\begin{document}
\baselineskip 0.6cm
\newcommand{\gsim}{ \mathop{}_{\textstyle \sim}^{\textstyle >} }
\newcommand{\lsim}{ \mathop{}_{\textstyle \sim}^{\textstyle <} }
\newcommand{\vev}[1]{ \left\langle {#1} \right\rangle }
\newcommand{\bra}[1]{ \langle {#1} | }
\newcommand{\ket}[1]{ | {#1} \rangle }
\newcommand{\Dsl}{\mbox{\ooalign{\hfil/\hfil\crcr$D$}}}
\newcommand{\nequiv}{\mbox{\ooalign{\hfil/\hfil\crcr$\equiv$}}}
\newcommand{\nsupset}{\mbox{\ooalign{\hfil/\hfil\crcr$\supset$}}}
\newcommand{\nni}{\mbox{\ooalign{\hfil/\hfil\crcr$\ni$}}}
\newcommand{\EV}{ {\rm eV} }
\newcommand{\KEV}{ {\rm keV} }
\newcommand{\MEV}{ {\rm MeV} }
\newcommand{\GEV}{ {\rm GeV} }
\newcommand{\TEV}{ {\rm TeV} }

\def\diag{\mathop{\rm diag}\nolimits}
\def\tr{\mathop{\rm tr}}

\def\Spin{\mathop{\rm Spin}}
\def\SO{\mathop{\rm SO}}
\def\O{\mathop{\rm O}}
\def\SU{\mathop{\rm SU}}
\def\U{\mathop{\rm U}}
\def\Sp{\mathop{\rm Sp}}
\def\SL{\mathop{\rm SL}}

\def\change#1#2{{\color{blue}#1}{\color{red} [#2]}\color{black}\hbox{}}


\begin{titlepage}
  
\begin{flushright}
  UT-11-02 \\
  IPMU11-0003 \\
  NSF-KITP-11-028 \\
\end{flushright}
   
\vskip 1cm
\begin{center}
  
 {\large \bf High--Energy Photon--Hadron Scattering in Holographic QCD}

\vskip 1.2cm
   
Ryoichi Nishio$^{1,2}$ and Taizan Watari$^2$
    
\vskip 0.4cm
 
{\it
  $^1$Department of Physics, University of Tokyo, Tokyo 113-0033, Japan  
   \\[2mm]
   
  $^2$Institute for the Physics and Mathematics of the Universe, University of Tokyo, Kashiwano-ha 5-1-5, 277-8583, Japan
   }
   
\abstract{
This article provides an in-depth look at hadron high energy
 scattering by using gravity dual descriptions of strongly coupled
 gauge theories. Just like deeply inelastic scattering (DIS) and deeply
 virtual Compton scattering (DVCS) serve as clean {\it experimental} probes 
into non-perturbative internal structure of hadrons, 
elastic scattering amplitude of a hadron and a (virtual) ``photon'' 
in gravity dual can be exploited as a {\it theoretical} probe. 
Since the scattering amplitude at sufficiently high energy (small 
Bjorken $x$) is dominated by parton contributions ($=$ Pomeron
 contributions) even in strong coupling regime, there is a chance 
to learn a lesson for 
generalized parton distribution (GPD) by using gravity dual models.   
We begin with refining derivation of Brower--Polchinski--Strassler--Tan 
(BPST) Pomeron kernel in gravity dual, 
paying particular attention to the role played by complex spin 
variable $j$. 
The BPST Pomeron on warped spacetime consists of a Kaluza--Klein tower
of 4D Pomerons with non-linear trajectories, and we clarify the 
relation between Pomeron couplings and Pomeron form factor. 
We emphasize that the saddle point value $j^*$ of the scattering
amplitude in the complex $j$-plane representation is a 
very important concept in understanding qualitative behavior of 
the scattering amplitude. The total Pomeron contribution to the
 scattering is decomposed into the saddle point contribution and 
at most a finite number of pole contributions, and when the pole
 contributions are absent (which we call saddle point phase), 
kinematical variable $(q,x,t)$ dependence of $\ln (1/q)$ evolution and 
$\ln(1/x)$ evolution parameters $\gamma_{\rm eff.}$ and 
$\lambda_{\rm eff.}$ in DIS and $t$-slope parameter $B$ of DVCS in HERA 
experiment are all reproduced qualitatively in gravity dual. 
All of these observations shed a new light on modeling of GPD.
Straightforward application of those results to other hadron high energy 
scattering is also discussed.} 
   
\end{center}
\end{titlepage}



\section{Introduction}

Despite plenty of data of hadron scattering, from which various 
qualitative features have been extracted, it remains difficult 
to derive and understand those features from the first principle, 
QCD, formulated as a perturbation theory.  
Gauge/gravity duality, however, can be exploited to study
non-perturbative aspects of ``hadron'' in strongly coupled gauge 
theories. Many papers along this line focus on static properties 
of hadrons, such as mass spectra and three point couplings, but 
nothing prevents us from using gravitational dual descriptions to 
study ``hadron'' scattering of strongly coupled gauge theories 
at arbitrary energy scale \cite{Polchinski2002,
PolchinskiJHEP0305:0122003, BrowerJHEP0712:0052007}.
Hadron--hadron scattering \cite{Polchinski2002}, 
total cross sections of deep inelastic scattering (DIS) 
\cite{PolchinskiJHEP0305:0122003}, form factors of various conserved 
currents \cite{Hong:2003jm, Hong:2004sa, Hong:2005np}, and 
saturation/unitarity \cite{Giddings:2002cd, Kang:2004jd,
AlvarezGaume:2006dw, Brower2007a, Hatta:2007, Brower2007,  
Cornalba:2007fs, AlvarezGaume:2008qs, Kovchegov:2009yj, Kovchegov:2010uk} 
are examples of non-perturbative 
observables that can be studied in gravitational dual, but 
potential power of gauge/gravity duality in hadron scattering 
is far from being fully exploited so far.

Although perturbative QCD can describe $q^2$ evolution of parton 
distribution functions (PDF) and generalized parton distributions (GPD), 
initial data of the evolution cannot be determined by perturbation theory. 
Such non-perturbative initial data for PDF can be obtained from DIS 
experiments (and have also been studied in gravitational dual 
models \cite{PolchinskiJHEP0305:0122003}; see \cite{Brower:2010wf} 
for a list of articles on DIS in gravitational dual), but GPD cannot be 
determined even from experimental data, without some theoretical
modeling of non-perturbative physics. 
GPD describes parton distribution in the transverse directions
\cite{Burkardt:2000za, Ralston:2001xs, Diehl:2002he, Burkardt:2002hr} and two parton 
correlation  in a hadron in general \cite{Ji:1998pc, 
Radyushkin:2000uy, Goeke:2001tz, DiehlPhys.Rept.388:41-2772003, Belitsky:2005qn, Boffi:2007yc}, and hence it is an interesting object 
on its own. In this article, we take this non-skewed GPD and more generally 
deeply virtual Compton scattering (DVCS) \cite{Ji:1998pc, Radyushkin:2000uy, Goeke:2001tz, 
DiehlPhys.Rept.388:41-2772003, Belitsky:2005qn, Boffi:2007yc} amplitude 
at small $x$ as examples of hadron scattering, and see that
gravitational dual descriptions can determine how those non-perturbative 
scattering amplitudes depend on kinematical variables such as 
center-of-mass energy, momentum transfer, impact parameter and 
photon virtuality.
Gauge/gravity dual also tells us how to think about various theoretical 
ideas that various models of GPD have been based on.

High energy behavior of elastic scattering amplitude $A(s,t)$ of two 
hadrons is characterized by poles and their residues of its partial wave
amplitude $A(j,t)$ on the complex angular momentum $j$-plane (e.g., 
\cite{Collins:1977, Forshaw:1997dc, BaronePredazzi}); 
the poles and residues depend on momentum transfer $t$. 
The poles in the $j$-plane have often been assumed to depend linearly 
in $t$, which is supported by the spectrum of mesons and hadron 
scattering cross sections at least for finite range of momentum transfer $t$. 
Given the fact that the earliest version of string theory was born 
out of efforts to describe hadron scattering, it is not surprising that 
some successful aspects of classical Regge theory are preserved 
in gravity dual, string theory on a warped background. 
Notable aspects of string theory on a warped spacetime, however, include 
i) a single ``Regge trajectory'' of string theory on 10 dimensions gives 
rise to a Kaluza--Klein tower of infinite ``Regge trajectories'' on 4
dimensions, and ii) those trajectories do not remain linear for 
arbitrary negative $t$ \cite{BrowerJHEP0712:0052007}.
The non-linear trajectories immediately result in non-Gaussian 
profile of GPD in the transverse directions (see section
\ref{ssec:b-space} of this article as well as \cite{Brower:2010wf}), 
although Gaussian profile of GPD is often assumed in phenomenological 
analysis.
We will also describe how the residues of the Regge poles are determined 
by holographic set-up, and also explain how the Kaluza--Klein tower of 
Regge trajectories organizes itself to become a single contribution 
with a form factor in momentum transfer $t < 0$.  

An extra energy scale $q$---photon virtuality---is available in 
photon--hadron scattering, in addition to the center of mass energy $W$
and confinement scale $\Lambda$.
This extra parameter makes theoretical understanding 
of the non-perturbative amplitude interesting. The scattering amplitude
is dominated by a contribution from a saddle point in the complex $j$-plane,
 not from a pole, for sufficiently large $q \gg \Lambda$.
The saddle point value $j^*$ depends on kinematical variables such as 
$W$, $q$ and $t$. We find, by following this dependence of $j^*$,
instead of naively taking small $x$ limit or large $q^2$ limit, that 
observables characterizing scattering amplitude such as $\ln (1/q)$-evolution 
parameter $\gamma_{\rm eff.}$, $\ln(1/x)$-evolution parameter 
$\lambda_{\rm eff.}$ and $t$-slope parameter $B$ show qualitatively 
the same behavior in the strong coupling regime (gravity dual) as 
expected in perturbative QCD or observed in HERA DVCS experiment. 
As the saddle point value $j^*$ and the leading poles are both given by 
the kinematical variables of the scattering, crossover 
from the saddle-point phase to the leading pole phase may also be expected, 
when the photon virtuality decreases to a smaller value.\footnote{
A similar crossover behavior has already been observed in the real part
to imaginary part ratio of the hadron--virtual ``photon'' scattering
amplitude in gravity dual \cite{Hatta:2007}. We will elaborate more also
on this crossover behavior in section \ref{ssec:real-part}.}

This article is organized as follows. Section \ref{sec:model} explains 
the set-up of gravitational dual for our calculation of photon--hadron 
scattering amplitude, while summarizing conventions and providing brief 
mini-reviews. Section \ref{sec:pomeron} explains how the
2-body--to--2-body scattering amplitudes are given in the gravitational 
dual set-up, and presents an explicit form of Pomeron kernel; this
section is largely a repetition of the contents 
of \cite{BrowerJHEP0712:0052007}, but we believe that some small 
improvements are also made and subtleties clarified in derivation and 
final expression of the amplitudes and kernel. The photon--hadron 
scattering amplitude is discussed for zero skewedness in 
sections \ref{ssec:t-space}--\ref{ssec:real-part}; 
section \ref{ssec:t-space} explains momentum-transfer $t$ dependence 
of the imaginary part of the scattering amplitude, while its impact 
parameter $b$ dependence (i.e., transverse profile) is described in
section \ref{ssec:b-space}. Section \ref{ssec:real-part} is devoted 
to the real part of the amplitude. 
The interpretation of the scattering amplitude on a warped spacetime in 
Froissart--Gribov--Regge language on 4 dimensions is given in detail 
in section \ref{sssec:regge}, while section \ref{sssec:slope} is devoted 
to the $t$-slope parameter of the scattering amplitude. 
We will see in section \ref{ssec:GPD} that GPD can be identified 
within the scattering amplitude even in the strong coupling regime, and
discuss the form factor of GPD. 
In section \ref{sec:real-world}, we address a question whether there is 
anything we can learn about GPD of the real world from the GPD 
calculation in the strong coupling regime. 
Section \ref{sec:discussion} briefly describes straightforward 
application of various results and observations in this article to 
other high-energy hadron scattering processes. 

\section{Model}
\label{sec:model}

Such gravitational backgrounds as \cite{ Polchinski2007, 
Klebanov2000d, Maldacena2001, Butti:2004pk} 
are an ideal
framework for holographic calculation of certain types of hadron 
interactions, as they have built-in confinement mechanism at IR, and 
allow renormalization group interpretation in terms of holographic radius. 
Those models have background geometry that are approximately 
$AdS_5 \times W$ with some compact 5-dimensional manifold $W$, apart 
from the region near the IR boundary. 
The hard wall model \cite{PolchinskiJHEP0305:0122003} replaces these 
geometries with an $AdS_5 \times W$ background\footnote{
In our convention, $\eta_{\mu\nu}= \diag (-1,1,1,1)$.
We use $M,N,\dots$ in labeling coordinates of ten dimensional spacetime,
$m,n,\dots$ for the $AdS_5$ part, and 
$\mu,\nu, \rho, \sigma, \kappa \dots$ for the coordinates of the 3+1 
dimensional Minkowski spacetime. $\theta^a, \theta^b, \cdots$ are
dimensionless coordinates of $W$, and $(g_W)_{ab} d\theta^a d\theta^b$
is the metric of $W$.} 
\begin{align}
 \label{eq:AdS5timesS5}
  ds^2  =G_{MN}dx^M dx^N&=g_{mn}dx^m dx^n +R^2 (g_{W})_{ab}d\theta^a d\theta^b,\\
 \label{eq:hwb}
  g_{mn}dx^m dx^n&=e^{2A(z)}(\eta_{\mu\nu}dx^\mu dx^\nu+dz^2), \;\;\;
 e^{2A(z)}=\frac{R^2}{z^2}.\;\;\; 
\end{align}
An IR boundary is introduced at $z = \Lambda^{-1}$ and boundary
conditions on various fields are set by hand instead. The dilaton vev is 
simply assumed to be constant, $e^\phi = g_s$. Such a background 
is not obtained as a stable solution to the Type IIB string theory, 
but Type IIB string calculations on such a background (while ignoring 
NS--NS tadpoles) are expected to maintain qualitative aspects of 
certain hadronic processes in the original holographic models. 
We use the hard wall model in the rest of this article, as it makes 
it possible to compute various physical quantities and study dynamics 
without consuming too much time. 

\begin{figure}[tbp]
 \begin{center}
  \includegraphics[scale=0.8]{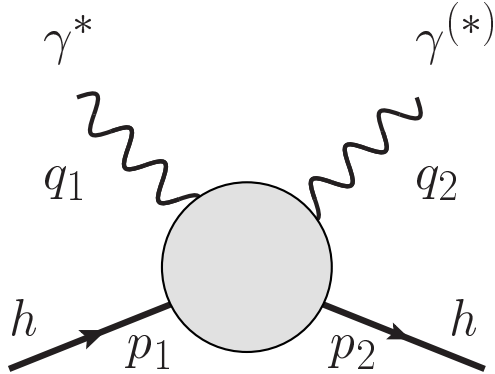} 
  \caption{\label{fig:DVCS-DDVCS} a cartoon picture showing elastic
  scattering of DVCS / DDVCS, with momentum labels on the external
  lines}
 \end{center}
\end{figure}

Deeply virtual Compton scattering (DVCS) and double deeply virtual
Compton scattering (DDVCS) are elastic scattering of a hadron $h$ and 
a (virtual) photon, 
$\gamma^\ast(q_1) + h(p_1)\rightarrow \gamma^{(\ast)}(q_2) + h(p_2)$, 
Figure \ref{fig:DVCS-DDVCS}, with the kinematics 
\begin{equation}
\label{eq:double deeply virtual}
 q_1^2 \gg \Lambda^2, \; q_2^2 = 0 \; ({\rm DVCS}), \qquad 
 q_1^2, q_2^2 \gg \Lambda^2 \; ({\rm DDVCS}).
\end{equation}
%
As the target hadron, we use one of normalizable modes in a scalar
degree of freedom $\phi(x,z)$ on $AdS_5$ in the holographic set up. 
In particular, we use a scalar field $\phi(x,z)$ originating from 
ten dimensional SUGRA fields upon Kaluza--Klein reduction on $W$, 
just like in \cite{Kim:1985ez}. 
In case of reduction of dilaton, for example,
\begin{align}
\label{eq:phiy}
 \phi(x,z,\theta)=\phi(x,z)Y(\theta)
\end{align}
with a non-trivial harmonic function $Y(\theta)$ on $W$, the target 
hadron $h$ corresponds to a glueball.\footnote{
To study DVCS and DDVCS of {\it baryons}, $D$-branes should be used in the
holographic set up, instead of a Kaluza--Klein mode of Type IIB SUGRA 
fields. 
We hope that there is still something to learn in an easier study with 
a scalar glueball target. 
} 

The holographic wavefunction of the target hadron $h$ is obtained 
by solving equation of motion derived from an effective action of the 
scalar field $\phi(x,z)$ on $AdS_5$. 
The bilinear part 
is
\begin{align}
\label{eq:bilinar action for scalar}
  S_\phi=\int d^4x dz \sqrt{-g}\left[
-\frac{c_\phi}{2\kappa_5^2}
 (\partial_m\phi \partial^m \phi^\ast +M^2\phi\phi^\ast)
\right],
\end{align}
where 
\begin{equation}
 \frac{1}{\kappa_5^2} = \frac{R^5 {\rm vol}(W)}{\kappa^2_{\rm IIB}} \sim
  {\cal O}\left( \frac{N_c^2}{R^3} \right),
\end{equation}
and a dimensionless constant\footnote{
It is defined as follows:
\begin{equation}
 c_\phi = \frac{1}{\int_W d^5 \theta \sqrt{g_W(\theta)}} 
  \int_W d^5 \theta \sqrt{g_W(\theta)} |Y(\theta)|^2.
\label{eq:def-c-phi}
\end{equation}
} $c_\phi$ of order unity and a mass 
parameter $M^2$ depend on the choice of $Y(\theta)$.
When a Dirichlet condition\footnote{
Qualitative aspects of our results will not change, when a Neumann
boundary condition is imposed, instead.} is imposed on $\phi(x,z)$ at the IR
boundary $z = \Lambda^{-1}$, the wavefunction\footnote{The wavefunction $\Phi_n(z)$ satisfies  
$\frac{1}{2\kappa_5^2}\int_0^{\Lambda^{-1}} d z \sqrt{-g} e^{-2A} 
 \Phi_n(z) \Phi_m(z) = \delta_{nm}$, so that the hadron field $h_n(x)$ in 
$\phi(x,z) = h_n(x) \Phi_n(z)$ has a canonical kinetic term upon
dimensional reduction to 3+1 dimensions. } of a hadron $h_n$ 
(corresponding to the $n$-th normalizable mode) with incoming 
momentum $p_\mu$ is given by 
\begin{align}
\phi(x,z) \rightarrow e^{i p \cdot x} \Phi_n(z), \qquad 
 \sqrt{c_\phi}\Phi_n(z) = 
 \sfrac{2\kappa_5^2}{R^3}^{1/2}\sqrt{2}\Lambda z^2 \frac{J_{\Delta-2}(j_{{\Delta-2},n}\Lambda z)}{J'_{\Delta-2}(j_{{\Delta-2},n})},
\label{eq:normalizable mode}
\end{align}
where $J_\mu(x)$ is the Bessel function, $j_{\mu,n}$ the $n$-th 
zero point of $J_\mu(x)$, and $\Delta = 2 + \sqrt{4 + M^2 R^2}$.
Mass of the hadron $h_n$ is given by 
\begin{align}\label{eq:masses}
 m_n=\Lambda j_{\Delta-2,n}.
\end{align}
We will not specify the excitation level $n$ of the target hadron
$h_n(x)$, as we will pay attention only to qualitative aspects of 
DVCS / DDVCS amplitudes, not to numerical details that depend 
on the excitation level $n$.

As the (virtual) photon probe of DVCS / DDVCS amplitudes, we gauge an
$R$-symmetry associated with $W$, and use it as a probe, just like in 
\cite{PolchinskiJHEP0305:0122003}. The type IIB metric field on 10 dimensions 
become a massless vector field on $AdS_5$ through 
\begin{align}
 \delta G_{ma}(x,z,\theta)=A_m(x,z) v_a(\theta),
\label{eq:G-Av}
\end{align}
where $v^a(\theta) \partial/\partial\theta^a$ is a Killing vector of
$W$, and $v_a = R^2 (g_{(W)})_{ab} v^b$.
The kinetic term of the effective action of $A_m(x,z)$ on $AdS_5$ 
is given by 
\begin{align}
 S_A=\frac{R^2 c_B c_A}{2\kappa_5^2}
  \int d^4x dz \sqrt{-g}\left[-\frac{1}{4}F_{mn}F^{mn}\right],
\end{align}
with dimensionless coefficients\footnote{
Arbitrary chosen normalization of the Killing vectors 
$v_i = v_i^a (\partial/\partial \theta^a)$ does not remain in $c_A$, 
as we define it by 
\begin{equation}
 T_R^{-1} {\rm tr}_R \left(t_i t_j \right) \; c_A = 
  \frac{1}{\int_W d^5 \theta \sqrt{g_W(\theta)}} 
   \int_W d^5 \theta \sqrt{g_W(\theta)} v^a_i v^b_j g_W(\theta); 
\end{equation}
the generators $t_i$'s are chosen so that they satisfy the same 
commutation relation $[t_i, t_j] = i f_{ij}^k t_k$ as the Killing vectors, 
$\{ v_i, v_j\} = f_{ij}^k v_k$, and the Cartan metric remain the same.
In the case a non-Abelian subalgebra of the Killing vectors of $W$ is
gauged, $F_{mn}F^{mn}$ in the kinetic term above should be understood as 
$T_R^{-1} {\rm tr}_R(F_{mn} F^{mn})$ using the generators we explained above.
} $c_A$ and $c_B$.
The non-normalizable wavefunction of the vector field $A_m(x,z)$ for the 
``photon probe'' with incoming spacelike momentum $q_\mu$ is given by 
\begin{eqnarray}
  \sqrt{c_A} F_{\rho\mu}(x,z) & \rightarrow & 
  i c_J (q_\rho \epsilon_\mu - \epsilon_\rho q_\mu) qz 
    \left\{K_1(qz) + \frac{K_0(q/\Lambda)}{I_0(q/\Lambda)}I_1(q z)
    \right\} e^{i q \cdot x}, \label{eq:wf for Fmn}\\
  \sqrt{c_A} F_{\rho z} (x,z) & \rightarrow & c_J (q^2 \epsilon_\rho
   - (q \cdot \epsilon) q_\rho ) z 
      \left\{ K_0(qz) - \frac{K_0(q/\Lambda)}{I_0(q/\Lambda)}I_0(q
       z)\right\}
    e^{i q \cdot x}, \label{eq:wf for Fmz}
\end{eqnarray}
where $c_J$ is a dimensionless constant of order unity. 
$q = \sqrt{q^2}$, and $\epsilon(q)_\mu$ is the polarization within 
the four dimensions. 
For sufficiently spacelike $q^2$ much larger than $\Lambda^2$, the second terms $I_1(qz)$ and
$I_0(qz)$ are negligible in (\ref{eq:wf for Fmn}, \ref{eq:wf for Fmz}), 
but the full expression needs to be used for the final state on-shell
photon in the DVCS.

The elastic scattering amplitude $A(\gamma^\ast h \rightarrow
\gamma^{(\ast)} h)$ of a hadron $h$ and a (virtual) photon probe
$\gamma^{(*)}$ is calculated in holographic model by using the world-sheet 
non-linear sigma model with the background metric(\ref{eq:AdS5timesS5},
\ref{eq:hwb}) and inserting vertex operators whose Born--Oppenheimer 
approximation \cite{Rohm:1985jv, Distler:1987ee} are specified 
by the wavefunctions (\ref{eq:phiy}, \ref{eq:normalizable mode}) and 
(\ref{eq:G-Av}, \ref{eq:wf for Fmn}, 
\ref{eq:wf for Fmz}) \cite{BrowerJHEP0712:0052007}.
The Compton tensor\footnote{In the real world QCD with QED probe, where 
only fermion partons are charged under the probe, 
\begin{equation}
(2\pi)^4 \delta(p_1 + q_1 - p_2 - q_2)\; i \; T^{\mu\nu} = - 
 \int d^4 x_2 \int d^4 x_1 e^{- i q_2 \cdot x_2} e^{i q_1 \cdot x_1} 
   \bra{h(p_2)}T \{ J^\nu(x_2) J^\mu(x_1) \} \ket{h(p_1)}. 
\end{equation}
 }
$T_{\mu\nu}$ is defined by removing polarization vectors,
\begin{align}
 A(\gamma^\ast h \rightarrow \gamma^{(\ast)} h)=\epsilon^\mu_1
 T_{\mu\nu}(\epsilon^\nu_2)^\ast, 
\end{align}
and the goal of this article is to determine five independent structure 
functions\footnote{Leading order perturbative QCD result in terms of
non-perturbative GPD is found in \cite{Radyushkin:2000ap} in the case 
of a scalar target hadron, and in \cite{Belitsky:2000vx} in the case 
of a fermion target hadron. 
See also \cite{Mankiewicz:1997bk, Belitsky:2001ns}.} 
$V_{1,2,3,4,5}$ (e.g., \cite{Marquet2010}) in 
\begin{align}
   T^{\mu\nu}=&V_1 P[q_1]^{\mu\rho} P[q_2]^{\nu}_{\rho}
           +V_2 (p\cdot P[q_1])^\mu (p\cdot P[q_2])^\nu
           +V_3 (q_2\cdot P[q_1])^\mu (q_1\cdot P[q_2])^\nu
\notag \\ \label{eq:structure functions of Compton tensor}
           &+V_4 (p\cdot P[q_1])^\mu (q_1\cdot P[q_2])^\nu
           +V_5 (q_2\cdot P[q_1])^\mu (p\cdot P[q_2])^\nu
           +A \epsilon^{\mu\nu\rho\sigma}q_{1\rho}q_{2\sigma}.
\end{align}
The last term with the coefficient function $A$ should vanish 
for a scalar target hadron $h$ in a parity-invariant theory. 
Here, we introduced a convenient notation 
\begin{align}
  P[q]_{\mu\nu}=\left[\eta_{\mu\nu}-\frac{q_\mu q_\nu}{q^2}\right].
\end{align}
Those structure functions, $V_{1,2,3,4,5}(x,\eta,t,q^2)$, should be 
expressed in terms of Lorentz invariant kinematical variables 
$x, \eta, t$ and $q^2$, where 
\begin{align}
\label{kinematics1}
  p^\mu &= \frac{1}{2}(p_1^\mu+p_2^\mu),& q^\mu &= \frac{1}{2}(q_1^\mu+q_2^\mu),& \Delta^\mu &=p_2^\mu -p_1^\mu = q_1^\mu - q_2^\mu,
\end{align}
\begin{align}
\label{kinematics2}
  x &= \frac{-q^2}{2p\cdot q},& \eta &= \frac{-\Delta \cdot q}{2 p\cdot
 q}, &s&= W^2 = -(p+q)^2, & t&=-\Delta^2,
\end{align}
just as in standard literature in perturbative QCD. 
The parameter $t$ is assumed to be small 
\begin{equation}
 |t| \ll q^2, W^2
\label{eq:Bjorken-2}
\end{equation}
throughout this article. The two conditions on the kinematical variables 
(\ref{eq:double deeply virtual}) and (\ref{eq:Bjorken-2}) combined is 
sometimes referred to as generalized Bjorken regime.
In this article, we focus on non-skewed DDVCS ($\eta=0$).
More general case ($\eta\ne 0$) including DVCS ($\eta=x$)
 requires farther analysis.
Holographic calculation of the DDVCS amplitude should reproduce 
the pure forward amplitude (whose imaginary part is the deeply inelastic
scattering (DIS) amplitude studied in \cite{PolchinskiJHEP0305:0122003})
when the skewedness $\eta$ and momentum transfer $t$ are set zero;
${\rm Im} \; V_1 (x,\eta, t, q^2) \rightarrow F_1(x,q^2)$ and 
$(q^2/(2x)) \times {\rm Im} \; V_2 (x,\eta, t, q^2) \rightarrow F_2(x,q^2)$.

For simplicity, we will study the sphere amplitude contribution to the 
four closed string external states. Although the sphere amplitude alone 
cannot discuss how the unitarity is maintained in the scattering, 
sphere amplitude is sufficient for large enough $N_c$ (or for not too
large $s = W^2$, for sufficiently large $q^2$, or for sufficiently large
impact parameter $b$). 
It is also possible to discuss with the sphere amplitude 
how the scattering approaches unitarity limit \cite{Hatta:2007}.
The pion cloud \cite{Sullivan:1971kd, Kumano:1997cy, Garvey:2001yq}
contribution to the impact parameter dependent profile 
\cite{Strikman:2009bd, Strikman:2010pu}, however, can be studied 
only in a more realistic holographic model containing 
pion \cite{Sakai:2004cn, Sakai:2005yt, Kuperstein2008, Dymarsky2009} 
by examining $\chi = 2 - 2g - h = -1$ amplitude. 

\section{Pomeron Contribution to the Structure Functions}
\label{sec:pomeron}

\begin{figure}[tbp]
 \begin{center}
\begin{tabular}{ccc}
  \includegraphics[scale=0.6]{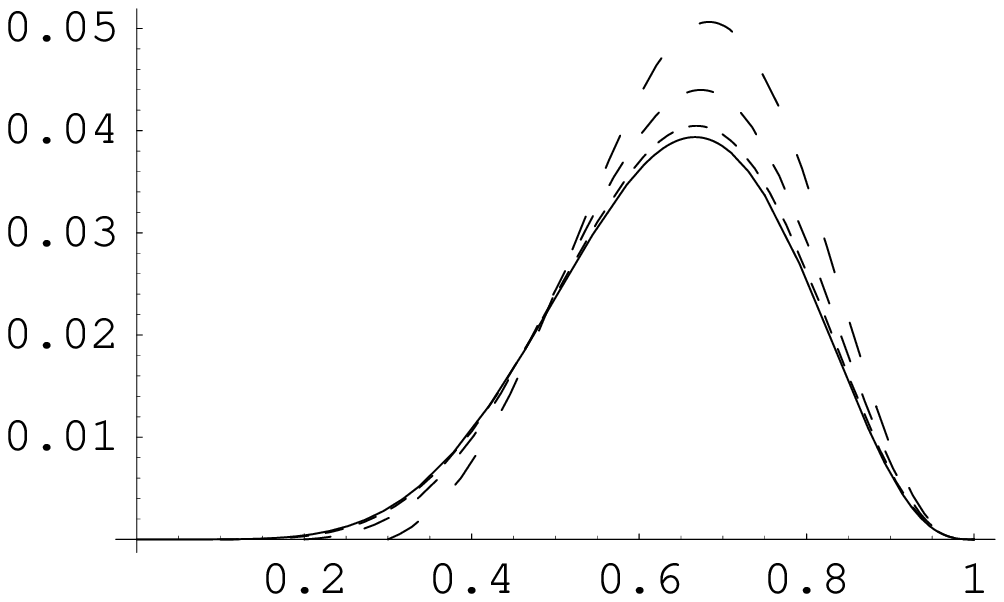} & &
  \includegraphics[scale=0.6]{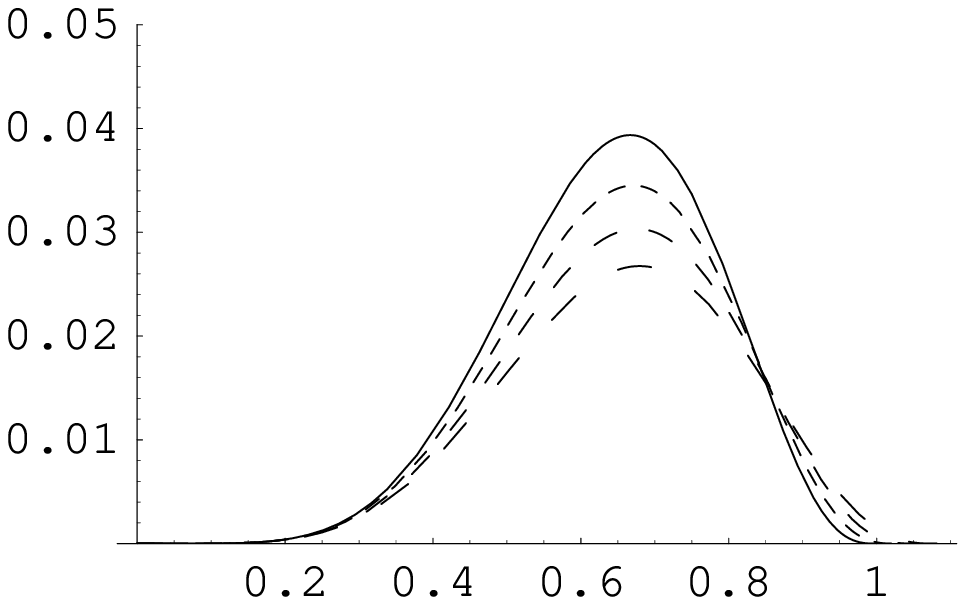} \\
  (a) & & (b)
\end{tabular}
\caption{\label{fig:sugra-resonance} Contribution to the DDVCS amplitude
  from double trace protected operators ($s$-channel sugra field
  resonances). $(c_J^2 2x/q^2)^{-1}{\rm Im} \; V_2$ is shown in (a) 
as functions of $x\in[\eta,1]$ for four different values of
  skewedness $\eta$. The solid line is for $\eta = 0$, and dashed 
lines are for $\eta = 0.1, 0.2$ and $0.3$. We set $t = 0$ in panel
  (a). On the other hand, the panel (b) shows the $(-t)$ dependence, 
while we set $\eta = 0$. The solid line is for $(-t)/q^2 = 0$, and three 
dashed lines are for $(-t)/q^2 = 0.1, 0.2$ and $0.3$, and are shown 
for the range $0 \leq x \leq (1+t/(4q^2))^{-1}$.
$q^2/\Lambda^2 = 10^2$ is used for both (a) and (b). As a target hadron,
the lowest normalizable mode of a scalar with $\Delta = 5$ was used.}
  \end{center}
\end{figure}

Holographic QCD has already been used to study DDVCS / DVCS amplitudes in 
the literature. Reference \cite{Marquet2010} calculated DDVCS / DVCS
amplitudes for scalar target hadron (a set-up that we explained in 
the previous section), with a single supergravity field in the
$s$-channel resonance (see Figure~\ref{fig:sugra-resonance}). 
This contribution to the amplitude corresponds to the resonant 
contribution to the deeply inelastic scattering in the $t = 0$ limit, 
and is known to be dominant in the large $q^2$ limit for moderately 
large 't Hooft coupling $\lambda$ and 
moderate $x$ \cite{PolchinskiJHEP0305:0122003}. 
We can certainly learn the structure functions of DDCVS / DVCS processes in this
way for the kinematical range mentioned above, but this sugra resonance
contribution in the holographic calculation corresponds to the higher 
twist double trace operators of weakly coupled gauge theories, and does 
not tell us much about the non-perturbative input of GPDs.

The operators that are approximately twist-2 in perturbative gauge
theories, on the other hand, have large corrections to the anomalous 
dimensions in strongly coupled regime (except the spin 2 operator). 
This ``twist-2'' contribution still dominates in the DIS amplitude 
for given $q^2$ at sufficiently small $x$
\cite{PolchinskiJHEP0305:0122003}, and are naturally 
expected to be so in the off-forward DDVCS amplitude as well.
In this article, we study this ``twist-2'' contribution in the small $x$
region by using holographic calculation, to get some hint on the 
non-perturbative input of the GPDs. 

The ``twist-2'' operators of gauge theories correspond to string states 
in graviton trajectory in holographic descriptions 
\cite{Gubser:2002tv, BrowerJHEP0712:0052007}. 
Contributions that involve such string states in the graviton trajectory 
to hadron scattering processes are called Pomeron amplitudes, and are 
known to be expressed as in (\ref{eq:two scalar amp}) 
\cite{PolchinskiJHEP0305:0122003, BrowerJHEP0712:0052007, Hatta:2007,
Brower2007}.
In this section, we begin with refining the derivation of 
(\ref{eq:two scalar amp}). 
Our derivation largely follows the ones in sections 2 and 3 
of \cite{BrowerJHEP0712:0052007} and combine 
various improvements already made in \cite{
Hatta:2007,Brower2007}, but 
we believe that the following presentation also made a couple of small 
improvements. All of the following issues are closely related: 
\begin{itemize}
 \item [a)] choice of integration contour in complex angular momentum plane, 
 \item [b)] validity of deformation of the integration contour, 
 \item [c)] origin of signature factor $[1 + e^{-\pi i j}]/\sin (\pi
       j)$, 
 \item [d)] absence of non-sense poles at negative integer angular 
momenta $j$, and
 \item [e)] the fact that the sphere amplitude of string theory is at best 
interpreted as a sum of $t$-channel and $u$-channel exchange of particles with various spins, not purely a sum of $t$-channel amplitudes.
\end{itemize}
It will be made clear how we should think\footnote{By no means, we
consider that the logical derivation in the following is the only possible one.
} about these issues a)--e) in writing down the Pomeron amplitude 
(\ref{eq:two scalar amp}); interpretation provided in section 4.3 of this article is affected by the understanding on the issue
e) we obtain here. 

Another improvement is to replace $\Delta_2$, a derivative operator 
on $AdS_5$ often used in Pomeron propagator in the literature, 
by $\Delta_j(t)$ in (\ref{eq:spin-j-deriv}) for complex
angular momentum $j$. The Pomeron wavefunction $\Psi^{(j)}_{i\nu}$ 
for spin $j \in \C$ is also introduced. Although this change does not 
leave a practical impact on the expression to be used as the 
Pomeron kernel (\ref{eq:pomeron kernel}), this extra conceptual 
clarification of the role played by complex angular momentum $j$ 
in the Pomeron kernel will enable us to provide clear theoretical 
understanding of the form factor associated with Pomeron-hadron-hadron 
coupling in section \ref{ssec:t-space}.

In section \ref{ssec:polarization}, we focus on  DDVCS amplitude and 
find explicit expressions of Pomeron contributions to the five independent 
structure functions. It may be an option for busy and practical readers 
with some familiarity to \cite{BrowerJHEP0712:0052007, 
Hatta:2007, Brower2007}
to skip section \ref{ssec:Pomeron-kernel} for the first reading.

\subsection{Pomeron Kernel from Sphere Amplitude}
\label{ssec:Pomeron-kernel}

Reference \cite{BrowerJHEP0712:0052007} derives Pomeron kernel 
in its section 2 by modifying Virasoro--Schapiro scattering amplitude 
of closed string on 10-dimensional flat spacetime so that it is understood
as an amplitude of scattering on a curved spacetime background 
(with small curvature $\alpha'/R^2 \ll 1$); we will combine it with 
the discussion in section 3 of \cite{BrowerJHEP0712:0052007} to fill 
a small gap in the process of modifying the scattering amplitude on 
flat 10-dimensions to that on the curved spacetime. 

In flat 10-dimensional spacetime, the sphere level scattering amplitude of 
two NS--NS closed strings has a factorized form 
\begin{align}
\label{eq:KandG}
 A(s_{10},t_{10})(2\pi)^{10}\delta^{10}(p_1+q_1-p_2-q_2) = K \; G,
\end{align}
where $s_{10}$ and $t_{10}$ are Mandelstam variables in 10-dimensions. 
The factor $G$ is a function of $s_{10}$ and $t_{10}$, and independent 
of polarizations of the external strings; 
\begin{align}\label{eq:Virasoro-Schapiro}
 G(s_{10},t_{10})=-\frac{\alpha'^3 s_{10}^2}{64}\prod_{\xi=s_{10},t_{10},u_{10}}\frac{\Gamma(-\alpha' \xi/4)}{\Gamma(1+\alpha' \xi/4)}.
\end{align}
The factor $K$, on the other hand, is given by wavefunctions---momentum
and polarization---of states involved in the scattering, 
whose explicit form is found in textbooks \cite{Polchinski:1998rr}.\footnote{   
We follow \cite{PolchinskiJHEP0305:0122003}, however, to use the factor 
$K$ that is $s_{10}^{-2}$ times that of \cite{Polchinski:1998rr}, and 
further includes delta function of momentum conservation; $s_{10}^{-2}$ 
can be factored out from the factor $K$ \cite{Polchinski:1998rr}, 
because we are interested only in the large $s_{10}$ small $|t_{10}|$ 
scattering, as in \cite{PolchinskiJHEP0305:0122003,
BrowerJHEP0712:0052007}.}
Normalization of the factor $K$ is 
\begin{align}\label{eq:flat scalar k}
 K & \sim \alpha'^4g_s^2 s_{10}^2
 (2\pi)^{10}\delta^{10}(p_1+q_1-p_2-q_2)
  \notag \\
  & \sim \frac{1}{2 \kappa_{\rm IIB}^2} 
 \int d^{10} x \; 
    [\alpha'^2 g_s e^{i p_1 \cdot x}]
    [\alpha'^2 g_s e^{i q_1 \cdot x}]
    [\alpha'^2 g_s e^{- i p_2 \cdot x}]
    [\alpha'^2 g_s e^{- i q_2 \cdot x}] s_{10}^2 
\end{align}
for dilation--dilaton scattering, up to a constant of order unity.
The factor $K$ for the case of our interest, dilaton--graviton
scattering as a holographic model of DDVCS, is a little more
complicated, because (dimensionless) polarizations of graviton external
states are involved.
We will first present the derivation of the scattering amplitude 
(\ref{eq:two scalar amp}) and explicit expression of Pomeron kernel (\ref{eq:pomeron kernel}), 
using the case of dilaton--dilaton scattering. 
Polarization dependent statements are deferred to 
section \ref{ssec:polarization}. 

We will study small $x \simeq q^2/s$ DDVCS amplitude, with momentum 
transfer $(-t) \ll q^2$, (\ref{eq:Bjorken-2}), that is not necessarily 
smaller than the hadronic scale $\Lambda^2$; here, $s = W^2$ and $t$ are 
Mandelstam variables of 4D kinematics of DDVCS. 
This means, as we explain later, that we need to examine the 
Virasoro--Schapiro amplitude in the kinematical region 
$\alpha' s_{10} \gg 1$, and $(-t_{10})/s_{10} \ll 1$, but not 
necessarily\footnote{This is a difference from the application to
DIS in \cite{PolchinskiJHEP0305:0122003}, where 
$\alpha' t_{10} \rightarrow {\cal O}(1/\sqrt{\lambda})$ can be ignored.} 
$|\alpha' t_{10}| \ll 1$.  
Ignoring terms that are suppressed by $(\alpha's_{10})$ or 
$(s_{10}/t_{10})$, one finds that 
\begin{align}
 &\prod_{\xi=s_{10},t_{10},u_{10}}\frac{\Gamma(-\alpha'
 \xi/4)}{\Gamma(1+\alpha' \xi/4)}
  \simeq   \frac{\pi}{\sin\left( \pi \alpha't_{10}/4\right)}
         \frac{1}{\Gamma^2\left(1+\alpha't_{10}/4\right)}  \notag \\
 & \qquad \qquad \qquad \qquad \times 
   \left[\cos\sfrac{\pi\alpha't_{10}}{4}+\cot\sfrac{\pi \alpha's_{10}}{4} \sin\sfrac{ \pi\alpha't_{10}}{4}\right]
\sfrac{\alpha's_{10}}{4}^{-2+\alpha't_{10}/2}
 \!\!\!\!\!\!\!\!\!\!\!\!\!\! . \label{eq:Regge-VS-10D}
\end{align}
In this expression, $\cot\sfrac{\pi \alpha' s_{10}}{4}$ 
contains $s$-channel poles of strings. As long as\footnote{
This is consistent with the well-known $i\epsilon$ prescription in 
the quantum field theory, which defines the physical amplitude in the
limit of $\arg s_{10}\rightarrow +0$.
Moreover one can understand (\ref{eq:cot -i}) as the average of 
$\cot\sfrac{\pi \alpha' s_{10}}{4}$ in $\alpha's_{10}\gg 1$ as 
in \cite{PolchinskiJHEP0305:0122003}.} 
$0<\arg s_{10}<\pi$,
in $|\alpha' s_{10}|\rightarrow \infty$,
\begin{align}\label{eq:cot -i}
\cot\sfrac{\pi \alpha' s_{10}}{4} \rightarrow -i.
\end{align}
As a result, one arrives at a well-known expression of the Regge
behavior of the Virasoro--Schapiro amplitude on flat 10-dimensional 
spacetime:
\begin{align}\label{eq:flat cal g}
 G(s_{10},t_{10})\simeq -\frac{\alpha'\pi}{4}\frac{1+e^{-i\pi \alpha' t_{10}/2}}{\sin(\pi \alpha' t_{10}/2)}\frac{1}{\Gamma^2(1+\alpha't_{10}/4)}\sfrac{\alpha's_{10}}{4}^{\alpha't_{10}/2}
\equiv {\cal G}(s_{10}, t_{10}).
\end{align}

There are a number of merits in seeing amplitudes in complex spin $j$-plane 
in classical Regge theory \cite{Collins:1977, Forshaw:1997dc,
BaronePredazzi}, as well as in perturbative 
QCD \cite{Ellis:1991qj, Forshaw:1997dc}, and furthermore, 
the $j$-plane description also has a couple of extra advantages 
in the present context.  
For one, we can clarify subtle points in how the derivation of Pomeron 
amplitude based on vertex operator OPE in section 3 of 
\cite{BrowerJHEP0712:0052007} is related to the somewhat heuristic 
modification of the flat spacetime amplitude in finding a scattering 
amplitude on a curved spacetime in section 2 of
\cite{BrowerJHEP0712:0052007}, as we explain shortly. Furthermore, 
we can find a clear guiding principle in the heuristic modification 
process of the scattering amplitude, although this process has not 
been crystal clear so far (in our eyes) in the literature. 

\begin{figure}[tbp]
\begin{center}
\begin{tabular}{ccc}
  \includegraphics[scale=0.65]{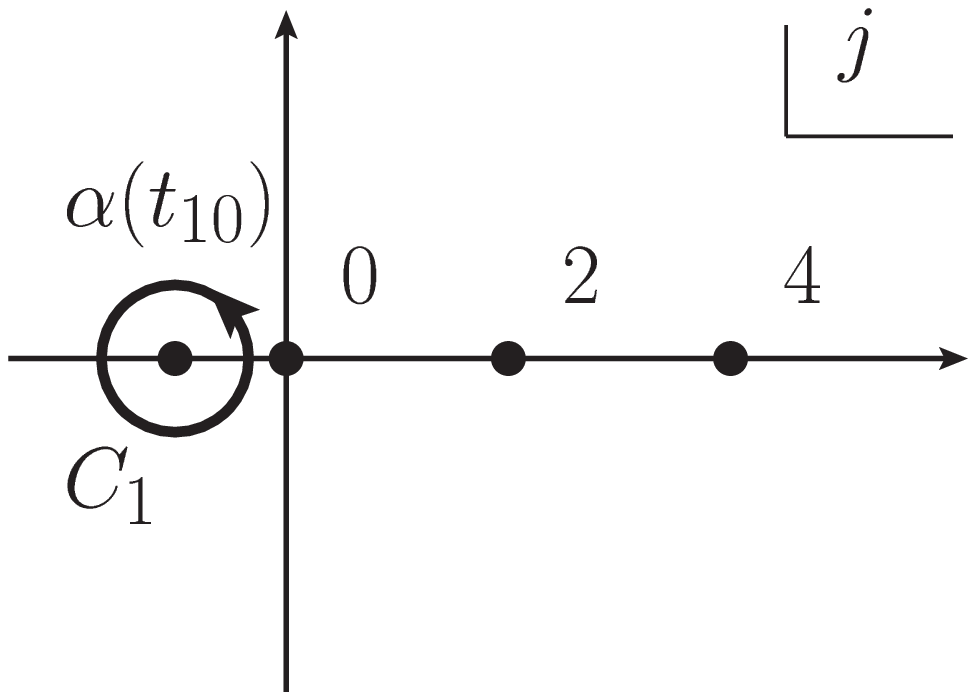} & &
  \includegraphics[scale=0.65]{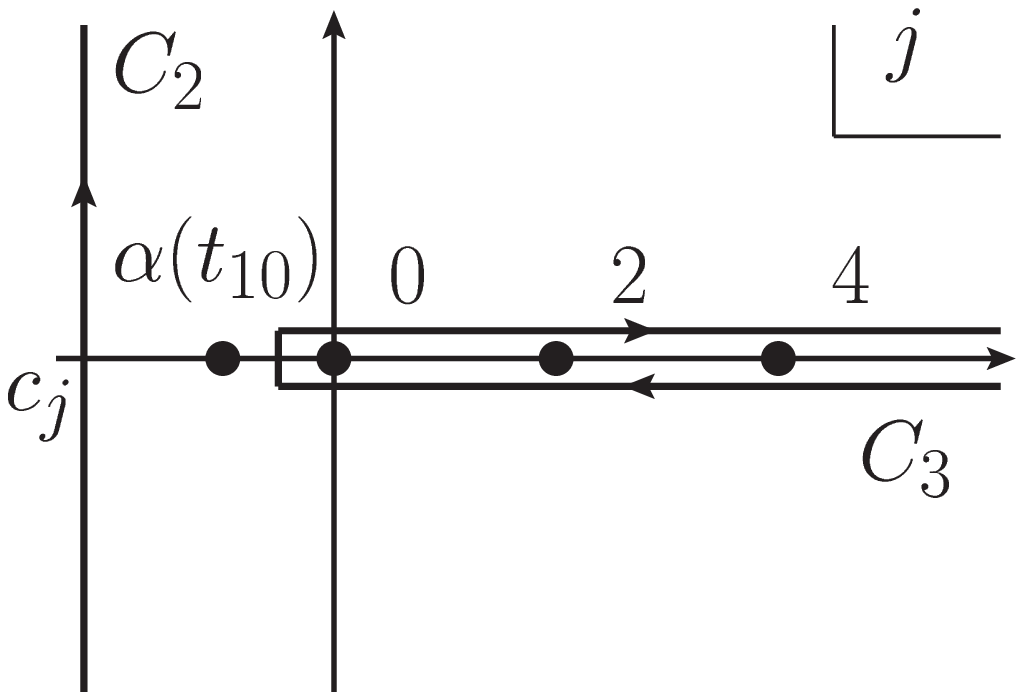}  \\
 (a) & & (b)
\end{tabular}
 \caption{\label{fig:contours} Integration contours $C_1$, $C_2$ and
$C_3$, and various singularities that appear in the complex $j$-plane.
The contour $C_2$ comes from $c_j - i \infty$, avoids the 
$j = \alpha(t_{10})$ singularity to the left, and goes to 
$c_j + i \infty$; $c_j \in \R$ is the real part value of the
 contour of $C_2$ in the asymptotic region (both ends). In order for the integral 
${\cal G}^{(C_2)}(s_{10}, t_{10}; \Omega^{j-\alpha(t_{10})})$ to return
a finite value, we need to take $c_j > -1$. }
\end{center}
\end{figure}

Let us first see, with the $j$-plane description, that 
${\cal G}(s_{10}, t_{10})$ can be decomposed into three pieces. 
First, note that the amplitude (\ref{eq:flat cal g}) is
the same as 
\begin{align}\label{eq:flat g in j}
 {\cal G}(s_{10}, t_{10})=
\frac{1}{2\pi i}\int_{C_1}dj \left(-\frac{\alpha'\pi}{4}\right)
\frac{1+e^{-i\pi j}}{\sin\pi j}
\frac{1}{\Gamma^2(j/2)}\sfrac{\alpha's_{10}}{4}^{j-2}\frac{1}{j-\alpha(t_{10})},
\end{align}
where $\alpha(t_{10}) \equiv 2+\alpha't_{10}/2$, which is the trajectory of graviton,
 and the contour of 
integration in complex $j$-plane $C_1$ is set as in Figure~\ref{fig:contours}.
Mathematically, one can add to the integrand a function of $j$ 
holomorphic at $\alpha(t_{10})$.
In order to see the explicit relation with vertex operator OPE,
we choose the following expression,
\begin{align}
  {\cal G}(s_{10}, t_{10})=
\frac{1}{2\pi i}\int_{C_1}dj \left(-\frac{\alpha'\pi}{4}\right)
\frac{1+e^{-i\pi j}}{\sin\pi j}
\frac{1}{\Gamma^2(j/2)}
\sfrac{\alpha(t_{10})}{j}^2
\sfrac{\alpha's_{10}}{4}^{j-2}\frac{\Omega^{j-\alpha(t_{10})}}{j-\alpha(t_{10})}.
\label{eq:flat g in j2}
\end{align}
We will explain their relation soon,
along with the meaning of $\Omega$,
which  is a real positive number.
Secondly, suppose for now that 
\begin{equation}\label{eq:region}
 \arg (s_{10})=\pi/2. 
\end{equation}
Then the integrand of (\ref{eq:flat g in j2}) vanishes fast enough 
when $|j|$ goes to infinity while satisfying ${\rm Re}\ j>-1$. It is,
thus, possible to change the contour from $C_1$ to $C_3 -
C_2$; see Figure~\ref{fig:contours}. The $\Gamma(j/2)$ factor in the 
denominator is crucial in the convergence of integration at large 
real positive $j$; the condition (\ref{eq:region}) and the choice 
$c_j > -1$ of $C_2$ are necessary for the convergence of integral 
along $C_2$. 
Finally, let us split the integral along $C_3$ into two pieces, 
one with the $\Omega^{j-\alpha(t_{10})}$ factor in the integrand 
replaced by $1$, and the other with the same factor replaced by 
$\left[\Omega^{j-\alpha(t_{10})} - 1\right]$. As a whole, 
${\cal G}(s_{10}, t_{10})$ is given by a sum of three pieces, 
\begin{align}
{\cal G}(s_{10},t_{10}) &
    = {\cal G}^{(C_1)}(s_{10}, t_{10}; \Omega^{j-\alpha(t_{10})}) \notag \\
  & = - {\cal G}^{(C_2)}(s_{10}, t_{10}; \Omega^{j-\alpha(t_{10})} )
    + {\cal G}^{(C_3)}(s_{10}, t_{10}; 1)
    + {\cal G}^{(C_3)}(s_{10}, t_{10}; \Omega^{j-\alpha(t_{10})} - 1).
\label{eq:3contrib}
\end{align} 

Physical meaning of the second term above will be clear. Replacing 
the $j$-plane integral along $C_3$ by residues at the poles 
$j =0,2,4, \cdots$, 
\begin{equation}
{\cal G}^{(C_3)} (s_{10}, t_{10}; 1) = 
  \frac{\alpha'}{2}
  \sum_{j=0,2,4,\cdots} \frac{1}{\Gamma^2(j/2)}\sfrac{\alpha(t_{10})}{j}^2
    \left( \frac{\alpha' s_{10}}{4} \right)^{j-2}
   \frac{1}{j-\alpha(t_{10})},
\label{eq:C3-t-chanel}
\end{equation}
which is regarded as $t$-channel exchange of spin $j$ string
on the graviton trajectory;\footnote{Although the factor $1/(j-\alpha(t_{10}))$
of $j = 0$ term in the summation seems to give rise to $\alpha(t_{10}) = 0$ tachyon pole (i.e., 
$\alpha' t_{10} = -4$), the amplitude ${\cal G}^{(C_3)} (s_{10}, t_{10}; 1)$ actually does not have such a tachyon pole.
 This is because a factor $\frac{1}{\Gamma^2(j/2)}\sfrac{\alpha(t_{10})}{j}^2$ 
cancels the $\alpha(t_{10})=0$ pole in $j=0$ term.}
 the $1/(j-\alpha(t_{10}))$ factor 
is understood as the propagator of a string with mass $m^2=4(j-2)/\alpha'$, 
and this contribution to the total amplitude $K \; G$ is proportional 
to $(s_{10})^j$, as the dominant contribution to the factor $K$ 
in $s_{10} \gg |t_{10}|$ is proportional to $s_{10}^2$. 

Vertex-operator OPE in \cite{BrowerJHEP0712:0052007} in world-sheet 
calculation also yields the same expression as (\ref{eq:C3-t-chanel}).
Let us use $(0, 0)$-picture vertex operators 
\begin{eqnarray}
 {\cal V}_1^{(0,0)}(w,\bar{w}) & = & 
 : \epsilon^{(1)}_{M_1 N_1} \!\! 
  \left[ i \partial X^{M_1} + \frac{\alpha'}{2}(q_1 \cdot \psi)\psi^{M_1}\right]
  \left[ i \bar{\partial} X^{N_1} + \frac{\alpha'}{2}
             (q_1 \cdot \tilde{\psi}) \tilde{\psi}^{N_1}
  \right] e^{i q_1 \cdot X(w)} :, \\  
 {\cal V}_2^{(0,0)}(0,0) & = & 
 : \epsilon^{(2)}_{M_2 N_2} \!\! 
  \left[ i \partial X^{M_2} - \frac{\alpha'}{2}(q_2 \cdot \psi)\psi^{M_2}\right]
  \left[ i \bar{\partial} X^{N_2} - \frac{\alpha'}{2}
              (q_2 \cdot \tilde{\psi}) \tilde{\psi}^{N_2}
  \right] e^{-i q_2 \cdot X(0)} :, 
\end{eqnarray}
for a massless incoming NS--NS string with $(q_1)_{10}^2 = 0$ and 
a massless outgoing NS--NS string with $(q_2)_{10}^2 = 0$. 
\begin{figure}[tbp]
\begin{center}
\begin{tabular}{ccc}
   \includegraphics[scale=0.45]{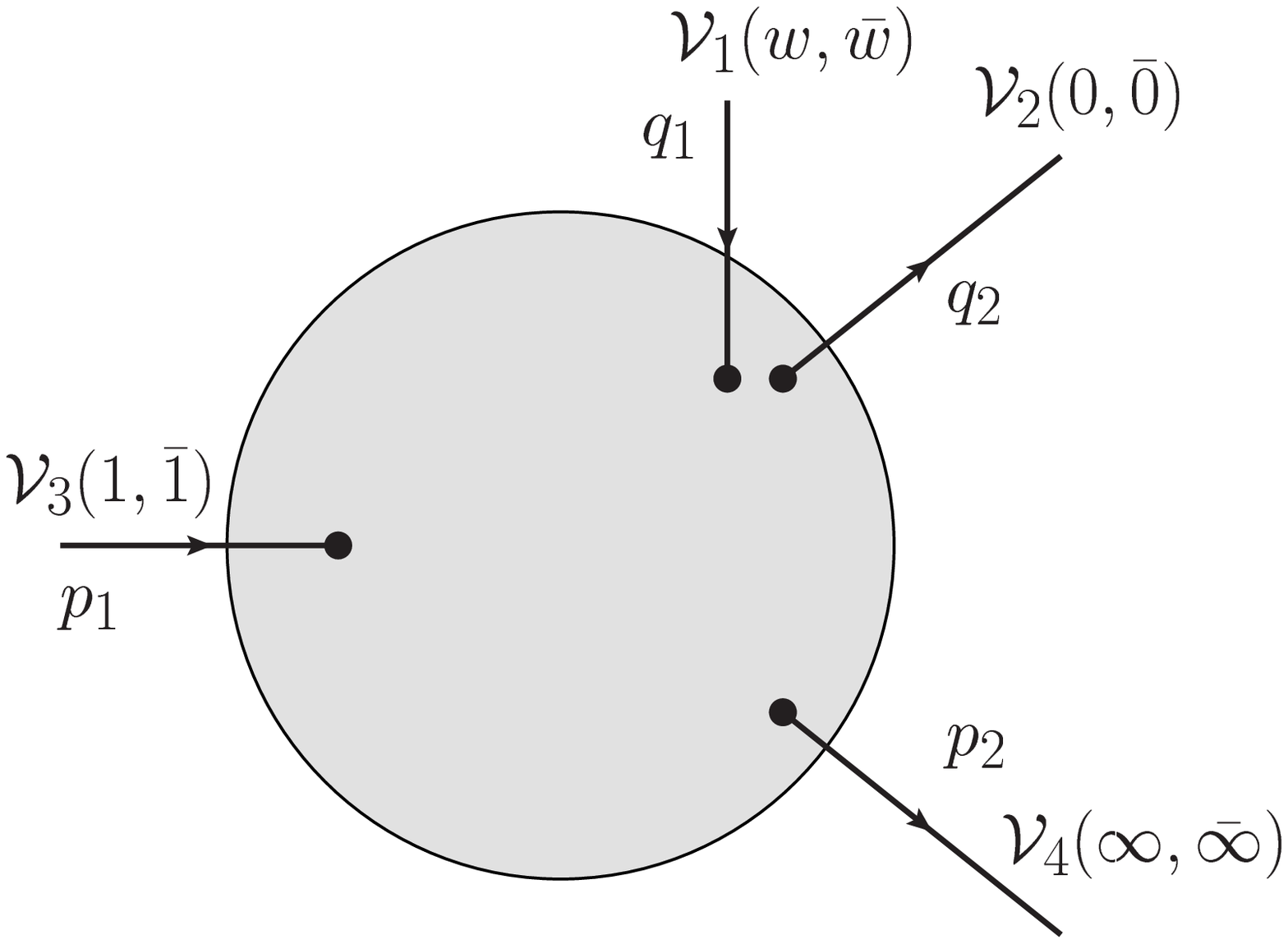} & &
  \includegraphics[scale=0.45]{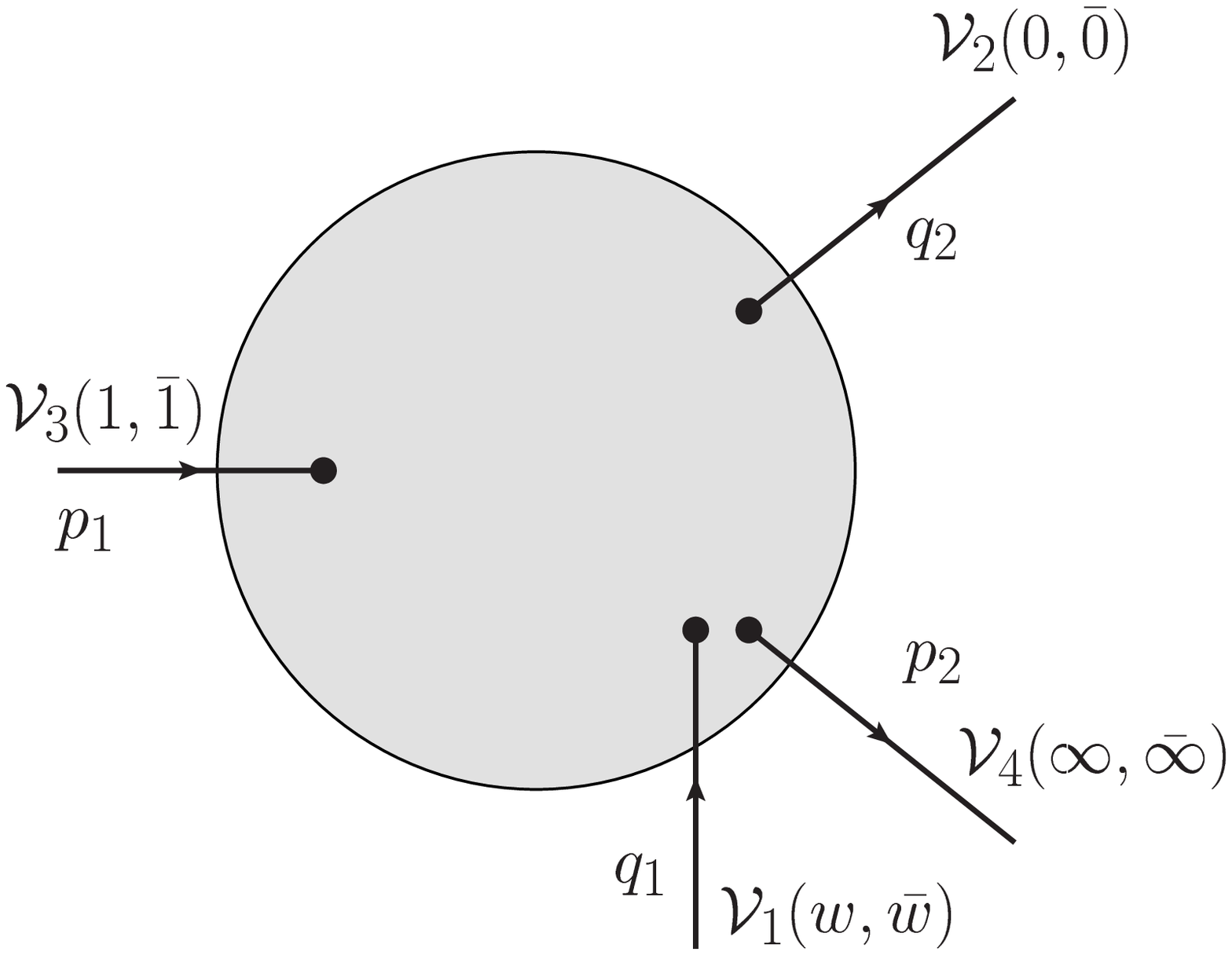}  \\
(a) & & (b)
\end{tabular}
 \caption{
A sphere amplitude describing scattering of four NS--NS string massless
states. Later on, the vertex operators
${\cal V}_3$ and ${\cal V}_4$ are used for incoming and outgoing 
hadron states, and ${\cal V}_1$ and ${\cal V}_2$ used for incoming 
virtual photon and outgoing virtual/real photon for DDVCS, 
respectively. 
${\cal V}_{2,3,4}$ are fixed at $w = 0,1,\infty$ on the $g = 0$ world sheet, 
and the sphere amplitude is obtained by integrating over the
complex coordinate $(w,\bar{w})$ of ${\cal V}_1$ over the sphere. 
Contribution from integration over the $0 \leq |w| \leq 1$ region (a)
 and the one over the $1 \leq |w| \leq \infty$ region (b) can be regarded 
as $t$-channel and $u$-channel amplitudes, respectively. 
} 
\end{center}
\end{figure}
The OPE of these two vertex operators includes the following 
series of operators
\begin{align}
 {\cal V}^{(0,0)}_1(w,\bar w) {\cal V}^{(0,0)}_2(0,\bar 0)  \sim
&
\left[
\left(\epsilon^{(1)}\cdot \epsilon^{(2)} \right)
\sfrac{\alpha '}{2}^2 
\sfrac{\alpha(t_{10})}{2}^2
|w|^{-4}
+\cdots
\right]
\notag \\
&\times|w|^{-\alpha't_{10}/2}
:
e^{iq_1\cdot(w\partial X+\bar w \bar \partial X)}
e^{i(q_1-q_2)\cdot X}
(0,\bar 0):.
\label{eq:vo-OPE}
\end{align}
In the first line, we omitted terms which are less singular than
$|w|^{-4}$ in the $w,\bar w \rightarrow 0$ limit.\footnote{
The $|w|^{-4}$ terms become those with the greatest power of
$s_{10}$, that is, $s_{10}^{\alpha(t_{10})}$.
Less singular terms of OPE, whose coefficients are 
$|w|^{-2}w^{-1}, |w|^{-2}\bar w^{-1}\dots$,
have lower power of $s_{10}$. This leading term in the OPE of string
scattering in 10 dimensions becomes the leading term in small $x$ 
(large $s = W^2$) in the structure function $V_1$ in 4 dimensions 
in section \ref{ssec:polarization}. Leading terms of other structure 
functions in small $x$, however, are determined in 
section \ref{ssec:polarization} without using such an OPE on world sheet.
}
Keeping only $|w|^{-4}$ terms for simplicity,  we obtain,
\begin{align}
  {\cal V}^{(0,0)}_1(w,\bar w) {\cal V}^{(0,0)}_2(0,\bar 0)  \sim
&
\left(\epsilon^{(1)}\cdot \epsilon^{(2)} \right)
\sfrac{\alpha '}{2}^2 
\sfrac{\alpha(t_{10})}{2}^2
\sum_{k,\tilde k\ge 0}
\frac{1}{k!\tilde k!}w^{L_0-1}\bar w^{\tilde L_0-1}{\cal O}_{k,\tilde k}(0,\bar 0),
\label{eq:OPE-of-vertex-operator-2}
\end{align}
where\footnote{We adopt a convention of \cite{Polchinski:1998rr}, where 
$L_0$ and $\tilde{L}_0$ vanish on physical states. 
Thus, $L_0$ and $\tilde{L}_0$ here corresponds to $L_0-1$ and
$\tilde{L}_0-1$ in \cite{BrowerJHEP0712:0052007}.} 
\begin{equation}
 L_0-1 = -\frac{\alpha'}{4} t_{10}^2 + k-2, \qquad 
 \tilde{L}_0-1 = -\frac{\alpha'}{4}t_{10}^2 + \tilde{k}-2,
\end{equation}
are weights of operator ${\cal O}_{k,\tilde k}$,
and 
\begin{align}
 {\cal O}_{k,\tilde k}(0,\bar 0)=:
     \left(i q_1 \cdot \partial X \right)^k
    \left(i q_1 \cdot \bar{\partial} X \right)^{\tilde{k}}
    e^{i (q_1 - q_2)\cdot X}(0,\bar 0):.
\end{align}
The sphere amplitude is given by integrating $w$ over the entire 
complex plane. Only the $L_0=\tilde L_0$ terms remain, and 
\begin{eqnarray}
 && 
   \int^{|w| \leq \Omega} d^2 w \; \langle 
   {\cal V}^{(0,0)}_1(w) \; {\cal V}^{(0,0)}_2(0) \; 
   {\cal V}^{(-1,-1)}_3(1) \; {\cal V}^{(-1,-1)}_4(\infty) 
     \rangle \times |\infty|^4  \label{eq:vo-OPE-4pt}\\
 & \sim & 
\left(\epsilon^{(1)}\cdot \epsilon^{(2)} \right)
\sfrac{\alpha '}{2}^2 
\sfrac{\alpha(t_{10})}{2}^2
  \int_0^{\Omega^2} \!\!\!\! d |w|^2 \sum_{k=0}^\infty 
   \frac{(|w|^2)^{L_0- 1}}{(k!)^2} 
      \langle {\cal O}_{k,k}(0) \; 
           {\cal V}_3(1) \; {\cal V}_4(\infty) \rangle \times |\infty|^4
   \nonumber \\
 & = & 
2\left(\epsilon^{(1)}\cdot \epsilon^{(2)} \right)
\sfrac{\alpha '}{2}^2 
\sum_{j =0,2,4,\cdots} 
  \frac{1}{\Gamma^2(j/2)}
\sfrac{\alpha(t_{10})}{j}^2
   \frac{\left[\Omega^{j-\alpha(t_{10})}-0\right]}
        {j-\alpha(t_{10})}
\label{eq:vo-OPE-4pt-2}
\\ \notag
  & &\quad \quad \quad \quad \quad \quad\quad \quad \quad\quad \quad \quad \quad \quad \quad\quad \quad \quad
      \times \langle {\cal O}_{j/2,j/2}(0) \; 
           {\cal V}_3(1) \; {\cal V}_4(\infty) \rangle \times
	   |\infty|^4.
 \nonumber
\end{eqnarray}
Because $\langle {\cal O}_{j/2,j/2}(0)\;{\cal V}_3(1) \; {\cal V}_4(\infty) \rangle 
\times|\infty|^4
\sim (\alpha's_{10})^j$,
we find that the $|w| \leq 1$ contribution to this world-sheet amplitude 
has the structure of ${\cal G}^{(C_3)}(s_{10}, t_{10}; 1)$ 
in (\ref{eq:C3-t-chanel}).

Now $\Omega$ in (\ref{eq:flat g in j}) has a clear meaning: the cut-off 
of integration of $|w|$. The third term in (\ref{eq:3contrib}), 
${\cal G}^{(C_3)}(s_{10}, t_{10}; \Omega^{j-\alpha(t_{10})} - 1)$, 
corresponds to the $1 \leq |w| \leq \Omega$ contribution of the world-sheet 
amplitude above. Obviously the cut-off $\Omega$ should be taken to
infinity. Thus, the sphere amplitude using 
${\cal V}_1$--${\cal V}_2$ OPE should be proportional to
\begin{equation} \label{eq:flat g t-u}
 {\cal G}(s_{10}, t_{10}) =  
\lim_{\Omega \rightarrow \infty} \left[
   {\cal G}^{(C_3)}(s_{10}, t_{10}; 1-0)
 + {\cal G}^{(C_3)}(s_{10}, t_{10}; \Omega^{j-\alpha(t_{10})}-1) \right].
\end{equation}
There is no contradiction between (\ref{eq:flat g t-u}) and 
(\ref{eq:3contrib}); whenever 
\begin{equation} \label{eq:region-a(t)}
 -1 < \alpha(t_{10}),
\end{equation}
the contour $C_2$ can be chosen as a straight line from 
$c_j - i \infty$ to $c_j + i \infty$ for some $-1 < c_j < \alpha(t_{10})$. 
Now, it is easy to see that 
\begin{equation}
 \lim_{\Omega \rightarrow + \infty} 
 {\cal G}^{(C_2)}(s_{10}, t_{10}; \Omega^{j - \alpha(t_{10})})
 = 0,
\label{eq:g-C2 gone}
\end{equation}
because of the $\Omega^{j-\alpha(t_{10})}$ factor.\footnote{
This also means that the $u$-channel contribution 
${\cal G}^{(C_3)}(s_{10}, t_{10}; \Omega^{j-\alpha(t_{10})} -1 )$ has 
a finite $\Omega \rightarrow + \infty$ limit. Moreover, 
because the total amplitude is also equal to 
\begin{equation}
{\cal G}^{(C_1)}(s_{10}, t_{10}; 1) = {\cal G}^{(C_3)}(s_{10}, t_{10};
 1)
 - {\cal G}^{(C_2)}(s_{10}, t_{10}; 1),
\end{equation}
one can see that the $u$-channel ($1\leq |w| \leq \infty$) contribution 
$\lim_{\Omega \rightarrow \infty} 
\left[{\cal G}^{(C_3)}(s_{10}, t_{10}; \Omega^{j - \alpha(t_{10})})
\right]$ is equal to $- {\cal G}^{(C_2)}(s_{10}, t_{10}; 1)$, at least 
mathematically. The $t$-channel and $u$-channel contributions are 
given by integral along $C_3$ and $-C_2$, respectively, and the total 
amplitude is given by $C_1$.
}
Although each piece of (\ref{eq:3contrib}) is well-defined and such
relations as (\ref{eq:3contrib}, \ref{eq:flat g t-u}, 
\ref{eq:g-C2 gone}) can be established only under the 
conditions (\ref{eq:region}, \ref{eq:region-a(t)}), the expression 
(\ref{eq:flat g in j}) is always well-defined, and it should be regarded 
as analytic continuation of the world-sheet amplitude\footnote{The same
story is in between the Virasoro--Schapiro 
amplitude (\ref{eq:Virasoro-Schapiro}) and 
$\int d^2 w\; |w|^{-\alpha't/2}|1-w|^{-\alpha' s/2}$.} 
(\ref{eq:flat g t-u}) off the kinematical 
constraints (\ref{eq:region}, \ref{eq:region-a(t)}).

We have so far discussed Regge limit of closed string scattering 
amplitude in flat 10-dimensional spacetime, but our true interest is in 
scattering in a curved spacetime $\simeq AdS_5 \times W$.
The prescription of section 2 of \cite{BrowerJHEP0712:0052007} is to rewrite 
the factor $K$ in (\ref{eq:flat scalar k}) by using wavefunctions of 
four external states in the scattering, and replacing $s_{10}$ and
$t_{10}$ in ${\cal G}(s_{10}, t_{10})$ by appropriate differential 
operators acting on the wavefunctions. We also follow this line of
argument, while making a couple of improvements. 

As the expression for $K$ becomes a little messy for dilaton--graviton 
scattering (holographic dual model of DDVCS amplitudes), we postpone 
working out the $K$ factor for this case until 
section \ref{ssec:polarization}. For an easier case, the $K$ factor for 
dilaton--dilaton scattering (holographic dual of elastic glueball
scattering) becomes 
\begin{align}
  K&\simeq \frac{c_s}{2 \kappa_{\rm IIB}^2} \int d^4 x dz d^5 \theta
   \sqrt{- G}
  (\Phi(z)Y(\theta))^2(\Phi'(z)Y'(\theta))^2 (e^{-2A}s)^2
    e^{i (p_1 + q_1 - p_2 -  q_2)\cdot x}, \notag \\
 & = (2\pi)^4\delta^4(p_1+q_1-p_2-q_2)  \notag \\
 & \qquad \qquad 
   \frac{c_s R^5}{2 \kappa_{\rm IIB}^2}
   \int dz \sqrt{-g(z)}\int d^5 \theta \sqrt{g_W(\theta)} 
  (e^{-2A}s)^2 (\Phi(z)Y(\theta))^2(\Phi'(z)Y'(\theta))^2, 
\label{eq:K-reduction-for-glueB-glueB}
\end{align}
where $\Phi(z)Y(\theta)$ and $\Phi'(z) Y'(\theta)$ are normalizable 
wavefunctions of the initial state hadrons, and $c_s$ is a dimensionless
constant of order unity. 
$s_{10}$ in (\ref{eq:flat scalar k}) is replaced by $(e^{-2A} s)$, including 
the warped metric.\footnote{Although $s_{10}$ also contains derivatives in $z$ and
$\theta$, such terms are ignored, because they are suppressed
relatively by of order $\Lambda^2/s$ and are negligible in the 
high-energy scattering.}  

In order to obtain the factor ${\cal G}(s_{10}, t_{10})$ for scattering 
in the curved spacetime $AdS_5 \times W$, $s_{10}$ is replaced 
by the Minkowski part $(e^{-2A} s)$ with the warped metric, just as above 
for the $K$ factor. As for $t_{10}$ in ${\cal G}(s_{10}, t_{10})$, 
on the other hand, one cannot drop such terms as $\partial_z^2$ or 
$\partial_\theta^2$, unless $|t| \gg \Lambda^2$
 \cite{BrowerJHEP0712:0052007}.
Although it is not immediately obvious which derivative operators should 
be used in curved spacetime, we adopt a prescription to go to 
$j$-plane description 
(\ref{eq:flat g in j2}, \ref{eq:C3-t-chanel}, \ref{eq:vo-OPE-4pt-2} \ref{eq:flat g t-u}),
first, and then replace $t_{10}$ in the spin $j$ partial wave by 
\begin{align}
 t_{10}\rightarrow \Delta_j(t) +R^{-2}\nabla^2_{W} + R^{-2}\delta_j,
 \label{eq:t-replace}
\end{align}
where  
\begin{align}
 \Delta_j(t)&=e^{-2A}t + 
     e^{jA}(-g)^{-1/2}\partial_z(-g)^{1/2}e^{-2A}\partial_z e^{-jA}, 
  \label{eq:spin-j-deriv} \\
 \nabla_{W}^2&=(-g_{W})^{-1/2}\frac{\partial}{\partial\theta^a}(-g_{W})^{1/2}(g_{W})^{ab}\frac{\partial}{\partial \theta^b}.
\end{align}
Since the $1/(j-\alpha(t_{10}))$ factor in (\ref{eq:C3-t-chanel}) is
regarded as a propagator, the first two terms in (\ref{eq:t-replace}) 
are the rank-2 differential operators appearing in the equation of motion 
of spin $j$ modes on the curved background.\footnote{
$\Gamma_{zz}^z = - 1/z$ and $\Gamma_{\mu z}^\nu = - \delta_\mu^\nu/z$ 
under the metric (\ref{eq:hwb}).} 
In addition to this reasoning based on local field theory intuition,
it is also possible to determine $j-\alpha(t_{10})(=2L_0)$ 
in direct calculation in worldsheet nonlinear sigma model.
The last term $R^{-2}\delta_j$ is a mass correction;
$\delta_j$ is a constant of order unity, 
and does not contain a derivative. 
The above prescription also can be applied to (\ref{eq:flat g in j}),
because (\ref{eq:flat g in j}) is equivalent to (\ref{eq:flat g in j2}).

Combining both the factors $K$ and ${\cal G}$, we obtain 
\begin{align}
 & A(s,t)  \simeq  \frac{c_s R^5}{2 \kappa_{\rm IIB}^2} 
    \int dz \sqrt{-g} d^5 \theta  \sqrt{g_W} 
 (e^{-2A}s)^2 (\Phi(z)Y(\theta))^2
{\cal G}(e^{-2A}s,t_{10})
(\Phi'(z)Y'(\theta))^2, \\
 & {\cal G}(e^{-2A}s, t_{10})  =  
\frac{1}{2\pi i}\int_{C_1}dj \left(-\frac{\alpha'\pi}{4}\right)
\frac{1+e^{-i\pi j}}{\sin\pi j}
\frac{1}{\Gamma^2(j/2)} \notag \\
 &  \qquad \qquad  \qquad \qquad  \qquad 
  \sfrac{\alpha'e^{-2A}s}{4}^{j-2}
  \frac{1}{j-\alpha(\Delta_j(t)+ R^{-2}(\delta_j +\nabla^2_{W}))}.
\end{align}
Note that all the derivatives $\partial_z$ in $t_{10}$ are placed on the 
right of all the $z$-dependent factors $\sqrt{-g}$ and
$(e^{-2A}s)^2$ in the first line, and $(e^{-2A}s)^{j-2}$ in the third
line; $\Delta_j(t)$ simply acts on the wavefunctions
$(\Phi'(z)Y'(\theta))^2$. This is not asymmetric in 
$[\Phi(z)Y(\theta)]^2$ and $[\Phi'(z) Y'(\theta)]^2$, because the $z$ 
derivative part of $\Delta_j(t)$ is Hermitian under the measure
$dz \sqrt{-g(z)} e^{-2j A(z)}$. 

We will drop $\nabla_W^2$ in the following; this is because $\nabla^2_W$
on a compact space $W$ has non-positive discrete spectrum. Apart from 
the constant mode, whose eigenvalue of $\nabla^2_W$ is zero,
$\nabla^2_W$ eigenvalues are negative and at least of order unity. Thus, 
at high-energy scattering, 
\begin{equation}
 \ln \left(s/\Lambda^2 \right) \gg \sqrt{\lambda}, 
\label{eq:high-e-scatter}
\end{equation}
a factor 
\begin{equation}
\left(\alpha' e^{-2A} s \right)^{\frac{\alpha'}{2}\frac{1}{R^2} \nabla^2_W }
\rightarrow  e^{- \frac{ \ln (s/\Lambda^2) +
                         \ln ((\Lambda z)^2/\sqrt{\lambda})}
                       {2\sqrt{\lambda}} {\cal O}(1)}
\end{equation}
in $(\alpha' e^{-2A}s)^j$ is always suppressed, unless $z$ is extremely
small. The condition (\ref{eq:high-e-scatter}) corresponds to 
exponentially small $x$ in DIS/DDVCS,  
\begin{equation}
 \ln \left(\frac{1}{\sqrt{\lambda}x}\right) \gg \sqrt{\lambda},
\label{eq:exp-small-x}
\end{equation}
which we will assume in later sections. The scattering amplitude now 
has an effective description in 5 dimensions, 
\begin{equation}
 A(s,t) \simeq \frac{c_s c_{\phi} c_{\phi'}}{2 \kappa_5^2}
  \int_0^{1/\Lambda} dz 
   \sqrt{-g(z)} (e^{-2A}s)^2 [\Phi(z)]^2 
    {\cal G}(e^{-2A} s, \Delta_j(t)+ \delta_j /R^2 ) [\Phi'(z)]^2,
\label{eq:5D-descr-glueB-glueB}
\end{equation}
where $c_{\phi}$ and $c_{\phi'}$ are defined in (\ref{eq:def-c-phi}).

The $(\alpha' e^{-2A}s)^j \propto (\alpha' e^{-2A}s)^{\alpha'
\Delta_j(t)/2}$ factor is non-local, and one can rewrite the 
factor ${\cal G}$ as a diffusion kernel by inserting a complete 
system of the operator $\Delta_j(t)$ \cite{BrowerJHEP0712:0052007}.
Eigenfunctions of $\Delta_j(t)$, 
\begin{align}
 \Delta_j(t)\Psi^{(j)}_{i\nu}(t,z)=-\frac{\nu^2+4}{R^2}\Psi^{(j)}_{i\nu}(t,z),
\label{eq:Delta-j-Psi-j}
\end{align}
are given by 
\begin{align}\label{psi}
 \Psi^{(j)}_{i\nu}(t,z)=ie^{A(j-2)}\sqrt{\frac{\nu}{2R\sinh\pi\nu}}\left[
\sqrt{\frac{ I_{-i\nu} (\sqrt{-t}/\Lambda)      }{I_{i\nu} (\sqrt{-t}/\Lambda)}}I_{i\nu}(\sqrt{-t}z)-
\sqrt{\frac{ I_{ i\nu} (\sqrt{-t}/\Lambda)      }{I_{-i\nu} (\sqrt{-t}/\Lambda)}}I_{-i\nu}(\sqrt{-t}z)
\right],
\end{align}
where $I_{\mu}(x)$ is the modified Bessel function.
Here, we imposed a Dirichlet boundary condition at IR in order to keep 
the expressions a little simpler, but essence 
will not change when a Neumann condition is imposed instead.\footnote{
The reflection coefficient $R(\nu, t)$ defined by 
\begin{equation}
 \Psi^{(j)}_{i\nu}(t,z) \propto 
  I_{i\nu}(\sqrt{-t}z) + R(\nu,t) I_{-i\nu} (\sqrt{-t}z)
\end{equation}
becomes $R(\nu, t) = - [I_{i\nu}(\xi_0)/I_{i\nu}(\xi_0)]
|_{\xi_0 = \sqrt{-t}/\Lambda}$ under the Dirichlet boundary condition.
It satisfies $R(-\nu,t) = 1/R(\nu, t)$ and $(1 + R(-\nu,t)) = 0$ for 
$i\nu \in \Z$. (If both $\sqrt{-t}$ and $\nu$ are real valued, 
$R(\nu, t)^* = R(-\nu, t)$, although we do not use this property in 
this article.) 
When the Neumann boundary condition 
$\partial_z [e^{-j A} \Psi^{(j)}_{i\nu}(t,z)] = 0$ is imposed 
instead \cite{BrowerJHEP0712:0052007}, $R(\nu, t)$ defined as above 
also has all of these properties.
} 
For negative\footnote{\label{fn:t-phase} We understand that 
$\sqrt{-t}$ is real positive for negative $t$, that is, 
${\rm arg} \; \sqrt{-t} = 0$. 
In its analytic continuation to real positive $t$ through the upper 
half-plane in complex $t$, then, ${\rm arg} \; \sqrt{-t} = - \pi/2$.} $t$, 
the eigenfunctions $\Psi^{(j)}_{i\nu}$ for eigenvalues $-(\nu^2+4)/R^2$
with $0 \leq \nu \in \R$ form a complete system. 
Replacing $[\Phi'(z)]^2$ by $\int dz' \delta(z-z') [\Phi'(z')]^2$ and 
using a relation 
\begin{align}\label{compl}
 \int_0^\infty d\nu \Psi^{(j)}_{i\nu}(t,z) \Psi^{(j)}_{i\nu}(t,z')=
 \left[(-g)^{1/2} e^{-2jA} \right]^{-1} \delta(z-z'), 
\end{align}
we find that 
\begin{align}
A(s,t) & \simeq \frac{c_s c_{\phi} c_{\phi'}}{2 \kappa_5^2} 
  \int_0^\infty d \nu \frac{1}{2 \pi i} \int_{C_1(\nu)} dj
  \int dz \sqrt{-g(z)} \int dz'\sqrt{-g(z')} e^{-2j A'}
    (e^{-2A}s)^2 [\Phi(z)]^2 \notag \\
  & \qquad \qquad 
  \left(-\frac{\alpha'\pi}{4}\right)
  \frac{1+e^{-i\pi j}}{\sin\pi j}
  \frac{1}{\Gamma^2(j/2)}
  \sfrac{\alpha'e^{-2A}s}{4}^{j-2} \notag \\
\label{eq:5D-descr-glueB-glueB-complete-system-inserted}
& \qquad \qquad \qquad \qquad 
     \frac{1}{j-\left(j_{\nu}+\delta_j/(2\sqrt{\lambda})\right)}
    \Psi^{(j)}_{i\nu}(t,z) \Psi^{(j)}_{i\nu}(t,z') [\Phi'(z')]^2,  
\end{align}
where $e^{A'}=e^{A(z')}$, and 
\begin{align}
 j_{\nu}=2-\frac{\nu^2+4}{2\sqrt{\lambda}}.
\end{align}
With a little more work, the expression above can be made completely 
symmetric for the two initial/final state hadrons; with a notation 
\begin{align}
 \tilde s=e^{-A-A'}s,
\end{align}
one finally arrives at the expression we are familiar with in 
\cite{BrowerJHEP0712:0052007, 
Hatta:2007, Brower2007}:
\begin{align}\label{eq:two scalar amp}
A(s,t) \simeq \frac{c_s}{2 \kappa_5^2}\frac{\pi}{2R^3}
\int dz \sqrt{-g(z)}P(z) \int dz' \sqrt{-g(z')}P(z') {\cal K}(s,t;z,z'), 
\end{align}
where
\begin{align}
 P(z)&=c_\phi (\Phi(z))^2, & P'(z')&=c_{\phi'}(\Phi'(z'))^2
\label{eq:glueB-glueB-IF}
\end{align}
for glueball--glueball elastic scattering.
The Pomeron kernel ${\cal K}(s,t; z, z')$ becomes 
\begin{align}
 {\cal K}(s,t;z,z')=
& -8R\sqrt{\lambda}
 \int_0^\infty d\nu \frac{1}{2\pi i}  \int_{C_1(\nu)}dj \; 
  \frac{1+e^{-i\pi j}}{\sin\pi j} \frac{1}{\Gamma^2(j/2)} \notag \\
\label{eq:pomeron kernel}
& \sfrac{\alpha'\tilde s}{4}^{j}
\frac{1}{j-\left(j_{\nu}+\frac{\delta_j}{2\sqrt{\lambda}}\right)} \; 
e^{-j A(z)}\Psi^{(j)}_{i\nu}(t,z) \; e^{-j A(z')}
 \Psi^{(j)}_{i\nu}(t,z').
\end{align}
Because
$e^{-j A(z)}\Psi^{(j)}_{i\nu}(t,z) = e^{-2 A(z)} \Psi^{(2)}_{i\nu}(t,z)$ 
(see (\ref{psi})), $[e^{-jA} \Psi^{(j)}_{i\nu}]$ in the kernel can be 
replaced by $[e^{-2 A} \Psi^{(2)}_{i\nu}]$ mathematically.
We should remark that this expression (\ref{eq:pomeron kernel})
is valid only in zero-skewedness scattering.

It should be emphasized that we have 
$[e^{-j A}\Psi^{(j)}_{i\nu}(t,z)][e^{-j A} \Psi^{(j)}_{i\nu}(t,z')]$ 
in the Pomeron kernel, not 
$[e^{-j A}\Psi^{(j)}_{i\nu}(t,z)][e^{-j A} \Psi^{(j)}_{i \nu}(t,z')]^*$
that uses complex conjugate.
This is a big difference, because the integrand in (\ref{eq:pomeron kernel})
can be regarded as a holomorphic function of $(j, \nu)$ (except some
pole loci), and the kernel itself is expressed as a holomorphic
integral in the spin--anomalous dimension $(j, \gamma)$
plane;\footnote{This is quite common in perturbative QCD; e.g., 
\cite{Jaroszewicz:1982gr, Ball:1997vf, Forshaw:1997dc, 
BrowerJHEP0712:0052007}. } 
note that $\gamma = \Delta - j -2 = i \nu - j$ \cite{BrowerJHEP0712:0052007}.
The kernel (\ref{eq:pomeron kernel}) is also holomorphic in the momentum
transfer $t$ of hadron scattering, and we will exploit this nature 
in later sections. 

The $j = 2,4, \cdots$ poles from $1/\sin (\pi j)$ correspond to the 
spin $j$ particle exchange, and contribute to the real part of the
kernel. The so called nonsense poles $j = 0,-2, -4, \cdots$ of 
$1/\sin (\pi j)$ in Regge theory are canceled,\footnote{The 
$1/\Gamma^2(j/2)$ factor also renders the total $t$-channel exchange 
contribution ${\cal G}^{(C_3)}(s_{10}, t_{10}; 1)$ finite, and makes 
the decomposition (\ref{eq:3contrib}) possible, as we have already 
mentioned.} and even become zeros 
in the $j$-plane due to the factor $1/\Gamma^2 (j/2)$ in the $j$-plane 
representation of the kernel. The absence of nonsense poles can easily
be traced back to the Virasoro--Schapiro 
amplitude (\ref{eq:Virasoro-Schapiro}) and its Regge limit 
(\ref{eq:flat cal g}). One can also trace the origin of 
$1/\Gamma^2(j/2)$ factor in the argument of vertex operator OPE; 
see (\ref{eq:OPE-of-vertex-operator-2}).

The remaining singularity in the $j$-plane comes from 
$1/(j - j_\nu - \delta_j/(2\sqrt{\lambda}))$.
$\delta_j$ vanishes 
at $j=2$ \cite{BrowerJHEP0712:0052007},
because massless graviton does not receive mass correction even in 
curved spacetime (put another way, energy momentum tensor has vanishing 
anomalous dimension). $j_\nu$ is a function of $\nu$, and hence the zero 
locus of the denominator determines the relation between $j$ and $\nu$.
The contour $C_1(\nu)$ of $j$ integration is around this
($\nu$-dependent) pole, and after integration, this pole locus
determines the large $s$ (high energy) behavior of the scattering
amplitude. 

The pole locus $j_r(\nu)$ in the $j$-plane is given approximately by 
\begin{equation}\label{jr1}
 j_r(\nu) = j_\nu + {\cal O} (\lambda^{-1} ) 
\end{equation}
for $|\nu|\lsim {\cal O}(1)$, and hence for 
$|j_\nu - 2| \lsim {\cal O}(1/\sqrt{\lambda})$. 
For $|\nu| \sim \lambda^{1/4}$, and hence for $|j_\nu| \sim {\cal O}(1)$, 
one can still find the pole locus $j_r(\nu)$ recursively 
\cite{Cornalba:2007fs}:
\begin{align}\label{jr2}
 j_r(\nu) = j_\nu + \frac{\delta_{j = j_\nu}}{2\sqrt{\lambda}} 
  +{\cal O}(\lambda^{-1})=j_\nu+{\cal O}(\lambda^{-1/2}).
\end{align}
This means that the pole locus $j_r(\nu)$ is shifted from $j_\nu$ due to
the $\delta_j$ correction, not by an amount as large as the leading 
${\cal O}(1)$ term, but by of order ${\cal O}(\lambda^{-1/2})$.
The ${\cal O}(1/\sqrt{\lambda})$ corrections 
are just as important as 
the $-2/\sqrt{\lambda}$ term in $j_\nu$.
Finally, for $|\nu| \gg \lambda^{1/4}$, and hence for 
$|j_\nu|\gg {\cal O}(1)$,
it even becomes impossible to try to
find the pole locus $j = j_r(\nu)$ recursively.
It is also known that for $j\gtrsim \sqrt{\lambda}$, $\gamma$-$j$ relation changes drastically \cite{Gubser:2002tv}.
To recap, the high-energy behavior of the Pomeron kernel is described 
fairly well by $s^{j_\nu}$ for $|\nu| \lsim {\cal O}(1)$, but the 
exponent $j_r(\nu)$ begins to deviate from $j_\nu$ 
quadratic in $\nu$, when $|\nu|$ becomes comparable to $\lambda^{1/4}$. 
This is equivalent to 
\begin{equation}
 \left(\frac{|\nu|}{R} \right)^2 \sim \frac{1}{\alpha'}; 
\label{eq:Pomeron-KK-go-stringy}
\end{equation}
loosely speaking, that is when $|\nu|/R$, ``the Kaluza--Klein momentum of 
Pomeron in the holographic radius direction'' becomes comparable to 
the string scale.\footnote{Another condition $|s_{10}/t_{10}|\gg 1$, 
which was used already at (\ref{eq:Regge-VS-10D}), is satisfied for
$\nu$ smaller than or saturating (\ref{eq:Pomeron-KK-go-stringy}); 
this is because $|\alpha' t_{10}|$ remains 
$|\alpha' \Delta_j(t)| \sim (\nu^2 + 4)/\sqrt{\lambda} \lesssim {\cal
O}(1)$, while we consider $\alpha' s_{10} \sim
[s/\Lambda^2]/\sqrt{\lambda} \gg e^{\sqrt{\lambda}}/\sqrt{\lambda} \gg 1$.
}

\subsection{Extracting Structure Functions of DDVCS Amplitudes}
\label{ssec:polarization}

Although the Pomeron kernel (\ref{eq:pomeron kernel}) is universal 
for all the hadron scattering processes at high energy (small $x$), 
the impact factors 
$P(z)$ and $P'(z')$ should be chosen for individual processes. 
$P(z)$ in (\ref{eq:normalizable mode}, \ref{eq:glueB-glueB-IF}) can 
be used for elastic scattering 
of two scalar glueballs \cite{BrowerJHEP0712:0052007, Brower2007}.
Two independent structure functions of DIS cross section are also 
expressed as in (imaginary part of) (\ref{eq:two scalar amp}), using 
(\ref{eq:normalizable mode}, \ref{eq:glueB-glueB-IF}) for the target 
hadron impact factor $P'(z')$; the impact factors $P(z)$ for virtual 
photon have also been determined for the two structure functions 
\cite{PolchinskiJHEP0305:0122003}. 
In the case of  DDVCS amplitudes for a scalar target hadron, there 
are five independent structure functions, and we need to determine 
the impact factors $P(z)$ for these individual structure functions. 

As a holographic model of DDVCS scattering, we use graviton--dilaton 
scattering, as we have announced in section \ref{sec:model}. In order to 
determine the impact factors for the DDVCS scattering, it is easiest 
to start from the known form of the factor $K$ in (\ref{eq:KandG}) 
for the graviton--dilaton scattering, and carry out the process 
corresponding to (\ref{eq:K-reduction-for-glueB-glueB}), 
the same strategy as in \cite{PolchinskiJHEP0305:0122003}.
For DDVCS, 
\begin{align} \label{eq:K in 6 dim}
 &K\simeq (2\pi)^4\delta^4(p_1+q_1-p_2-q_2)\notag \\
&\frac{c'_s}{2 \kappa_{\rm IIB}^2}
 \int dz \sqrt{-g}\int d^5\theta R^5\sqrt{g_{W}}v_a(\theta)v^a(\theta)
 F_1^{\rho m}(z)F_{2\;m}^\sigma(z)\{(p_1)_\rho(p_2)_\sigma+ (p_2)_\rho (p_1)_\sigma\}(\Phi(z)Y(\theta))^2,
\end{align}
where $c'_s$ is a constant of order unity; 
$F_1$ and $F_2$ should be understood as wavefunctions 
in (\ref{eq:wf for Fmn}, \ref{eq:wf for Fmz}) except the plane wave part 
already taken into account in the four momentum conservation. 
For the $F_1$ for incoming virtual photon and $F_2$ for outgoing virtual 
photon, the wavefunction with $q_\mu = (q_1)_\mu$ and the one with 
$q_\mu = (-q_2)_\mu$ should be used, respectively. 
We have kept only the terms where four momenta of graviton $(q_\rho)$'s 
and dilaton $(p_\rho)$'s are contracted in (\ref{eq:K in 6 dim}), because 
such terms dominate in high-energy scattering, as 
in \cite{PolchinskiJHEP0305:0122003}. 
$p_1$ and $p_2$ are made symmetric in $\{\cdots\}$ in (\ref{eq:K in 6 dim}),
 because the polarization of the graviton propagator is symmetric.
Inserting the Regge limit process-universal amplitude ${\cal G}$ to the 
right of $\sqrt{-g}$ and $(e^{-2A} q \cdot p)^2$, we find that 
\begin{align}\label{eq:T no2}
  \epsilon_1^\mu T_{\mu\nu} (\epsilon_2^\nu)^\ast
 \simeq &\frac{c'_s R^5}{2 \kappa_{\rm IIB}^2}
  \int dz \sqrt{-g}\int_W d^5\theta\sqrt{g_W}
     v_a v^a  g^{mn}(F_1)_{\rho' m}(z)(F_2)_{\sigma' n}(z) \notag\\
&  e^{-2A}\eta^{\rho' \rho} e^{-2A} \eta^{\sigma' \sigma}
   \{(p_1)_\rho(p_2)_\sigma+ (p_2)_\rho (p_1)_\sigma\}
   {\cal G}(e^{-2A}s, t_{10})(\Phi(z)Y(\theta))^2.
\end{align}
For DDVCS at exponentially small 
$x = - q^2/(2 q \cdot p) \sim - q^2/(p+q)^2 = q^2/s$, (\ref{eq:exp-small-x}),  
only the constant mode on $W$ contributes, and the expression above 
is reduced to an effective description on 4+1 dimensions, just like 
in (\ref{eq:5D-descr-glueB-glueB}, \ref{eq:two scalar amp});
$c_s c_\phi c_{\phi'}$ is now replaced by $c'_s c_A c_\phi$. 

We are now ready to read off the impact factors for the DDVCS amplitude.
The impact factor for the target hadron is the same as in the elastic 
scattering of the target hadrons, and $P'(z') = P_{hh}(z')$ is the same 
as (\ref{eq:glueB-glueB-IF}). The impact factor for the virtual photon 
is 
\begin{align}
 P(z)= c_A \; R^2 s^{-2} g^{mn} (F_1)^{\rho}_{\; m} (z)(F_{2})^\sigma_{\; n}(z)
    \{(p_1)_\rho(p_2)_\sigma+ (p_2)_\rho (p_1)_\sigma\}.
\end{align}
It can be decomposed into the five structure functions 
of (\ref{eq:structure functions of Compton tensor}); the result is as follows:  
\begin{align}
 V_1&\simeq \frac{1}{2} I_1, &
 V_2&\simeq \frac{2x^2}{q^2}(I_0+I_1), & 
 V_3&\simeq \frac{x^2}{2q^2}(I_0+I_1), \notag \\
 V_4&\simeq \frac{x}{q^2}I_1,&V_5&\simeq \frac{x}{q^2}I_1,& & 
\label{eq:DVCS-polarization}
\end{align}
where we treated all of $(\Delta^2/q^2 = -t/q^2)$ and $x$ 
to be much smaller than unity.\footnote{We have already assumed $\eta =
0$.} Here, 
\begin{equation}
 I_i (x, \eta, t, q^2) \simeq \frac{c'_s}{2 \kappa_5^2}\frac{\pi}{2 R^3} 
  \int dz \sqrt{-g(z)} \int dz' \sqrt{-g(z')} 
   P^{(i)}_{\gamma^\ast\gamma^\ast}(z) \; {\cal K}(s,t,z,z') \; 
   P_{hh}(z'),
\label{eq:Ii-def}
\end{equation}
with 
\begin{eqnarray}
 P^{(1)}_{\gamma^\ast\gamma^\ast}(z) & = & c_J^2 R^2 e^{-2 A(z)}
   [(q_1z)(K_1(q_1z)][(q_2 z)K_1(q_2 z)], \label{eq:P1-def}\\
P^{(0)}_{\gamma^\ast\gamma^\ast}(z) & = & \frac{c_J^2 R^2 e^{-2 A}}{q^2}
   [(q^2_1z)(K_0(q_1z)][(q^2_2 z)K_0(q_2 z)].  \label{eq:P0-def}
\end{eqnarray}
When one ignores terms that are suppressed by powers of $x$, 
as we have done so far, the five structure functions are in fact given by 
only two contributions, $I_0$ and $I_1$.
Although the longitudinal and transverse polarization of photon in the 
incoming $\gamma^*(q_1) + h(p_1)$ beam axis is different from 
those in the outgoing $\gamma^*(q_2) + h(p_2)$ beam axis in the presence
of non-zero momentum transfer, they become approximately the same in small 
angle (and hence in small $x$) scattering; the two beam axes are
precisely the same in the $t = 0$ limit.
In this sense, $I_0$ and $I_1$ correspond to the amplitude of the
virtual photons with ``longitudinal'' and ``transverse'' 
polarizations, respectively. 

\section{DDVCS Amplitude and GPD at Small $x$ in Gravity Dual}
\label{sec:amplitude}

Now that concrete expressions are given to the (Pomeron contribution 
to the) structure functions of  DDVCS 
amplitude (\ref{eq:DVCS-polarization}, \ref{eq:Ii-def}), with an explicit 
expression for the Pomeron kernel (\ref{eq:pomeron kernel}), 
let us evaluate the integrals to get physics out of them. 
The momentum transfer $t$ dependence\footnote{
For very large momentum transfer $-t \gg \Lambda^2$ in the QCD in the
real world, however, perturbative QCD can be used to argue rough scaling 
behavior \cite{Brodsky:1973kr, Matveev:1973ra, Brodsky:1974vy, 
Sivers:1975dg, Donnachie:1979yu} of the  DDVCS amplitude. } 
of  DDVCS amplitude at small momentum transfer 
$0 \leq -t \lesssim \Lambda^2$ is highly non-perturbative information, 
and this is where gauge/gravity dual can play an important role.

We will first focus on {\it scattering amplitude} (i.e., structure
functions) in sections \ref{ssec:t-space}--\ref{ssec:real-part}; 
the imaginary part of the amplitude is studied in 
sections \ref{ssec:t-space} and \ref{ssec:b-space}, which 
sheds a light on non-perturbative form of the parton
distribution in the transverse direction at small $x$ \cite{Burkardt:2002hr}.
The real part of the amplitude is described in section
\ref{ssec:real-part}. We will argue in section \ref{ssec:GPD}
that {\it GPD} can be defined as an inverse Mellin transformation of 
operator matrix element and is calculable even in gravity dual; the 
structure functions and the scattering amplitude as a whole are 
given in the convolution form involving GPD, just like in the QCD 
factorization formula. 

It is useful in studying the DDVCS amplitude in the generalized 
Bjorken regime (\ref{eq:double deeply virtual}, \ref{eq:Bjorken-2}) 
to write down the Pomeron kernel explicitly as follows.
Carrying out $j$ integral around the pole $j = j_r(\nu)$, and 
using the explicit form of $\Psi^{(j)}_{i\nu}(t,z)$ in (\ref{psi}) 
for the hard wall model, we obtain
\begin{align}
  {\cal K}(s,t;z,z')  &= 4\sqrt{\lambda}e^{-2A-2A'}
\int^\infty_{-\infty} d\nu 
 \left[ - \frac{1+e^{-i\pi j_\nu}}{\sin\pi j_\nu} \right]
  \frac{1}{\Gamma^2(j_\nu/2)}
  \left(\frac{\alpha' \hat s}{4}\right)^{j_\nu} \notag \\
    \frac{\nu}{\sinh \pi\nu}&\left[
      I_{i\nu}(\sqrt{-t}z)I_{-i\nu}(\sqrt{-t}z')
    - \frac{I_{-i\nu}(\sqrt{-t}/\Lambda)}{I_{i\nu}(\sqrt{-t}/\Lambda)}
          I_{i\nu}(\sqrt{-t}z)I_{i\nu}(\sqrt{-t}z')
                               \right], 
\label{eq:kernel in hw1} 
\end{align}
\begin{align}
\qquad \qquad &=
4\sqrt{\lambda}e^{-2A-2A'}
\int^\infty_{-\infty} d\nu 
 \left[ - \frac{1+e^{-i\pi j_\nu}}{\sin\pi j_\nu} \right]
  \frac{1}{\Gamma^2(j_\nu/2)}
  \left(\frac{\alpha' \tilde s}{4}\right)^{j_\nu} \notag \\
    \frac{2i\nu}{\pi}&I_{i\nu}(\sqrt{-t}z)
\left[K_{i\nu}(\sqrt{-t}z')-\frac{K_{i\nu}(\sqrt{-t}/\Lambda)}{I_{i\nu}(\sqrt{-t}/\Lambda)}I_{i\nu}(\sqrt{-t}z')\right].
\label{eq:kernel in hw2}
\end{align}
Although both of the second lines of 
(\ref{eq:kernel in hw1}, \ref{eq:kernel in hw2}) are equivalent to 
(\ref{eq:kernel-in-hw-3}) when integrated over $\nu$, 
those in (\ref{eq:kernel in hw1}, \ref{eq:kernel in hw2}), which is not
symmetric under the exchange of $z$ and $z'$, turn out to be a little 
more convenient than the $z \leftrightarrow z'$ symmetric expression
(\ref{eq:kernel-in-hw-3}) in evaluating the DDVCS 
amplitude;\footnote{The second line of 
(\ref{eq:kernel in hw1}, \ref{eq:kernel in hw2}) 
can also be written as 
\begin{equation}
\frac{2\nu}{\pi^2} \left[ 
  \sinh (\pi \nu) K_{i\nu}(\sqrt{-t}z) K_{i\nu}(\sqrt{-t} z') 
 - i \pi 
  \frac{K_{i\nu}(\sqrt{-t}/\Lambda)}{I_{i\nu}(\sqrt{-t}/\Lambda)}
  I_{i\nu}(\sqrt{-t}z) I_{i\nu}(\sqrt{-t}z')
                   \right]. 
\label{eq:kernel-in-hw-3}
\end{equation}
The second term vanishes in the $\Lambda \rightarrow 0$ limit for 
fixed $z, z', s, t$, and the Pomeron kernel on $AdS_5$ 
in \cite{Brower2007a} is reproduced; the $\Lambda\rightarrow 0$ limit of the Pomeron kernel
in this article is different from the one in \cite{Brower2007a} only 
by a factor of $(4\sqrt\lambda)^{j-2}/\Gamma^2(j/2)$, which becomes 
$1$ at $j = 2$. 
} dominant contribution to the amplitudes comes from small $z$ region ($z
\lesssim 1/q \ll 1/\Lambda$) because of the virtual photon wavefunctions 
localized toward UV boundary, and for such a small value of
$(\sqrt{-t}z)$, the $I_{i\nu}(\sqrt{-t}z)$ and $I_{-i \nu}(\sqrt{-t}z)$
terms in $\Psi^{(j)}_{i\nu}(t,z)$ have quite different behavior as a
function of $(i\nu)$. $I_{i\nu}(\sqrt{-t}z)$ decreases rapidly toward 
positive ${\rm Re }\: (i\nu)$, while $I_{-i\nu}(\sqrt{-t}z)$ toward negative
${\rm Re }\: (i\nu)$. This is why the $I_{-i\nu}(\sqrt{-t}z)$ term has been turned into 
$I_{i\nu}(\sqrt{-t}z)$ in (\ref{eq:kernel in hw1}, 
\ref{eq:kernel in hw2}) by relabeling $-\nu$ by $\nu$.

\subsection{Momentum Transfer Dependence of the Imaginary Part}
\label{ssec:t-space}

The imaginary parts of the structure functions simply come from 
the imaginary parts of the integrals $I_1$ and $I_0$ 
(\ref{eq:Ii-def}), and their imaginary parts come from the imaginary 
part of $[1+e^{-i \pi j}]$ in the Pomeron kernel.
It must be straightforward to substitute the expression of the kernel 
above into (\ref{eq:Ii-def}) and evaluate them for 
kinematical variables of our interest; References \cite{
PolchinskiJHEP0305:0122003, Hatta:2007} (see the reference list of 
\cite{Brower:2010wf} for other articles) have done that for purely 
forward case $t = 0$, and we can just carry out a similar procedure 
of calculation for $ 0 \leq -t$ case as well. 
Before doing so, however, we find it worthwhile to write down the 
amplitude in the following form, which leads us to a better theoretical
understanding of the $t \neq 0$ amplitude.

Exploiting the kinematical constraint of the generalized Bjorken regime 
(\ref{eq:Bjorken-2}), we expand $I_{i\nu}(\sqrt{-t}z)$ in the Pomeron 
kernel (\ref{eq:kernel in hw2}) in a power series and keep only the 
first term. Contributions to $I_i$ from the higher order terms are 
suppressed by powers of $(-t/q^2)$, because the integration over the 
holographic coordinate $z$ is dominated by the region $z \lesssim 1/q$. 
Ignoring the higher order terms is like dropping higher twist
contributions in perturbative QCD. One can then see that the 
structure functions can be written as 
\begin{align}
 I_i (x, \eta = 0, t, q^2) & \simeq
  \sqrt{\lambda} \int^\infty_{-\infty} d\nu 
  \left[ - \frac{1 + e^{-\pi i j_\nu}}{ \sin \pi j_\nu}\right]
  \frac{1}{\Gamma^2(j_\nu/2)} \; 
  C^{(i)}_{i \nu} \; A_{hh}, 
  \label{eq:Im-Ii-cm} 
\end{align}
where 
\begin{eqnarray}
 C^{(i)}_{i\nu} & = & \frac{1}{R^3} \int dz \sqrt{-g(z)} 
   P^{(i)}_{\gamma^* \gamma^*}(z) e^{-2A(z)} 
   \left(\frac{z}{R}\right)^{i\nu} (R z)^{j_\nu}, 
 \label{eq:C(i)-def} \\
 A_{hh} & \simeq  &  
 \frac{c'_s}{\kappa_5^2} \int dz' \sqrt{-g(z')} P_{hh}(z')
   \left[\frac{ e^{-2A(z')} W^2}{4 \sqrt{\lambda}}\right]^{j_\nu}
   \nonumber \\
  & & \qquad 
   \left[\frac{e^{(j_\nu - 2)A(z')}}{K_{i\nu}(\sqrt{-t} R)} 
      \left( K_{i\nu}(\sqrt{-t}z')
             - \frac{K_{i\nu}(\sqrt{-t}/\Lambda)}
                    {I_{i\nu}(\sqrt{-t}/\Lambda)}
               I_{i\nu}(\sqrt{-t}z')\right)\right]. 
\label{eq:Ahh-def}
\end{eqnarray}
We will see in section \ref{sssec:regge} that $C^{(i)}_{i\nu}$ 
and $A_{hh}$ correspond to OPE coefficient and matrix element of 
a ``twist-2'' spin $j = j_\nu$ operator\footnote{
Because of the factor $1/K_{i\nu}(\sqrt{-t}R)$ in the integrand of (\ref{eq:Ahh-def}),
the second line in (\ref{eq:Ahh-def}) is normalized
to be unity at UV boundary $z'=R \:(\ll 1/\sqrt{-t})$.
A careful reader may notice that ${\rm Re}\: (i\nu)>0$ is assumed in the expression of (\ref{eq:Ahh-def}).
This assumption gives rise to no problem, since throughout this article
we always estimate the $\nu$ integral (\ref{eq:Im-Ii-cm})
in the lower half $\nu$-plane.

}. 

The two factors $C^{(i)}_{i\nu}$ and $A_{hh}$ depend on kinematical 
variables $q^2$ and $x$ as follows:
\begin{equation}
 C^{(i)}_{i\nu} \simeq c_J^2 \frac{1}{(q R)^{\gamma_\nu}} 
   \frac{1}{(q^2)^{j_\nu}} \bar{c}^{(i)}_{i\nu}, 
\end{equation}
where $\gamma_\nu = i \nu - j_\nu$, and 
\begin{align}
 A_{hh}  \simeq  
  c'_s \left(\frac{W^2}{4\sqrt{\lambda}}\right)^{j_\nu}
  \left(R \Lambda \right)^{\gamma_\nu} 
  g^h_{i\nu}(\sqrt{-t}/\Lambda)
   \simeq 
  c'_s
  \left(\frac{1}{4\sqrt{\lambda}x}\right)^{j_\nu} 
  \left(q^2 \right)^{j_\nu}
  \left(R \Lambda \right)^{\gamma_\nu} 
  g^h_{i\nu}(\sqrt{-t}/\Lambda); \label{eq:Ahh--gh}
\end{align}
$\bar{c}^{(i)}_{i\nu}$ is a constant of order unity that depends only 
on $\nu$ (when $\eta = 0$, so that $q_2 = q_1 \simeq q$), while 
the $t$ dependence of the structure functions $I_i$ now remains 
only in a dimensionless factor $g^h_{i\nu}(\sqrt{-t}/\Lambda)$ 
whose definition can be read out from (\ref{eq:Ahh-def}, \ref{eq:Ahh--gh}).
Therefore, the imaginary part of the structure functions are given by 
\begin{align}
{\rm Im} \; I_i & \simeq
   c'_s \sqrt{\lambda} \int^\infty_{-\infty} d\nu 
  \frac{c_J^2}{\Gamma^2(j_\nu/2)} 
  \sfrac{1}{4\sqrt{\lambda}x}^{j_\nu}\sfrac{\Lambda}{q}^{\gamma_\nu}
  \bar{c}^{(i)}_{i\nu} \:g^h_{i\nu}(\sqrt{-t}/\Lambda).
\label{eq:Im-Ii}
\end{align}
%

\subsubsection{Small momentum transfer: $-t \lesssim \Lambda^2$}
\label{sssec:small-(-t)}

Now let us evaluate the amplitudes (\ref{eq:Im-Ii}), first, for the 
case with small momentum transfer 
$0 \leq -t \lesssim \Lambda^2$. 
The purely forward amplitude (i.e., one for the DIS total cross section)
is a part of this story. 
The $\nu$ integration can be evaluated by the saddle point method 
for exponentially small $x$ (\ref{eq:exp-small-x}), just like in  
\cite{Hatta:2007}.
The factor $g^h_{i\nu}(\sqrt{-t}/\Lambda)$ has a non-zero finite 
(dimensionless and ${\cal O}(1)$) limit when $-t \rightarrow 0$, and 
$g^h_{i\nu}(\sqrt{-t}/\Lambda)$ is a slowly changing 
function of $(i\nu)$, unless $\Lambda^2 \ll -t$. 
Thus, large $\nu$ dependence comes only from 
\begin{align}
 \sfrac{1}{\sqrt{\lambda}x}^{j_\nu}
\sfrac{\Lambda}{q}^{\gamma_\nu}
\label{eq:saddle-point-determinating-part}
\end{align}
in (\ref{eq:Im-Ii}). The saddle point $\nu^*$ is determined 
by the kinematical variables\footnote{
As we discussed around (\ref{jr2}),
the expression of the kernel (\ref{eq:kernel in hw1})
is valid for $|\nu|\lesssim \lambda^{1/4}$.
Therefore, the kinematical variables are restricted within the following region:
\begin{align}
 |i\nu^\ast|&\lesssim \lambda^{1/4} \Leftrightarrow 
\lambda^{-1/4}\ln\sfrac{1}{\sqrt{\lambda}x}\gtrsim \ln\sfrac{q}{\Lambda}.
\label{eq:cond-for-inuast<lmabda^1/4}
\end{align}
}
 $x$ and $q^2$ as in 
\begin{align}\label{eq:nu ast for small -t}
 i\nu^\ast \left(q / \Lambda, x, -t \lesssim \Lambda^2 \right)
   = {\sqrt{\lambda}}
             \frac{
                \ln(q/\Lambda)
                  }
                  {
                \ln\sfrac{q/\Lambda}{\sqrt{\lambda}x}
}, 
\end{align}
and the amplitudes approximately become 
\begin{align}
 {\rm Im}\: I_i \simeq 
   c'_s \sqrt{\lambda} \frac{c_J^2}{\Gamma^2(j_{\nu^\ast}/2)}
   \left(\frac{\ln\frac{q/\Lambda}{\sqrt{\lambda}x}}{2\pi \sqrt{\lambda}
}\right)^{-1/2}
\sfrac{1}{4\sqrt{\lambda}x}^{j_{\nu^\ast}}\sfrac{\Lambda}{q}^{\gamma_{\nu^\ast}}
\bar{c}^{(i)}_{i{\nu^\ast}}\:g^h_{i{\nu^\ast}}(\sqrt{-t}/\Lambda).
\label{eq:Ii-saddle-point}
\end{align}
This leading order expression of the saddle point approximation 
(\ref{eq:Ii-saddle-point}) can be improved by going to higher order; 
those terms would give rise to corrections that are suppressed
relatively by powers of 
$\sqrt{\lambda}/\ln[(q/\Lambda)/(\sqrt{\lambda}x)]$.

Ignoring all the factors of order unity and the Gaussian measure 
of the saddle point approximation, we find that the DDVCS amplitude 
for small momentum transfer is roughly of the form
\begin{align}\label{eq:J for small -t}
 {\rm Im} \; I_i(x,\eta=0, -t \lesssim \Lambda^2, q^2)\sim 
    \sfrac{1}{\sqrt{\lambda}x}^{j_{\nu^\ast}}
    \sfrac{\Lambda}{q}^{\gamma_{\nu^\ast}}
    =
    \sfrac{1}{\sqrt{\lambda}x}^{j_0}
    \sfrac{q}{\Lambda}^{j_0}
    e^{- \frac{\sqrt{\lambda} [ \ln(q/\Lambda)]^2}
              {2 \ln ((q/\Lambda)/\sqrt{\lambda} x) }}. 
\end{align}
These results can also be used for the DIS structure functions, 
with $F_1(x,q^2) = {\rm Im} \; I_1/2$, 
and $F_2(x,q^2) = {\rm Im} \; [x (I_0 + I_1) ]$ in the purely forward
limit $t = 0$ and $\eta = 0$.

To characterize the $q^2$ dependence and $x$ dependence of 
the DDVCS and DIS structure functions, let us introduce effective 
exponents, as often done in phenomenological analysis of structure 
functions.
\begin{align}
 \gamma_\text{eff.}(x,t,q^2)&=\frac{\partial \ln
 [x \; I_i(x,\eta=0,t,q^2)]}{\partial \ln (\Lambda/q)}, \qquad &
 \lambda_\text{eff.}(x,t,q^2)&=\frac{\partial \ln [x
 I_i(x,\eta=0,t,q^2)]}{\partial \ln (1/x)}.
\label{eq:def-exponents}
\end{align}
From (\ref{eq:J for small -t}), we find that 
\begin{align}
 \gamma_\text{eff.}(x,t,q^2)&=\gamma_{\nu^\ast}, & \qquad 
\lambda_\text{eff.}(x,t,q^2)&=j_{\nu^\ast} -1.
\label{eq:effective-exponent}
\end{align}
Both $\gamma_{\rm eff.}$ and $\lambda_{\rm eff.}$, and hence the 
$q^2$ and $\ln (1/x)$ evolutions, are controlled by the saddle point
value $i \nu^*$; the saddle point value $i\nu^* \in \R_{\geq 0}$ in 
(\ref{eq:nu ast for small -t}) becomes large for large $q^2$ and 
decreases to zero for smaller $x$. \label{page:gamma-lambda}

The effective anomalous dimension $\gamma_{\rm eff.} = i\nu^* - j_{\nu^*} \sim 
(i\nu^* - 2)$ for a given $q^2$ is positive for larger $x$, 
and negative for smaller $x$. This means that the (generalized) 
parton density decreases in $q^2$ evolution for larger $x$
and increases for smaller $x$. For a given value of $x$, 
$\gamma_{\rm eff.}$ turns from negative to positive as $q^2$ increases, 
and the (generalized) parton density at that value of $x$ begins to decrease;
the parton splitting from $x$ to smaller $x'$ becomes faster
than the splitting from larger $x'$ to $x$.
This is precisely the behavior expected in
\cite{PolchinskiJHEP0305:0122003}.
Note that the essence of seeing this expected behavior is in 
keeping the $(q,x)$ dependence of the saddle point value $\nu^*$ in 
small $x$ ($\ln(1/x) \gg \sqrt{\lambda}$) and large $q^2$ ($q^2 \gg
\Lambda^2$) region, without naively taking 
$q^2 \rightarrow \infty$ limit. Parton picture still remains even in the 
strong coupling regime, although the parton contributions do not
dominate in the DIS structure functions at moderate
$x$ ($\sqrt{\lambda}^{-1} \lesssim x$), and the DGLAP evolution is very fast. 

Similarly, for a fixed value of $q^2$, $\lambda_{\rm eff.}$ becomes 
smaller than 1 for sufficiently small $x$, rendering the $x$-integration
for $j=2$ moment convergent at $x = 0$ \cite{Hatta:2007}.\footnote{
One can also see the convergence directly, when the scattering amplitude
is given by a complex $j$-plane integral; the $x$-integration for 
$j = n$ moments converge for all $n$ larger than the largest real part
of the singularities in the $j$-plane.} 
For a given value of $x$, $\lambda_{\rm eff.} = j_{\nu^*} - 1$ increases 
for larger $q^2$, implying that the (generalized) parton distributions 
rise more steeply toward $x = 0$ for higher $q^2$ (see 
Figure~\ref{fig:transition} (a)).
This observation has already been made in the purely forward $t = 0$ 
case \cite{Brower:2010wf}.

The discussion so far clearly shows the conceptual importance of 
the saddle point value of $i\nu^*$ and hence of $j_{\nu^*}$.
The Pomeron amplitude corresponds to a sum of contributions from 
various states/operators of spin $j \in 2 \N$, and the sum can be 
expressed as a holomorphic integral in the complex $\nu$-plane.
Equivalently, this integral can also be expressed in the complex $j$-plane,
via the relation $j=j_\nu$.
The saddle point value $j^\ast (= j_{\nu^*})$ reflects the ``center 
of weight'' of contributions from various $j \in 2\N$.
The whole amplitude approximately shows the $x$-evolution and 
$q^2$-evolution of ``spin $j^\ast$ operator'', as clearly shown 
in (\ref{eq:effective-exponent}).
We will see later in this article that not just the $W^2$ and $q^2$ 
dependence of $\gamma_{\rm eff.}$ and $\lambda_{\rm eff.}$ 
in (\ref{eq:effective-exponent}) but that of almost all the observable
parameters ($t$-slope parameter in section \ref{sssec:slope} and
real part to imaginary part ratio in section \ref{ssec:real-part}) 
of the photon--hadron scattering amplitude are governed by the saddle 
point value $j^\ast$.

The essence of the saddle point approximation is in the following 
expression
\begin{align}
  \int \frac{dj}{2\pi i} x^{-j}\left(\frac{\Lambda}{q}
			      \right)^{\gamma(j)},
\label{eq:general-j-int}
\end{align}
where we already assume that we are interested in the small $x$ and 
large $q^2$ region, and ignored various factors the are irrelevant to 
the determination of the saddle point. 
The anomalous dimension $\gamma(j)$ as a function of complex spin
variable $j$ is \cite{BrowerJHEP0712:0052007}
\begin{align}
 \gamma(j)=\gamma_{\nu_j}
=i\nu_j-j=\{2\sqrt{\lambda}(j-j_0)\}^{1/2}-j
\label{eq:gamma(j)}
\end{align}
in the hard wall model, where $\nu_j$ is the inverse function of $j=j_\nu$.
We obtain an approximation 
\begin{align}
  \int \frac{dj}{2\pi i} x^{-j}\left(\frac{\Lambda}{q}
			      \right)^{\gamma(j)}
  \sim x^{- j_* + \gamma(j_*) \frac{\ln(q/\Lambda)}{\ln (1/x)}}, \qquad 
  \left. \frac{\partial \gamma(j)}{\partial j} \right|_{j = j_*} 
     = \frac{\ln (1/x)}{\ln (q/\Lambda)}.
\label{eq:saddle-p-QCD}
\end{align}
The saddle point value of $j$ is a function\footnote{
A little more careful discussion should be given in non-conformal 
theories. It will not be difficult, however, to incorporate the running 
coupling effect within weakly coupled gauge theories or separately 
within gravity dual descriptions.}
 of $[\ln (q/\Lambda)]/[\ln(1/x)]$, 
just like in (\ref{eq:nu ast for small -t}) and in (\ref{eq:saddle-p-QCD}).
The saddle point approximation at the leading order ignores terms that
are suppressed by $1/\ln(1/x)$, but keeps full
$[\ln(q/\Lambda)]/[\ln(1/x)]$ dependence at all order. Higher order
corrections in the saddle point approximation take account of
$1/\ln(1/x)$ suppressed corrections. 

One will notice that this argument does not rely on detailed form 
of the anomalous dimension $\gamma(j)$ very much. Indeed, exactly 
the same line of argument has been used in perturbative QCD for the 
study of behavior of PDF in the small $x$ and large $q^2$ 
region \cite{Ellis:1991qj}. The anomalous dimensions $\gamma(j)$
of the twist-2 series of operators in weak coupling gauge theories 
(with some variations in the approximation scheme (e.g., double leading
log approximation (DLLA))) are not the same 
as those in gravity dual models such as (\ref{eq:gamma(j)}), but they
can be continuously deformed from one to the other by changing the value 
of the 't Hooft coupling \cite{BrowerJHEP0712:0052007}.
Discussion so far makes it clear i) that contribution associated with 
the ``twist-2'' series of operators (parton contribution) does exist 
in the weak coupling and strong coupling regimes alike, 
ii) that the kinematical variable dependence of the parton contribution 
in the small $x$ and large $q^2$ region can be captured by the saddle 
point approximation (\ref{eq:general-j-int}, \ref{eq:saddle-p-QCD}), and 
iii) that the $(q,x)$ evolution of the parton contribution remains 
qualitatively the same in the both regimes, despite the difference 
in the anomalous dimensions.  
This similarity of the parton component of a hadron in the both regimes 
is an encouraging factor in trying to take advantage of gravity dual 
descriptions to study non-perturbative aspects of partons in a hadron 
of a confining gauge theory (like the real world QCD). 
We will elaborate more on this in section \ref{sec:real-world}.

\subsubsection{Large momentum transfer: $\Lambda^2 \ll -t$}
\label{sssec:large-(-t)}

Let us now evaluate the holographic DDVCS amplitudes (\ref{eq:Im-Ii}) 
for large momentum transfer, $\Lambda^2 \ll -t \ll q^2$.
In this case, the second term of the integrand in (\ref{eq:Ahh-def}) is negligible
for almost all the range of $z' \leq 1/\Lambda$, and moreover, 
the range of $z'$ integration is effectively limited by 
$z'\lesssim (-t)^{-1/2}$, because the modified Bessel function 
$K_{i\nu}(\sqrt{-t}z')$ falls off exponentially for $z' \gg (-t)^{-1/2}$.
In the region, $0<z'\lesssim (-t)^{-1/2}$, the wavefunction of hadron 
shows the power law behavior, $\Phi(z')\propto (z\Lambda)^\Delta$.
Hence, $g^h_{i\nu}(\sqrt{-t}/\Lambda)$ is approximately,
\begin{align}
 g^h_{i\nu}(\sqrt{-t}/\Lambda) \simeq 
 \sfrac{\Lambda}{\sqrt{-t}}^{-\gamma_{\nu}+2\Delta-2} \tilde g^h_{i\nu},
\label{eq:g-h-tilde}
\end{align}
where $\tilde g^h_{i\nu}$ is independent of $t$, and is of order unity.
Because the factor $(\Lambda/\sqrt{-t})^{-\gamma_\nu}$ 
in (\ref{eq:g-h-tilde}) is a rapidly changing function of $i\nu$ for
$\Lambda^2 \ll -t$, this factor has an impact on the saddle point value 
$\nu^*$ of $\nu$ integration in (\ref{eq:Im-Ii}).
The saddle point now depends on $t$, as in 
\begin{align}
 i\nu^\ast \left( q/\Lambda, x, -t \gg \Lambda^2 \right)=
   \sqrt{\lambda}\frac{\ln(q/\sqrt{-t})}
                      {\ln\sfrac{q/\sqrt{-t}}{\sqrt{\lambda}x}}, 
\label{eq:nu ast for large -t}
\end{align}
and the saddle point approximation of the DDVCS amplitudes at leading
order is given
by 
\begin{align}
 {\rm Im}\: I_i \simeq 
  \frac{c'_s c_J^2 \sqrt{\lambda}}{\Gamma^2(j_{\nu^\ast}/2)}
  \left(\frac{\ln\frac{\sqrt{-t}/\Lambda}{\sqrt{\lambda}x}}{2\pi \sqrt{\lambda}}\right)^{-1/2} \!\!\!\!\!\!\!\!
  \sfrac{1}{4\sqrt{\lambda}x}^{j_{\nu^\ast}}
\sfrac{\Lambda}{q}^{\gamma_{\nu^\ast}}
\sfrac{\Lambda}{\sqrt{-t}}^{-\gamma_{\nu^\ast}+2\Delta-2}
\bar{c}^{(i)}_{i{\nu^\ast}} \:
\tilde g^h_{i{\nu^\ast}}.
\label{eq:Ii-saddle-point2}
\end{align}
Ignoring factors of order unity and the Gaussian measure of the saddle
point approximation,\footnote{The effective exponents 
introduced in (\ref{eq:def-exponents}) are still given by 
$\gamma_{\rm eff.} = \gamma_{\nu^*}$ and 
$\lambda_{\rm eff.} = j_{\nu^*} - 1$; the saddle point value of $\nu^*$
is now given by (\ref{eq:nu ast for large -t}).
Thus, the discussion in p.\pageref{page:gamma-lambda} holds true also 
for this case without modification. 
} 
\begin{align}
 {\rm Im} \; I_i(x,\eta=0,t,q^2)
&\sim
\sfrac{1}{\sqrt{\lambda}x}^{j_{\nu^\ast}}\sfrac{\Lambda}{q}^{\gamma_{\nu^\ast}}\sfrac{\Lambda}{\sqrt{-t}}^{-\gamma_{\nu^\ast}+2\Delta-2}, \\
&=
\sfrac{1}{\sqrt{\lambda}x}^{j_0}\sfrac{q}{\Lambda}^{j_0}\sfrac{\Lambda}{\sqrt{-t}}^{2\Delta-2+j_0}e^{-\frac{\sqrt{\lambda}}{2}\frac{(\ln(q/\sqrt{-t}))^2}{\ln((q/\sqrt{-t})/\sqrt{\lambda}x)}}.
\label{eq:J for large -t}
\end{align}

It is customary in perturbative QCD to describe the momentum transfer
$t$ dependence of GPD in terms of a ``form factor'' $F$ defined by
\begin{equation}
 H(x,\eta, t, q^2) = H(x, \eta=0, t=0, q^2) F(x,\eta, t, q^2).
\label{eq:ratio-GPD}
\end{equation}
Since the GPDs become PDFs $H(x,q^2)$ for 
$\eta = 0$ and $t = 0$, the form factor $F(x,\eta, t, q^2)$ should be
$1$ in that limit. This form factor takes account of finite size of hadrons, 
and is non-perturbative in nature. There is no way calculating 
the form factor in perturbative gauge theories,\footnote{In the approach
of \cite{Brodsky:1973kr, Matveev:1973ra, Brodsky:1974vy, Sivers:1975dg, 
Donnachie:1979yu}, power-law scaling in energy can be concluded for 
large $(-t)$ region, but still non-perturbative wavefunctions of partons 
are required for quantitative results.} 
and experimentally measurable form factors (such as the electromagnetic 
form factor) are sometimes used for theoretical modeling of GPD (though 
it should depend on $(x,\eta,q^2)$) and for fitting of DVCS experimental data. 
With the holographic set up, however, it can be calculated from first 
principle. 
\begin{figure}[tbp]
 \begin{center}
\begin{tabular}{ccc}
   \includegraphics[scale=0.6]{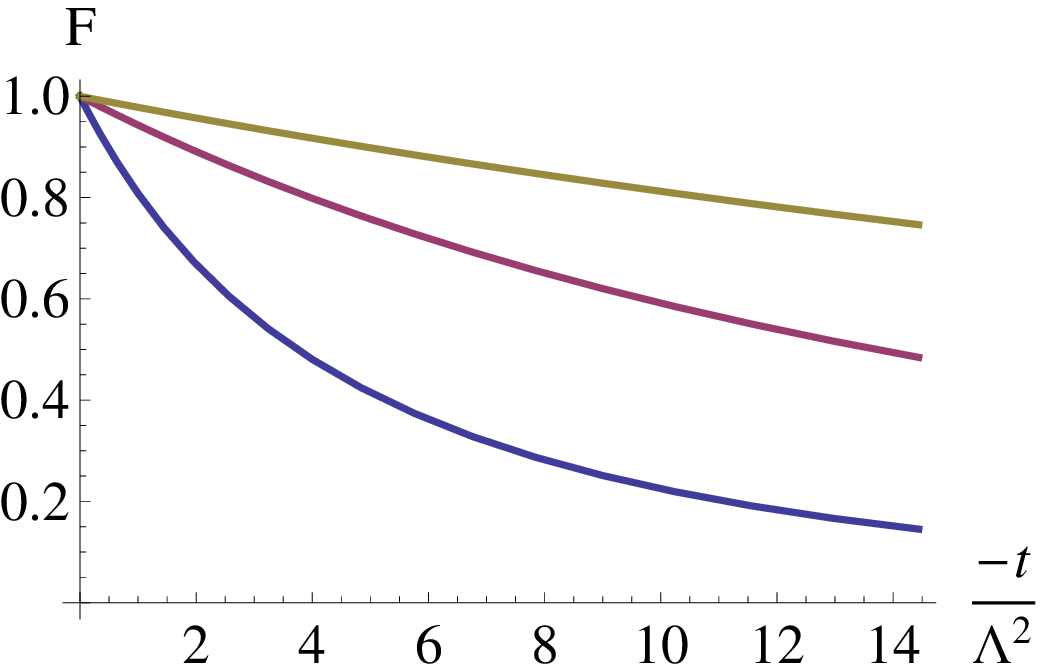} & &
   \includegraphics[scale=0.7]{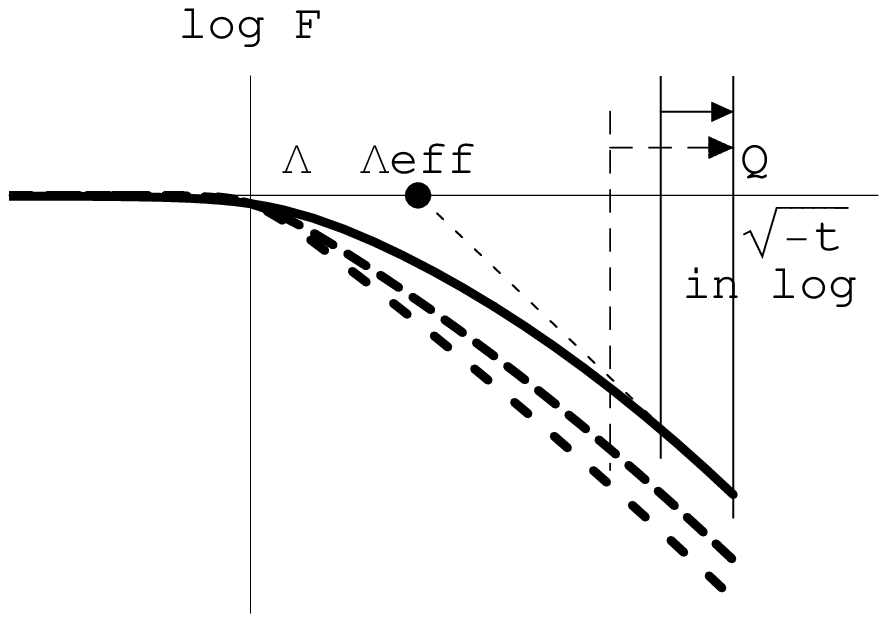} \\
  (a) & & (b)
\end{tabular}
\caption{\label{fig:GPD-form-factor} (color online) 
The panel (a) shows the form factor $F$ (\ref{eq:GPD-form-factor}) 
as a function of $-t/\Lambda^2$ in low momentum transfer 
($-t\lesssim \Lambda^2$) region. We used the scalar target hadron 
with $\Delta = 5$, $n = 1$ explained in section \ref{sec:model}. 
The three curves in the panel (a) correspond to 
finite $x$, $i\nu^\ast=4$, smaller $x$, $i\nu^\ast=2$, and small $x$ limit, $i\nu^\ast=0$
from above (yellow/light) to below (blue/dark).
The panel (b) is a schematic picture of form factor for large momentum 
transfer. Dashed curve corresponds to a smaller $x$ than the one for 
the solid curve, and short dashed curve to an even smaller $x$.
The form factor eventually shows a power-law behavior for sufficiently 
large momentum transfer; this power behavior is reached for smaller
momentum transfer for smaller $x$ (see discussion 
around (\ref{eq:cond4simple-power})), as indicated in the figure. }
 \end{center}
\end{figure}

Just one common form factor $F$ is necessary for all the structure 
functions\footnote{Here, we talk of form factors describing $t$
dependence in the {\it structure functions}, rather than the $t$ dependence 
of {\it GPD}. Characterization of GPD in strong coupling regime is given in
section \ref{ssec:GPD}, where we will see that the form factor
of {\it structure functions} (\ref{eq:GPD-form-factor}) can also be taken as 
that of {\it GPD} in the sense of (\ref{eq:ratio-GPD}).} 
$V_{1,\cdots,5}$, because the expressions (\ref{eq:J for small -t}, \ref{eq:J for large -t})
 are both not much different for $I_0$ and $I_1$, at least for small $x$ we have 
assumed so far. Taking ratio 
of (\ref{eq:Ii-saddle-point}, \ref{eq:J for small -t}, 
\ref{eq:J for large -t}) to $I_i(x,\eta = 0,t=0,q^2)$, 
we find that 
\begin{equation}
 F(x,\eta = 0, t, q^2) \simeq \left\{
\begin{array}{ll}
 g^h_{i \nu^*}(\sqrt{-t}/\Lambda) / g^h_{i \nu^*}(0) 
     & (0 \leq -t \lesssim \Lambda^2), \\
   \left(\frac{\Lambda}{\sqrt{-t}}\right)^{2 \Delta - 2 + j_0}
  e^{- \frac{\sqrt{\lambda}}{2} 
        \frac{[\ln(q/\sqrt{-t})]^2}{\ln((q/\sqrt{-t})/\sqrt{\lambda}x)}
     }
    e^{+ \frac{\sqrt{\lambda}}{2}\frac{ [ \ln(q/\Lambda)]^2}
              { \ln ((q/\Lambda)/\sqrt{\lambda} x) }} & 
  ( \Lambda^2 \ll -t \ll q^2).
\end{array}\right. 
\label{eq:GPD-form-factor}
\end{equation}
The form factor largely shows power-law dependence on $(-t)$ at large 
momentum transfer, and for sufficiently large momentum transfer,  
\begin{equation}
 \ln \left( \frac{q}{\sqrt{-t}} \right) \ll  
\frac{1}{\sqrt{\lambda}} \ln \left(\frac{1}{\sqrt{\lambda}x}\right), 
\label{eq:cond4simple-power}
\end{equation}
the exponent of $\sqrt{-t}$ is approximately given by 
$-(2\Delta -2 + j_0)$. 

The Regge behavior in the Virasoro--Schapiro 
amplitude $(s_{10})^{\alpha' t_{10}/2}$ does not lead to exponential 
dependence on the momentum transfer $t$ of DDVCS processes, 
because this factor works as an exponential cut-off 
$(\tilde{s})^{- \nu^2/2\sqrt{\lambda}}$ for contributions from 
Pomeron with large Kaluza--Klein momentum in the direction of 
holographic radius. Pomeron with small $\nu$ still contributes, 
but the Pomeron wavefunction 
$\Psi_{i\nu}^{(j)}(t, z) \sim K_{i\nu}(\sqrt{-t}z)$ cuts off 
the IR $z \gsim 1/\sqrt{-t}$ region of $AdS_5$ for spacelike momentum 
transfer \cite{BrowerJHEP0712:0052007}. This exponential cut-off and 
the power-law wavefunction (\ref{eq:normalizable mode}) of the target 
hadron combined results in the power-law $\sqrt{-t}^{-(2\Delta -2 +
j_0)}$ dependence of the form factor, as we have already seen 
in (\ref{eq:g-h-tilde}).\footnote{This mechanism is similar
to the way the power-law $q^2$ dependence of DIS cross section was 
obtained for moderate $x$ \cite{PolchinskiJHEP0305:0122003}, although 
the integration along the holographic radius was limited by the 
non-normalizable wavefunctions of virtual photons 
$P^{(i)}_{\gamma^* \gamma^*}(z)$ (\ref{eq:P1-def}, \ref{eq:P0-def}) in 
the DIS case, not by the Pomeron wavefunction $\Psi^{(j)}_{i\nu}(t,z)$ 
as in this case. Gravitational form factor---the $j = 2$ moment 
of non-skewed GPD---also shows power-law dependence on the momentum 
transfer \cite{Hong:2003jm, Abidin:2008ku, 
Abidin:2008hn, Brodsky:2008pf, Carlson:2008ha}. }

Note that this power-law dependence of the form factor is a generic
consequence of asymptotically conformal theories, and is independent 
of details of IR geometry in holographic models. The power is determined 
by the conformal dimension $\Delta$ of the bulk field the hadron belongs 
to. The power-law behavior in the UV conformal holographic models 
is due to the cost of squeezing the entire hadron into a size 
of $1/\sqrt{-t}$. This power-law behavior in holographic picture, however,
should not be taken as an explanation for the power-law behavior 
observed in the real world hadron--hadron elastic scattering data 
at large momentum transfer $\Lambda^2 \ll -t$, because the holographic 
models assuming large 't Hooft coupling even at high energy is not 
truly dual to the real world QCD. The power-law behavior expected from 
the naive power counting based on the number of valence partons \cite{
Brodsky:1973kr, Matveev:1973ra, Brodsky:1974vy, Sivers:1975dg, 
Donnachie:1979yu} should be close to the truth of the power law 
in the $\Lambda^2 \ll -t$ region. Despite this difference, it is 
an encouraging fact that the form factor $F(x,t,q^2)$ of holographic 
models generically matches on to something semi-realistic (power-law
behavior) at large momentum transfer $\Lambda^2 \ll -t$, rather than 
to something totally different. There is a long tradition of using 
a GPD form factor with a power-law behavior in $t$ at the $\Lambda^2 \ll -t$
region (e.g., \cite{White:1973, Frankfurt:2002ka, Diehl:2005wq}), 
and it might be 
possible to provide a better theoretical foundation to such form factors. 
See also section \ref{sec:real-world}.
 
To learn more about the structure of the non-perturbative form 
factor (\ref{eq:GPD-form-factor}) in the strong coupling regime, let us examine the kinematical
variable dependence of the form factor (\ref{eq:GPD-form-factor}) 
more carefully. First, the $x$ dependence and $(-t)$ dependence are 
not completely factorized. Although 
the form factor gradually approaches to power-law in $\sqrt{-t}$ 
with a common power $-(2\Delta -2 + j_0)$, it does so under 
$x$-dependent condition (\ref{eq:cond4simple-power}). 
Even within the region (\ref{eq:cond4simple-power}),
the coefficient 
of the power-law $(\Lambda/\sqrt{-t})^{2\Delta-2 + j_0}$ also 
depends on $x$ and $q^2$; we find that an effective energy scale 
$\Lambda_{\rm eff}$ in 
$F \sim [\Lambda^2_{\rm eff}/(-t)]^{\Delta - 1 + j_0/2}$
is given by
\begin{equation}
 \Lambda_{\rm eff.}^2 = \Lambda^2 \times 
    e^{\frac{\sqrt{\lambda}[\ln(q/\Lambda)]^2}
            {(2\Delta-2+j_0) \ln[(q/\Lambda)/\sqrt{\lambda}x]}}
%
\end{equation}
and in particular, $\Lambda_{\rm eff.}^2$ decreases for small $x$ 
(large $W^2$). Note also that this form factor and 
$\Lambda^2_{\rm eff.}$ properly take account of $q^2$ dependence, not 
just $(x,t)$ dependence at a fixed value of $q^2$.

In the extremely small $x$ limit (for a given value of $q^2$), however, 
$\Lambda_{\rm eff.}$ does not become arbitrarily small, but approaches 
a finite value, $\Lambda^2$. In this limit, the form factor greatly 
simplifies, to become 
\begin{equation}
 F(x,\eta=0, t,q^2) \simeq 
   \left(\frac{\Lambda}{\sqrt{-t}}\right)^{2\Delta-2+j_0} \qquad
   (\Lambda^2 \ll -t \ll q^2), 
\end{equation}
from which $x$ and $q^2$ dependence has disappeared. 
This is when $i\nu^* = 0$.

\subsubsection{Regge theory revisited}
\label{sssec:regge}

Regge theory and the dual resonance model had a close relation  
in description of hadron scattering in late 1960's--early 1970's. 
Certainly the dual amplitude on 3+1 dimensions were not able to 
explain power-law behavior in the fixed angle high-energy scattering 
on one hand, and the theoretical consistency of string theory hinted 
a spacetime dimension higher than 3+1 on the other. 
But, string theory on a warped spacetime has resurrected as a 
theoretical framework of high energy scattering of hadrons in 
strongly coupled gauge theories \cite{Polchinski2002, 
PolchinskiJHEP0305:0122003, BrowerJHEP0712:0052007}.
When the scattering amplitudes of holographic QCD are seen as 
an amplitude of hadrons in four dimensions, some part of the 
Regge behavior of the scattering amplitude of strings in ten 
dimensions still seem to remain. The Pomeron trajectories predicted 
by holographic QCD are not linear, on the other hand, and the 
complex $j$-plane description of hadron scattering amplitude of 
holographic QCD may be a little different from what one naively 
imagines from traditional Regge theory with some linear trajectories. 
The question is, then, how they are different. 

Preceding literatures \cite{BrowerJHEP0712:0052007, 
Hatta:2007, Brower2007} have already made efforts in providing complex $j$-plane 
description of the scattering amplitudes of holographic QCD. 
Our discussion in sections
\ref{sssec:small-(-t)}--\ref{sssec:large-(-t)} already shows 
the importance of the saddle point value $j = j_{\nu^*}$ of the 
scattering amplitude in $j$-plane in extracting kinematical variable
dependence of the scattering amplitudes. It is thus worthwhile to 
take a moment in this section \ref{sssec:regge} and elaborate more on 
$j$-plane description of the hadron scattering amplitudes in holographic QCD.
It is known that i) holographic QCD gives rise to Kaluza--Klein towers 
of Pomeron trajectories, and that ii) they are not linear. From these
properties, we will see in the following that scattering amplitudes 
are described better by treating those trajectories individually for 
some kinematical range, while they are better described by treating 
a Kaluza--Klein tower as a whole for some other kinematical regions. 
Such transitions are triggered by the location of the saddle point 
in the $j$-plane relatively to other singularities of the amplitude. 

\begin{figure}[tbp]
  \begin{center}
\begin{tabular}{ccc}
  \includegraphics[scale=0.65]{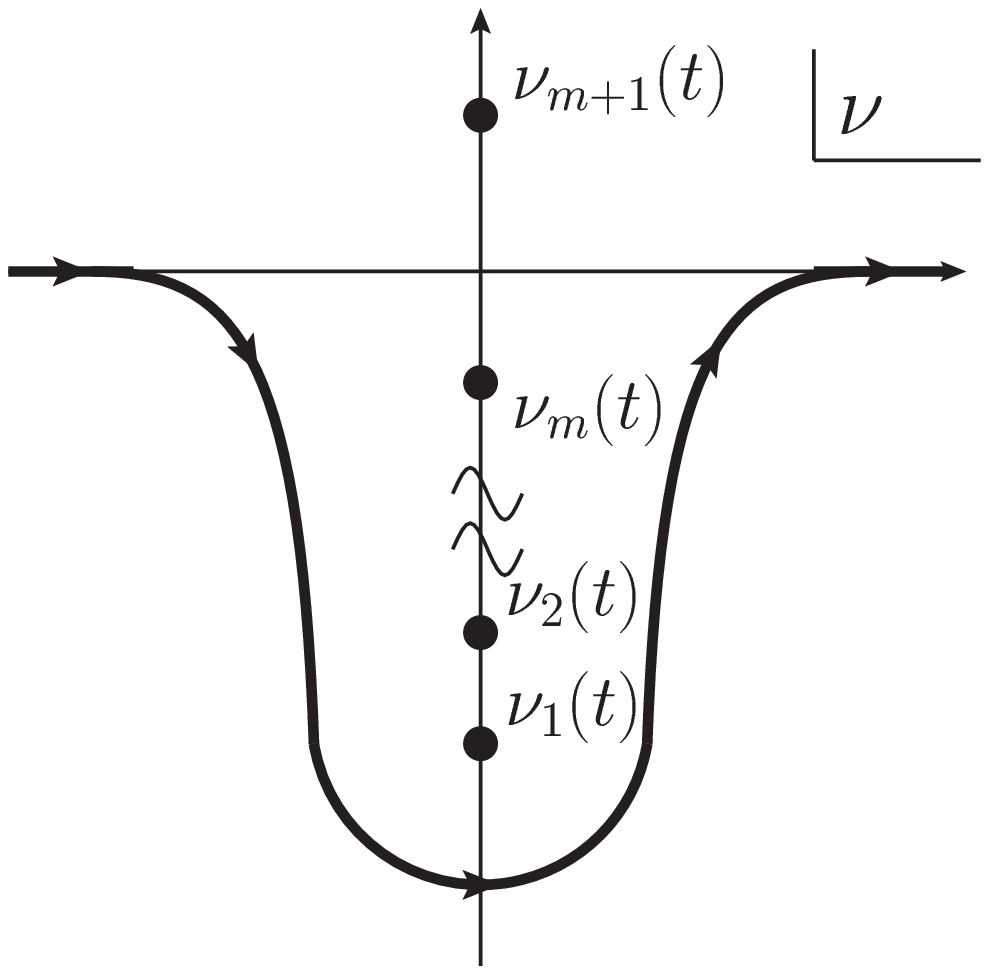} & $\qquad$ &
  \includegraphics[scale=0.65]{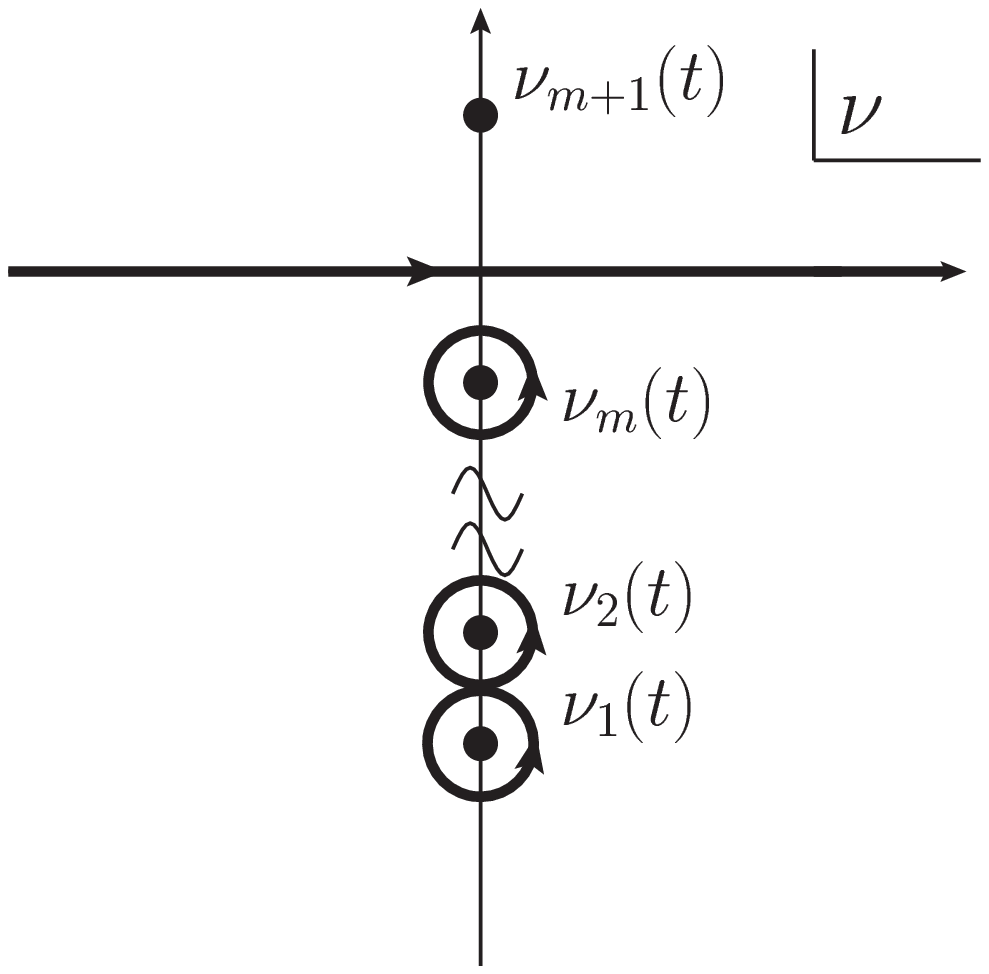} \\
  (a) &  & (b)
\end{tabular}
\caption{\label{fig:nu-plane} integration contour in $\nu$-plane
}
  \end{center} 
\end{figure}

Let us begin with studying behavior of the DDVCS amplitude, not just 
for physical region of $(s,t)$ plane, but for region including real 
positive $t$.
The integrand of (\ref{eq:kernel in hw1}) can be regarded as 
a holomorphic function of $\nu$ and $t$.
The denominator of the second term, $I_{i\nu}(\sqrt{-t}/\Lambda)$,  
becomes zero if
\begin{align}\label{eq:poles-of-integrand-of-kernel}
 t=\Lambda^2 (j_{i\nu,n})^2, \;\; n = 1,2, \cdots, \in \N, 
\end{align}
and these zeros in the denominator can be regarded as poles in 
the $\nu$-plane for a given kinematical variable $t$. 
As we analytically continuate the integrand from real negative 
$t \ll -\Lambda^2$ in the physical kinematical region
to real positive $t  \gg \Lambda^2$ through the upper-half complex plane
of $t$, the poles in the $\nu$-plane move and some of them show up 
in the region with real positive $i\nu$; 
see (\ref{eq:poles-of-integrand-of-kernel}). 
We introduce a notation $t_{c,n} = (\Lambda j_{0,n})^2$; if 
$t_{c,m}< t < t_{c,m+1}$, there are $m$ poles in the lower-half 
$\nu$-plane.
The positions of such poles satisfying
(\ref{eq:poles-of-integrand-of-kernel}) are denoted by $\nu_n(t)$.
Thus, for $\Lambda^2 \ll t$, the integration contour in the $\nu$-plane 
needs to be deformed as in Figure~\ref{fig:nu-plane}~(a), which can then be 
rearranged as in Figure~\ref{fig:nu-plane}~(b).

Contributions from the poles $\nu = \nu_n(t)$ can be written
as\footnote{
Although we obtained
(\ref{eq:Regge-Pole-contribution-to-GG-amplitude-2}) 
by analytical continuation of the amplitude from $t <0$ into 
$t >0$ through the upper half plane, there is another derivation.
In section \ref{ssec:Pomeron-kernel}, we derived the Pomeron kernel
by inserting the complete system (\ref{compl}) 
into (\ref{eq:5D-descr-glueB-glueB}), but we implicitly assumed that 
$t < 0$. For $\Lambda^2 \ll t$, however, (\ref{compl}) is not a complete
system, as the ``Schr\"{o}dinger equation'' 
$- \Delta_j(t) \Psi^{(j)} = (E_\nu/R^2) \Psi^{(j)}$ may have discrete 
spectrum as well \cite{BrowerJHEP0712:0052007}. 
Indeed, the wavefunctions (\ref{eq:psi-tilde:discrete-eigenfunction-of-deltaj})
 become  the discrete spectrum of $\Delta_j(t)$
by replacing  $i\nu_j$ with $i\nu_n(t)$ 
and
 $m_{j,n}$ with $\sqrt{t}$, respectively.
The pole contributions (\ref{eq:Regge-Pole-contribution-to-GG-amplitude-2}) 
come from the discrete spectrum part of the inserted complete system.
}  
\begin{equation}
 I_i \simeq \sum_{n=1}^m 
   \left[ - \frac{1 + e^{- \pi i j_{\nu_n(t)}}}
                 { \sin (\pi j_{\nu_n(t)})}
   \right]
   \frac{1}{\Gamma^2 (j_{\nu_n(t)}/2)}
   \beta_n(\alpha_{\P,n}(t), t; q^2) 
   \left( \frac{q^2}{4\sqrt{\lambda} x \Lambda^2} \right)^{j_{\nu_n(t)}},
\label{eq:Regge-Pole-contribution-to-GG-amplitude-2}
\end{equation}
where
\begin{equation}
 j_{\nu_n(t)} = 
 \left. \left[ j_0 + \frac{(i\nu)^2} {2 \sqrt{\lambda}} \right]
 \right|_{\nu = \nu_n(t)} \equiv \alpha_{\P,n}(t), 
\label{eq:pomeron-trajectory}
\end{equation}
and 
\begin{equation}
\beta_n(\alpha_{\P,n}(t),t; q^2) = \frac{c'_s}{2}2\pi\sqrt{\lambda} 
  \left.   
    \left[\frac{4 \mu}{j_{\mu,n} \frac{\partial j_{\mu,n}}
                                      {\partial \mu}}
    \right]
 \right|_{\mu = i \nu_n(t)}  \!\!\!\!\!\!\!\!\!\!
 \gamma_{\gamma^* \gamma^* \P n}(\alpha_{\P , n}(t)) \; 
 \gamma_{h h \P n}(\alpha_{\P , n}(t)), 
\label{eq:beta-n(t)} 
\end{equation}
\begin{eqnarray}
\gamma_{h h \P n}(j) & = & \frac{1}{\kappa_5^2} 
   \int dz \sqrt{-g} P_{hh}(z) (z\Lambda)^j e^{-2 A(z)}
   \frac{J_{i\nu_j}( m_{j,n} z)}{J'_{i\nu_j}(j_{i\nu_j, n})}
   \left[\frac{\kappa_5^2}{R^3}\right]^{1/2}
\label{eq:gamma'-n(t)} \\
  & = & 
   \frac{1}{\kappa_5^2} 
   \int dz \sqrt{-g} P_{hh}(z) (R \Lambda)^j 
     e^{-2 j A} 
     \left[e^{(j-2) A} \sfrac{J_{i\nu_j} (m_{j,n}z)}
         {J'_{i\nu_j}(j_{i\nu_j,n})}    
           \left[\frac{\kappa_5^2}{R^3}\right]^{1/2}
     \right].
\label{eq:gamma'-n(t)-2}
\end{eqnarray}
Under the conditions that 
\begin{equation}
 j = j_\nu \quad ({\rm equiv.~} \nu=\nu_j) 
 \quad {\rm and} \quad
 \nu = \nu_n(t) \quad ({\rm equiv.~} \sqrt{t}/\Lambda = j_{i\nu,n}),
\label{eq:j-nu-t}
\end{equation}
(from which $j = \alpha_{\P, n}(t)$ follows), $\nu = \nu_j = \nu_n(t)$,
and a $j$ dependent mass parameter $m_{j,n} = m^{(\nu)}_n$ is defined by 
\begin{equation}
 m_{j,n} \equiv \Lambda j_{i\nu_j, n} = \sqrt{t} = \Lambda j_{i\nu,n} 
   \equiv m^{(\nu)}_n. 
\end{equation}

The factor $\gamma_{\gamma^* \gamma^* \P n}(j)$ can also be defined
similarly to (\ref{eq:gamma'-n(t)}, \ref{eq:gamma'-n(t)-2}) by replacing 
$P_{hh}(z)$ 
by $P^{(i)}_{\gamma^* \gamma^*}(z)$ ($i = 0,1$) in 
(\ref{eq:P1-def}, \ref{eq:P0-def}).
Because the wavefunction of photon depends on $q^2$,
it has dependence on $q^2$.
Explicitly, one can find 
 $\gamma_{\gamma^* \gamma^* \P n}(\alpha_{\P, n}(t))\sim (\Lambda/q)^{\gamma(\alpha_{\P,n}(t))+2\alpha_{\P,n}(t)}$,
and
\begin{align}
\label{eq:Regge-Pole-contribution-to-GG-amplitude-3}
 I_i \simeq \sum_{n=1}^m 
   \left[ - \frac{1 + e^{- \pi i j_{\nu_n(t)}}}
                 { \sin (\pi j_{\nu_n(t)})}
   \right]
\frac{1}{\Gamma^2 (j_{\nu_n(t)}/2)}
\tilde \beta_n(t)
   \sfrac{\Lambda}{q}^{\gamma(\alpha_{\P,n}(t))}\sfrac{1}{x}^{\alpha_{\P,n}(t)},
  \end{align}
where $\tilde \beta_n(t)=
\beta_n(\alpha_{\P, n}(t), t; q^2)(q/\Lambda)^{\gamma(\alpha_{\P,n})+2\alpha_{\P,n}}$
 is independent of $q$.

The expression (\ref{eq:Regge-Pole-contribution-to-GG-amplitude-2})
is precisely in the form assumed in traditional Regge theory:
\begin{align}
 A_\text{Regge}(s,t)=-\oint_{j=\alpha(t)} \frac{dj}{2\pi i} \frac{1+e^{-i\pi j}}{\sin \pi j}\frac{\beta(j,t)}{j-\alpha(t)}\sfrac{s}{s_0}^{j}
                    =-\frac{1+e^{-i\pi \alpha(t)}}{\sin\pi\alpha(t)}\beta(\alpha(t),t)\sfrac{s}{s_0}^{\alpha(t)}.
\label{eq:traditional-Regge-amplitude}
\end{align}
Traditional theory was only able to 
assume a linear form 
\begin{equation}
\alpha(t) = \alpha_{\P, 0} + \alpha'_{\P} t  
\label{eq:linear-traj}
\end{equation}
for simplicity or from fit to experimental data, but holographic QCD 
predicts a trajectory (\ref{eq:pomeron-trajectory}); it is approximately 
linear for $\Lambda^2 \ll t$ (but not too large $-t$), 
but it is not unless $\Lambda^2 \ll t$. 
The residues $\beta_n(\alpha_{\P,n}(t), t; q^2)$ satisfy factorization 
condition; they factorize into 
$\gamma_{\gamma^* \gamma^* \P n}(\alpha_{\P, n}(t))$
and $\gamma_{hh \P n}(\alpha_{\P, n}(t))$ holomorphically 
in $j = \alpha_n(t)$.
This factorization is necessary for unitarity of hadronic scattering 
processes in 3+1 dimensions, which is not guaranteed a priori in a
theory that is not based on a local field theory on 3+1 dimensions 
\cite{Collins:1977, Gribov-cpx-ang-mom, Donnachie:2002en}.\footnote{
Factorization predicts relations among differential cross sections:
$d\sigma_{\rm el}(A+B)/d\sigma_{\rm el}(A+C) 
= d\sigma_{\rm el}(D+B)/d\sigma_{\rm el}(D+C)$ \cite{BaronePredazzi}.}
Scattering amplitudes of dual resonance model and superstring theory in 
10-dimensions have the factorization property \cite{Bardakci:1969gr},
and the factorization (\ref{eq:beta-n(t)}) in 4-dimensions is remnant 
of the factorization in string theory on higher dimensions. 

We have so far dealt with the (analytic continuation of the) scattering 
amplitude $I_i$, but closer connection to the classical Regge theory 
can be established by clarifying physical meaning of 
$\alpha_{\P,n}(t)$ in (\ref{eq:pomeron-trajectory}) and of the factorized 
residues $\beta_n \propto \gamma_{hh\P n} (j=\alpha_{\P, n}(t))$ 
(\ref{eq:beta-n(t)}, \ref{eq:gamma'-n(t)-2}).
The Pomeron--hadron--hadron coupling 
$\gamma_{hh \P n}(t)$ (\ref{eq:gamma'-n(t)-2}) looks like an overlap 
integration of three wavefunctions; two of them 
$P_{hh}(z) = c_\phi [\Phi(z)]^2$ are those of the target hadron, and 
the last one satisfies
\begin{align}\label{eq:J-inu(sqrt-t-z)-is-eigenfunc-of-Delta-j}
\left[ \Delta_j(t) + \frac{\nu^2+4}{R^2}\right]
  \left[e^{A(j-2)}J_{i\nu}(\sqrt{t}z) \right] =0.
\end{align}
For a special case as in (\ref{eq:gamma'-n(t)-2}), when $\nu$, $t$ and $j$
are related by $\nu = \nu_j$ and $t = m^2_{j,n}$, this means 
\begin{align}
 \left[ \Delta_j(m^2_{j,n}) -\frac{2}{\alpha'}(j-2)\right]
 \left[ e^{A(j-2)}J_{i\nu_j}(m_{j,n} z) \right] =0, 
\label{eq:J-inu(sqrt-t-z)-is-eigenfunc-of-Delta-j-at-j=j-nu}
\end{align}
which is the equation of motion for spin $j$ string state\footnote{
The mass of the string will differ from one in the flat space.
But we have shown in section \ref{ssec:Pomeron-kernel} that the mass shift 
$\delta_j$ can be consistently neglected within our approximation.} 
on the graviton trajectory in $AdS_5$, when $j \in 2\N$; this
 momentum in the 3+1-dimensional Minkowski spacetime is on the 
mass shell $t = m^2_{j,n}$ in this wavefunction, when the IR boundary 
condition is satisfied. 
Therefore, $j = \alpha_{\P, n}(t)$ describes the spin--mass relation 
of 4D hadrons corresponding to the $n$-th Kaluza--Klein mode in the 
spin $j$ string state in the graviton trajectory; 
$j = \alpha_{\P, n=1}(t)$ is the leading trajectory, and 
$j = \alpha_{\P, n}(t)$ with $2 \leq n \leq m$ become daughter 
trajectories \cite{BrowerJHEP0712:0052007}.
Normalizable wavefunctions of all the 4D hadrons in the trajectory 
$j = \alpha_{\P, n}(t)$ (\ref{eq:j-nu-t}) are reproduced from a single 
wavefunction
\begin{equation}
\label{eq:psi-tilde:discrete-eigenfunction-of-deltaj}
  \psi^{(j)}_n(z) = e^{(j-2) A}
      \frac{J_{i\nu_j}(m_{j,n} z)}{J'_{i\nu_j}(j_{i\nu_j,n})}
      \left[\frac{\kappa_5^2}{R^3}\right]^{1/2}
\end{equation}
defined for $j \in \C$; the wavefunction for $j \in 2\N$ is simply 
the special case of the one above. One might refer 
to~(\ref{eq:psi-tilde:discrete-eigenfunction-of-deltaj}) as a
wavefunction of the $n$-th Pomeron trajectory.
The Pomeron--hadron--hadron coupling $\gamma_{hh \P n}(j)$ is given 
by an overlap integral of two $\Phi$'s along with this 
wavefunction $\psi^{(j)}_n(z)$ of the $n$-th trajectory. 
That is, 
\begin{equation}
 \gamma_{hh \P n}(j) = \frac{1}{\kappa_5^2} \int dz \sqrt{-g} \; 
   P_{hh}(z) \; e^{- 2 j A} \; \psi^{(j)}_n(z) \times
   (R \Lambda)^{j}.
\label{eq:P-h-h-3pt-coupling}
\end{equation}

\begin{figure}[tbp]
  \begin{center}
\begin{tabular}{ccc}
  \includegraphics[scale=0.32]{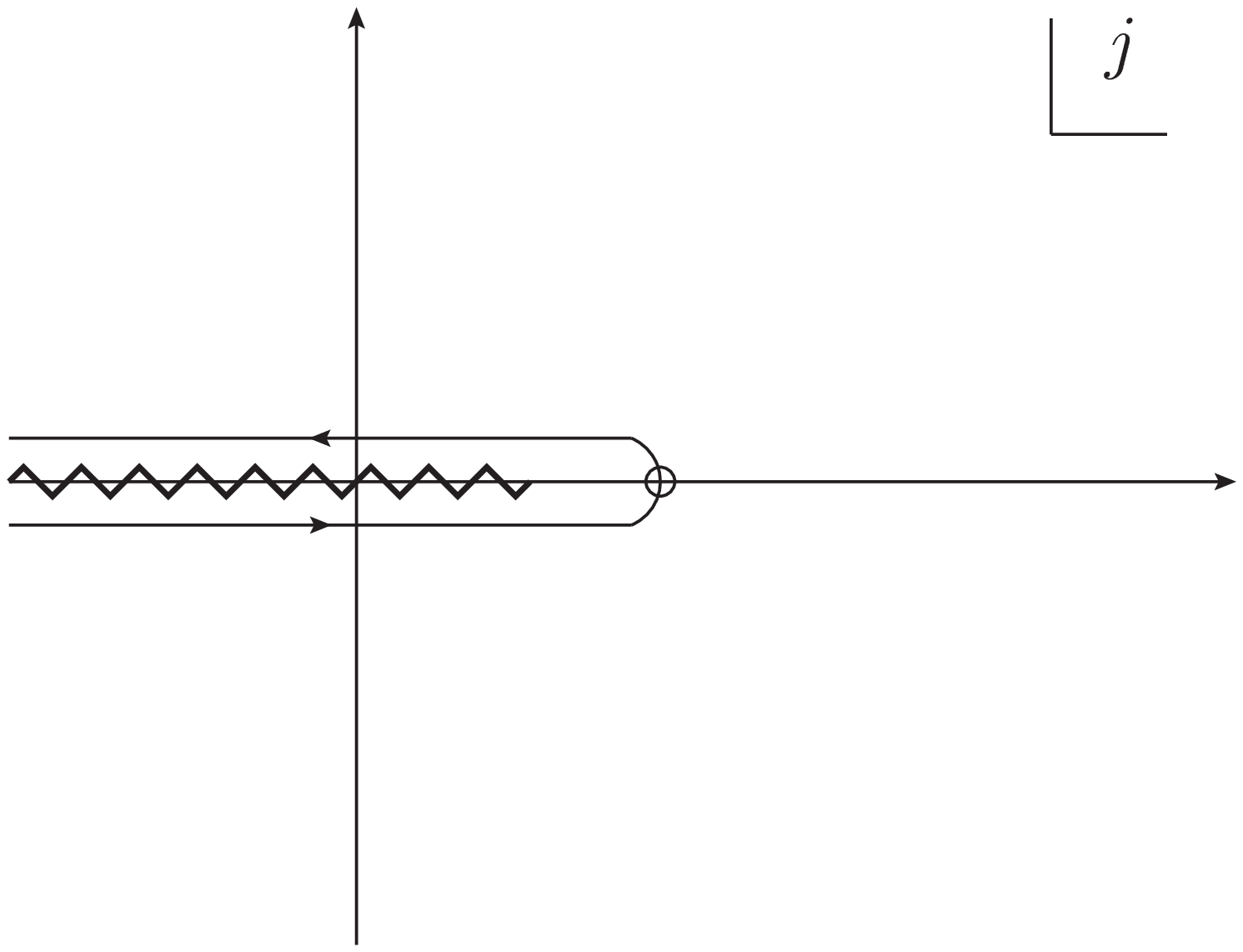} & 
  \includegraphics[scale=0.32]{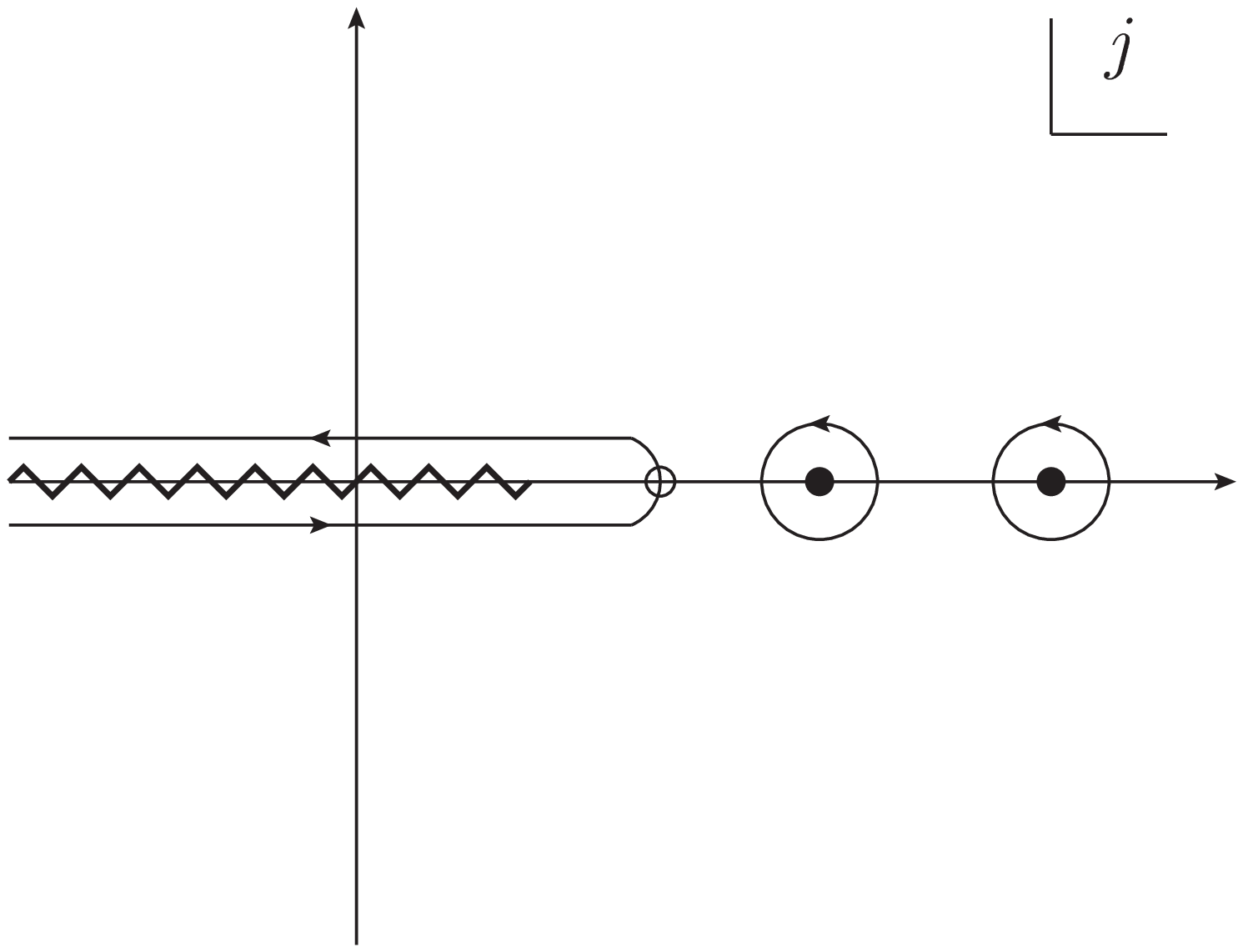} & \\
  (a) &  (b) & \\
\hline
  \includegraphics[scale=0.32]{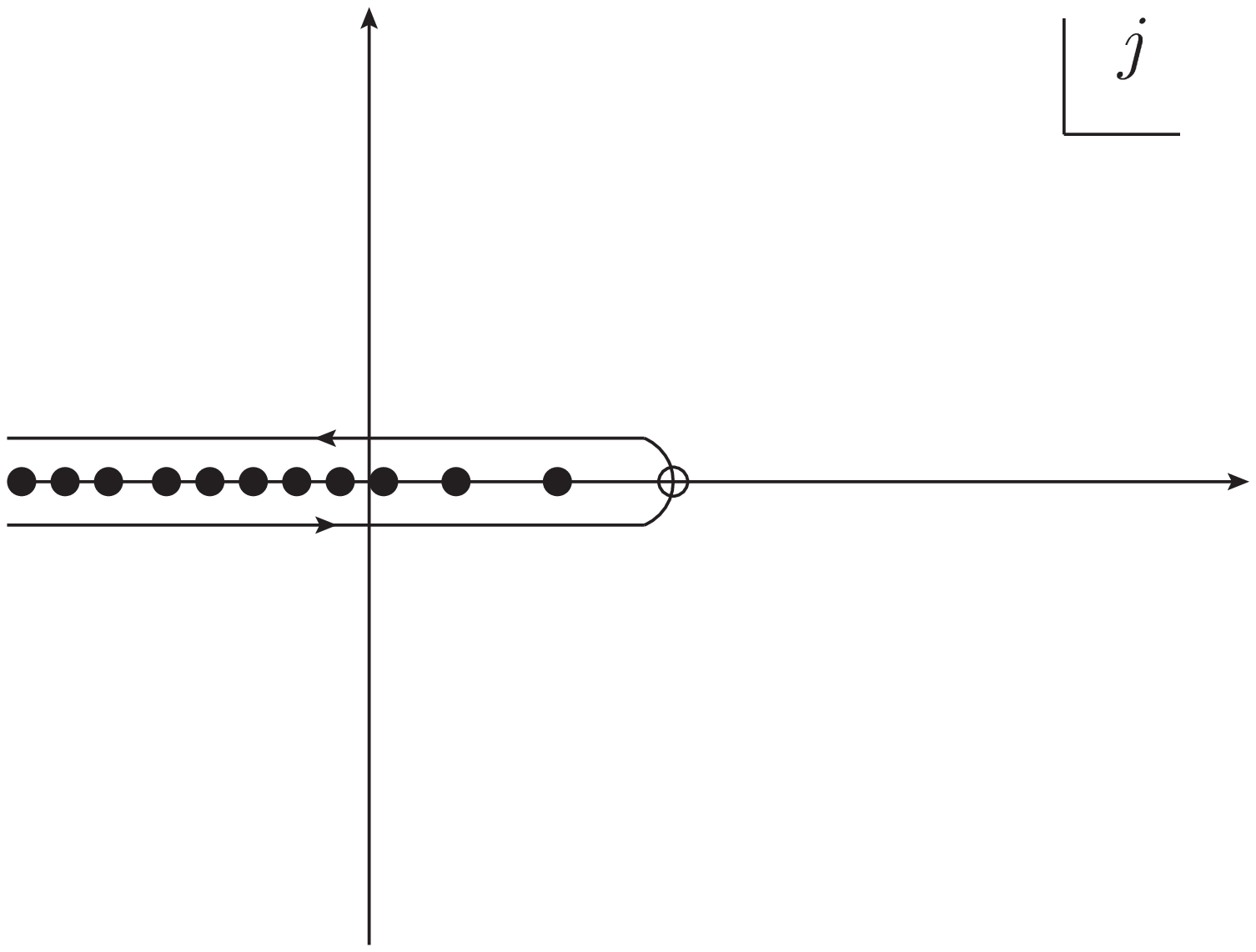} & 
  \includegraphics[scale=0.32]{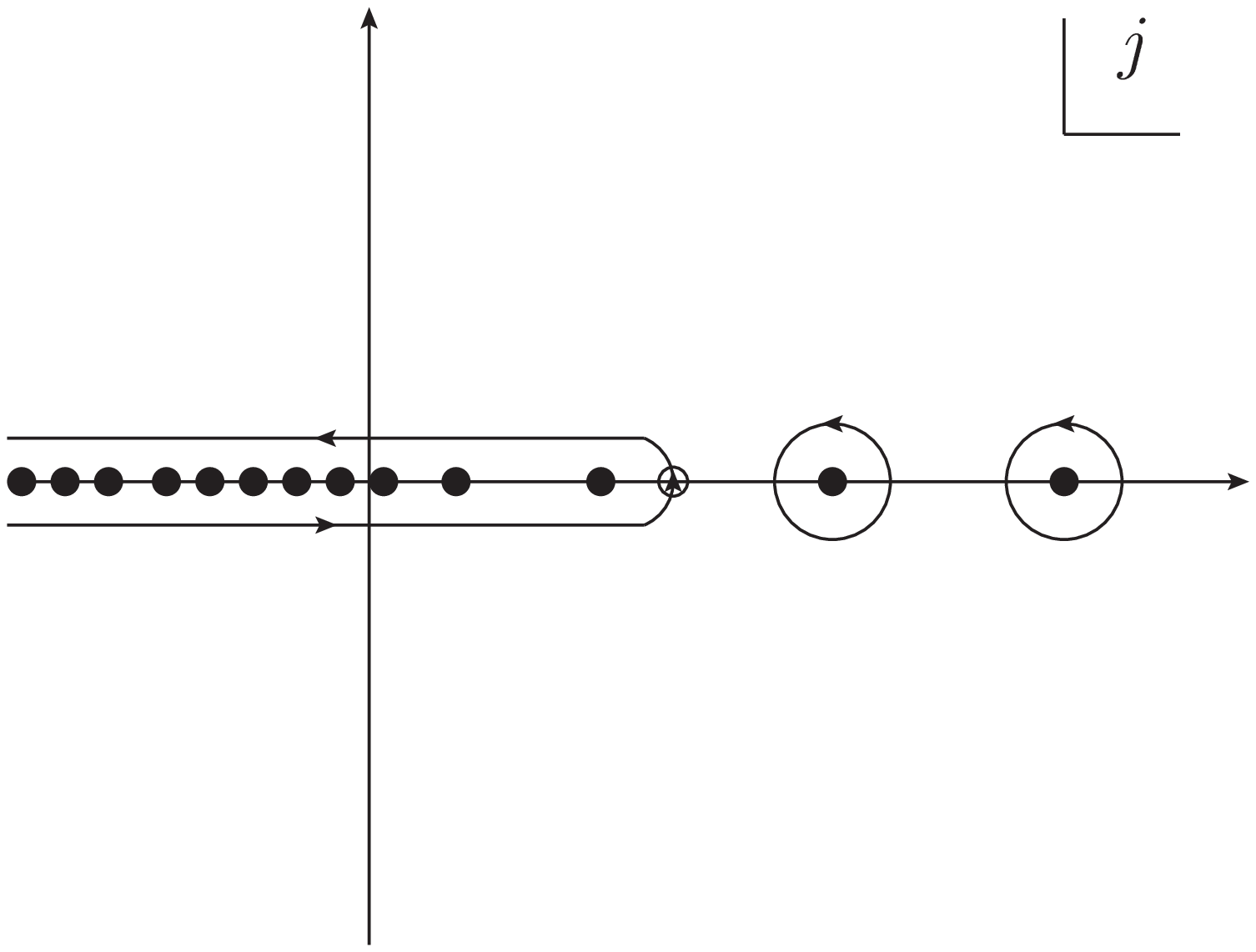} & 
  \includegraphics[scale=0.32]{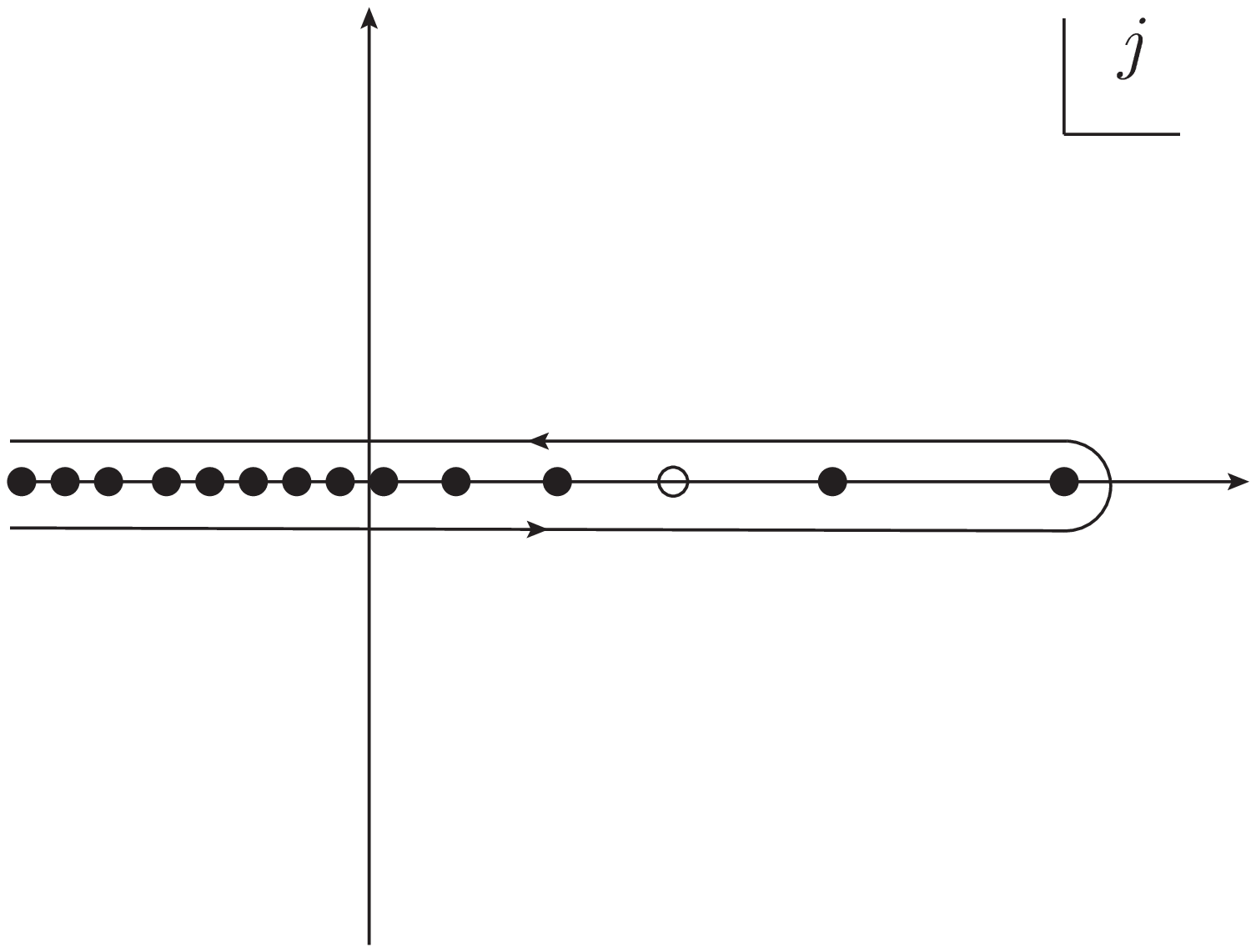} \\
(c) & (d) & (e)
\end{tabular}
\caption{\label{fig:j-plane-2} Singularities and integration contours in 
complex $j$-plane. Hard-wall model is assumed for  (a) $t \ll t_{c,1}$
and (b) $t_{c,1} \ll t$, while a holographic model for asymptotic free 
running coupling is assumed (c) with a smaller $t$ and (d, e) with a larger $t$.
The panels (d) and (e) differ only in the choice of contour. 
Black dots are poles, wiggling lines in (a, b) are branch cuts, and 
open circles in (a--e) denote saddle points of the amplitude on the 
complex $j$-plane.
}
  \end{center}
\end{figure}

When $t<t_{c,1}$, on the other hand, all the poles in the lower half 
$\nu$-plane have moved back to the upper half plane, and the scattering 
amplitude $I_i$ is given by a continuous integration of $\nu$ real axis.
The $\nu$ integration in (\ref{eq:Im-Ii}) can be converted into $j$
integration\footnote{The integration contour becomes the one in 
Figure~\ref{fig:j-plane-2}~(a).} through a change of variables $j = j_\nu$, but there is 
no pole in this $j$ integration, unlike in the traditional 
form of Regge theory amplitude (\ref{eq:traditional-Regge-amplitude}); 
the DDVCS amplitude (\ref{eq:Im-Ii-cm}, \ref{eq:Im-Ii}, 
\ref{eq:Ii-saddle-point}) that we studied in 
sections \ref{sssec:small-(-t)}--\ref{sssec:large-(-t)} for 
$t< t_{c,1}$ do not seem to be based on the Pomeron {\it pole} 
exchange idea, at least apparently.\footnote{
It should also be kept in mind that the subleading contribution 
comes from the exchange of the second Pomeron trajectory 
(the $n =2$ term in (\ref{eq:Regge-Pole-contribution-to-GG-amplitude-2}) 
for $t_{c,2} < t$, which is power suppressed, 
$\times (\Lambda^2/W^2)^{\alpha_1(t) - \alpha_2(t)}$.
Corrections to (\ref{eq:Ii-saddle-point}, \ref{eq:Ii-saddle-point2})
on the other hand, are suppressed only by 
$\sqrt{\lambda}/\ln (W^2/\Lambda^2)$.} 
If so, while the coefficient of the 
$(W^2)^{j=\alpha_n(t)}$ factor for $t_{c,n} < t$, $\gamma_{hh \P_n}(j)$, 
can be characterized as hadron--hadron--[Pomeron $j = \alpha_n(t)$] 
three point coupling (\ref{eq:P-h-h-3pt-coupling}), how should we 
characterize the coefficient of the $(W^2)^{j_{\nu^*}}$ factor for 
$t < t_{c,1}$, $g^h_{i\nu^*}(\sqrt{-t}/\Lambda)$, in the
absence of Pomeron {\it exchange} picture?

Field theory dual language is useful in characterizing 
the factors $C^{(i)}_{i\nu}$, $A_{hh}$ and 
$g^h_{i\nu}(\sqrt{-t}/\Lambda)$ 
in (\ref{eq:Im-Ii-cm}--\ref{eq:Ahh-def}). 
Let us define 
\begin{equation}
 \left[C^{(i)}(j,q) \right]_{1/\epsilon} = 
 C^{(i)}_{i\nu_j} \times \left(\frac{\epsilon}{R}\right)^{-\gamma(j)}, 
  \qquad
 \left[A_{hh}(j, W^2, t)\right]_{1/\epsilon} = 
 A_{hh} \times \left(\frac{\epsilon}{R}\right)^{\gamma(j)}
\end{equation}
for a parameter $\epsilon$ that has a dimension of [length].
Then one can see that the two factors $[C^{(i)}(j,q)]_{1/\epsilon}$ and 
$[A_{hh}(j, W^2, t)]_{1/\epsilon}$ correspond to OPE coefficient of a
``twist-2'' spin $j$ operator renormalized at $\mu = 1/\epsilon$, and 
matrix element of the spin $j$ operator renormalized at $\epsilon^{-1}$
\cite{WittenAdv.Theor.Math.Phys.2:253-2911998, Gubser1998b, Mueck1998}.
To be more precise,
we can define
\begin{eqnarray}
 \left[\Gamma_{hh \P^*}(j, t)\right]_{1/\epsilon} 
 & \equiv & \left[g^h_{i\nu_j}(\sqrt{-t}/\Lambda)\right] \times 
    \left(\epsilon \Lambda \right)^{\gamma(j)}, \nonumber \\
 & = & \frac{1}{\kappa_5^2} \int dz \sqrt{-g} P_{hh} e^{-2jA} 
      (R\Lambda)^j \left( \frac{\sqrt{-t}}{2\Lambda}\right)^{i\nu_j} 
      \frac{2}{\Gamma(i\nu_j)} (\epsilon \Lambda)^{\gamma(j)} \nonumber \\
 & & \qquad \times \left[e^{(j-2)A} \left(
          K_{i\nu_j}(\sqrt{-t}z) - 
          \frac{K_{i\nu_j}(\sqrt{-t}/\Lambda)}
               {I_{i\nu_j}(\sqrt{-t}/\Lambda)}
          I_{i\nu_j}(\sqrt{-t}z)    \right)
                   \right]
\label{eq:spin-j-form-factor}
\end{eqnarray}
by pulling out $(W^2)^j\sim (2q\cdot p)^j$ from $\left[A_{hh}(j,W^2,t)\right]_{1/\epsilon}$.
It is easy to see that the short distance scale $R$ drops out from (\ref{eq:spin-j-form-factor}),
and $\left[\Gamma_{hh\P^\ast}(j,t)\right]_{1/\epsilon}$ can be expressed only in terms of low-energy data and renormalization scale $\mu=1/\epsilon$.
Now one can see that
\begin{align}
 \sum_{j \in 2 \N} 
  [C^{(i)}(j,q)]_{1/\epsilon} \left[A_{hh}(j, W^2,t)\right]_{1/\epsilon} 
& \sim \sum_{j \in 2 \N} 
   \frac{1}{(\epsilon q)^{\gamma(j)}}\frac{1}{(q^2)^j} (q\cdot p)^j 
   \left[\Gamma_{hh\P^\ast}(j,t)\right]_{1/\epsilon} \notag\\
& \sim \sum_{j\in 2\N}
 \frac{1}{(\epsilon q)^{\gamma(j)}}\frac{1}{(q^2)^j}
   \left[q_{\mu_1} \cdots q_{\mu_j}\right]
   \bra{h(p_2)} \left[T^{\mu_1 \cdots \mu_j} \right]_{1/\epsilon}
   \ket{h(p_1)}.  
  \end{align}
Therefore, $\left[\Gamma_{hh\P^\ast}(j,t)\right]_{1/\epsilon}$ is regarded as the coefficient
of $[p^{\mu_1}\cdots p^{\mu_j}]$ 
of the matrix element of the target hadron $h$ with insertion
 of a twist-2 spin $j$ operator $[T^{\mu_1 \cdots \mu_j}]_{1/\epsilon}$
(that is, spin $j$ form factor or reduced matrix element).
$\Gamma_{hh \P^*}(j, t)$ is now defined for arbitrary $j \in \C$ in the 
expression above, not just for $j \in 2 \N$.

Using Kneser--Sommerfeld expansion of modified Bessel 
functions,\footnote{
For $0 \leq w \leq W \leq 1$ and arbitrary 
complex numbers $\mu$ and $\tau$, 
\begin{equation}
 \frac{\pi}{4}\frac{J_\mu(\tau w)}{J_\mu(\tau)}
 \left[ J_\mu (\tau) Y_\mu (\tau W) - Y_\mu (\tau) J_\mu (\tau W)
 \right] 
= \sum_{n = 1}^{\infty} \frac{1}{\tau^2 -( j_{\mu,n})^2}
  \frac{J_\mu(w j_{\mu,n}) J_\mu(W j_{\mu,n})}
       {[ J'_\mu (j_{\mu,n}) ]^2}.
\end{equation}
This is regarded as the Green function on $AdS_5$ with an infrared
cut-off, seen as a propagation of a 5D field (left hand side), or 
of a Kaluza--Klein tower of 4D fields (right hand side).  
} one can see for arbitrary complex $j$ and $t$ that 
\begin{equation}
 \left[\Gamma_{hh \P^*}(j, t)\right]_{1/\epsilon} 
  =  \sum_{n=1}^{\infty} \frac{-2}{t-m^2_{j,n}}
      \frac{\gamma_{hh \P n}(j)}{\Lambda^{j-2}}
      \frac{\left(\frac{m_{j,n}}{2}\right)^j 
      \left(\frac{m_{j,n} \epsilon}{2}\right)^{\gamma(j)}}
           {[j'_{i\nu_j}(j_{i\nu_j,n})]}
      \frac{2}{\Gamma(i\nu_j)}
      \left[\frac{R^3}{\kappa_5^2}\right]^{1/2}.
\label{eq:Pomeron-form-factor-3pt-decay}
\end{equation}
The spin $j$ form factor $\Gamma_{hh \P^*}(j,t)$ to be used in 
describing the DDVCS amplitude for $t < t_{c,1}$ and the 
hadron--hadron--[Pomeron $j = \alpha_n(t)$] three point couplings 
$\gamma_{hh \P n}(j)$ are related as above. Such a relation
between a form factor $F(t)$ and three point couplings of 
Kaluza--Klein hadrons $g_{hhn}$,   
\begin{equation}
 F(t) = \sum_{n=1}^{\infty} \frac{1}{t - m_n^2} g_{hh n} F_n,
\end{equation}
has been known for conserved currents \cite{Hong:2003jm, 
Hong:2004sa, Hong:2005np}. Here,
(\ref{eq:Pomeron-form-factor-3pt-decay}) 
establishes such a relation simultaneously for arbitrary $j \in \C$, 
not just for a given spin, and that makes it possible to 
``sum up'' Pomeron exchange amplitudes labeled by the Kaluza--Klein 
excitation level $n$.

\begin{center}
 * $\qquad \qquad \qquad \qquad \qquad $ * 
 $\qquad \qquad \qquad \qquad \qquad $ *
\end{center}

Now we know that the amplitude is given (approximately) by Kaluza--Klein 
tower of Pomeron pole exchange\footnote{As we are talking about {\it closed} 
string amplitude, contributions from a single trajectory cannot be 
regarded as their $t$-channel exchange; they can be regarded as a sum 
of $t$-channel and $u$-channel exchange, however. 
See the review material in section \ref{ssec:Pomeron-kernel}.}
for large positive $t$ ($t_{c,1} \ll t$), while the entire amplitude 
as a whole is approximated by the saddle point method for large negative $t$ 
($t \lesssim t_{c,1}$). Since $t_{c,1} = (\Lambda j_{0,1})^2$ in the 
hard wall model is positive, the latter should be applied for all the 
physical region $t \leq 0$. Such subtle things, however, depend on 
details of infrared geometry of holographic models, and cannot be 
regarded as a robust prediction of holographic QCD.

Imagine, for example, a holographic model corresponding to an
asymptotic free running coupling constant (e.g. \cite{Aharony:1998xz}). 
With the $AdS_5$ curvature changing 
over the holographic radius, the entire spectrum of the Schr\"{o}dinger 
equation 
$- \Delta_j(t) \Psi^{(j)}_{i\nu}(t,z) = (E_\nu/R^2) \Psi^{(j)}_{i\nu}(t,z)$ 
becomes discrete \cite{BrowerJHEP0712:0052007}.\footnote{
Although we talk of ``asymptotic free'' running,
it is safer to consider gravity dual models corresponding to gauge theories
with asymptotic free running only up to some energy scale,
above which they become strongly coupled conformal theories (e.g. \cite{Aharony:1998xz}).
In this case, although there are many bound states in the spectrum of the Schr\"{o}dinger
equation, the continuous spectrum still remains.
Thus, the densely packed poles are still followed by a branch cut along
the real $j$ axis in the small ${\rm Re}\: j$ region.
As long as one pays attention to a certain region in the $j$-plane,
rather than to formal difference in the literally large negative ${\rm Re}\: j$ limit,
the such models pass for ``asymptotic free'' gravity dual models.
Whenever we refer to asymptotic free gravity dual models in this article,
this must be understood. \label{fn:asymp-free}
}
It is now more convenient to write the amplitude in complex $j$-plane, 
rather than in $\nu$-plane. 
The branch cut 
at $j \leq j_0$ in the hard wall model is replaced by a densely 
packed poles along the real $j$ axis, and there is no special value 
like $t_{c,1}$ for some critical change in the spectrum. 
The spectrum may change for different values of $t$, like in 
Figure~\ref{fig:j-plane-2}~(c) and \ref{fig:j-plane-2}~(d, e), but 
they are not different qualitatively. 
Thus, there is nothing wrong a priori in choosing the integration 
contour as in (d) or (e), even for a given spectrum. 

There exists a convenient choice of the contour in the $j$-plane, 
however. That is to let the contour to pass the saddle point
$j = j_{\nu^*}$, and treat all the poles to the right of the saddle
point (i.e., $j_{\nu^*} < {\rm Re} \; j$) as isolated discrete
contributions. All the rest are treated as if they formed 
a continuous spectrum as in the hard wall model.  
As we will see in the following, whether there are some poles remaining 
to the right of the saddle point (e.g., Figure~\ref{fig:j-plane-2}~(b,
d, e)) or not still works as a criterion for various physical
transitions, independently of detailed difference in various holographic
models. 

The saddle point in the $j$-plane is understood as a consequence of 
two competing effects in the DDVCS.
One is the $(W^2/\Lambda^2)^j \simeq (q^2/x \Lambda^2)^j$ factor, which is
large for large real $j$. The coupling of spin $j$ string state 
with a virtual photon, on the other hand, behaves as 
$(\Lambda/q)^{2j + \gamma(j)}$, where $\gamma(j)$ is the 
anomalous dimension of the operator corresponding to the string state. 
This second factor is small for large real $j$. These two factors combined, 
(\ref{eq:general-j-int}), 
 forms a saddle point $j^*$ in the $j$-plane for $W\gg q \gg \Lambda$ (meaning $x\ll 1$);
 it is quite generic for holographic models with 
$R^2\ll \alpha'$ that the large $j$ behavior\footnote{We are not talking
about $j$ as large as $\sqrt{\lambda}$, however. } 
of the anomalous dimension 
is $\gamma(j) \sim [\sqrt{\lambda} j]^{1/2}$.
(We neglect here correction from the factor of power of $\sqrt{t}/\Lambda$.)
Thus, in the right-hand side of the saddle point $j^\ast$ in the $j$-plane, 
$j^* < {\rm Re} \, j$, 
the larger the real part of a pole, the larger the contribution of the
pole is. Writing down the amplitude as a sum of these poles, starting 
from the one with the largest real part of ${\rm Re}\: j$
 to the ones with smaller ${\rm Re}\: j$, we obtain 
a finite term sum that gives a good approximation to the amplitude. 
Individual contributions from the poles in the left hand side of the saddle point $j^\ast$,
 on the other hand, becomes larger and larger as the 
real part of $j$ of the poles become smaller. Thus, their sum does not make sense\footnote{
In ``asymptotic free'' gravity dual models in the sense of footnote \ref{fn:asymp-free},
the series of poles stop at a certain small value of ${\Re }\: j$,
and  a branch cut starts toward large negative ${\rm Re}\: j$.
Thus, the contour can be chosen so that the amplitude is given by a sum of all
the individual pole contributions and an integral around the branch cut.
The sum is thus, formally, well-defined in this sense.
It will be obvious, because of the discussion in the main text, however,
that the resulting many-term summation contains large cancellation
between the cut contribution and the pole contributions,
and is not as practically useful an expression as what we described in the main text.
}; 
 all contribution in the left hand side should be treated as 
an integration on a contour as in Figure~\ref{fig:j-plane-2}~(d). All of 
these contributions combined can be evaluated by the saddle point
method, as we have presented by using the hard wall model, and their 
contribution is 
\begin{align}
 \sfrac{1}{\sqrt{\lambda}x}^{j^\ast }
   \sfrac{\Lambda}{q}^{\gamma_{j^\ast}}.
\label{eq:saddle-point-amplitude}
\end{align}
This contribution is even smaller than the one from a pole whose real
part is even slightly larger than $j^*$.
This is why it is convenient to take the contour
in $j$-plane so that it passes the saddle point as shown 
in Figure~\ref{fig:j-plane-2}~(d).

\begin{figure}[tbp]
  \begin{center}
\begin{tabular}{ccc}
  \includegraphics[scale=0.6]{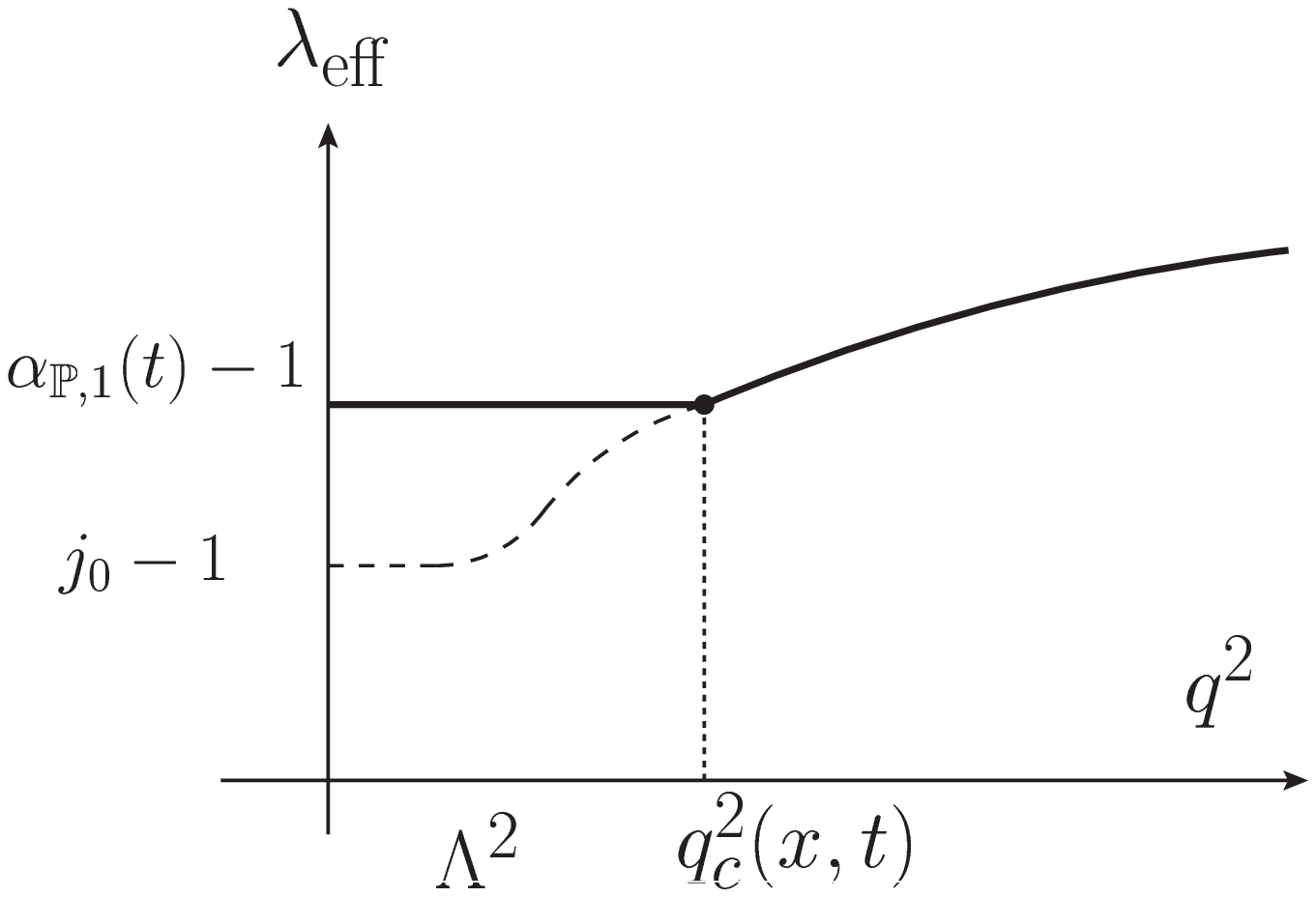} & &
  \includegraphics[scale=0.55]{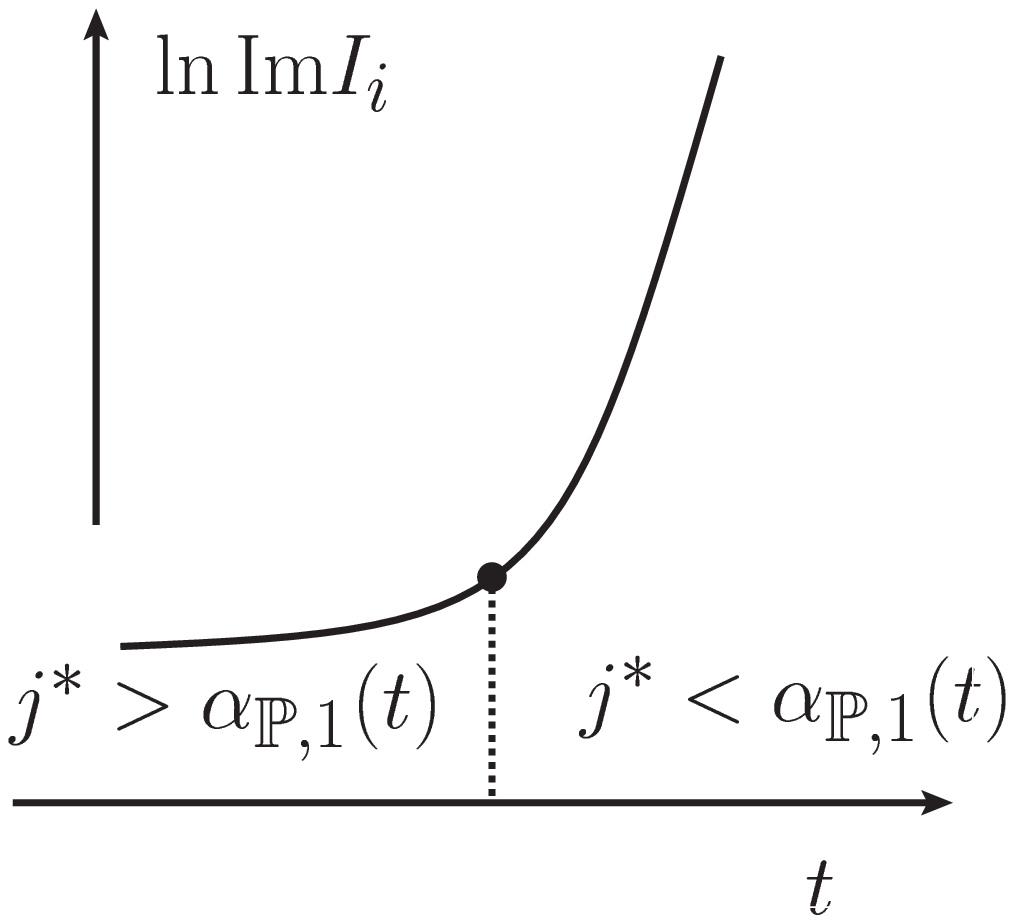} \\
(a) & & (b)
\end{tabular}
\caption{\label{fig:transition} Transitions (crossover, to be more
   precise) in (a) $\lambda_{\rm eff.}$ as a function of $q^2$ for 
fixed $(x, t)$, and (b) in [$d [{\rm Im} \; I_i]^2 /dt$] as a function 
of $t$ for fixed $(x, q^2)$.
}
  \end{center}
\end{figure}

\begin{figure}[tbp]
 \begin{center}
    \includegraphics[scale=0.2]{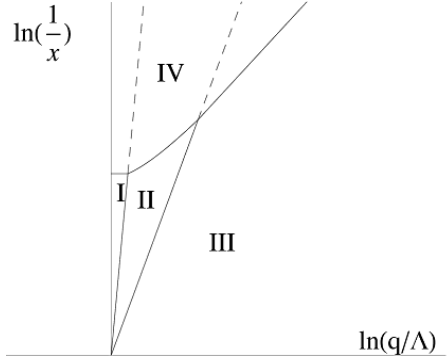}  
  \caption{\label{fig:phasediagramSC} Phase diagram (at a given value of
  $t$) in the strong coupling regime. 
I: leading singularity phase, 
II: saddle point phase 
 with $\gamma_{\rm eff.} < 0$ and 
III: saddle point phase (spin-2 phase) with $0< \gamma_{\rm eff.}$. 
A region where higher genus amplitudes are just as important 
as the sphere amplitude, phase IV, is characterized by the 
condition $|\chi^\text{sphere}(b \sim 1/\Lambda)| \sim {\cal O}(1)$, 
just as in \cite{Hatta:2007}. The leading singularity phase I is absent 
in the forward scattering $t=0$ in the hard wall model, 
as in \cite{Hatta:2007}.
This phase diagram in the strong coupling regime looks quite similar 
to that of QCD. The boundary lines between the phase I and II, 
and II and III, however, are nearly vertical in the strong coupling 
regime, rather than being nearly horizontal as in the weak coupling regime. }
 \end{center}
\end{figure}

Then we can understand that there is a transition depending on 
whether the saddle point value $j^*(x,q,t)$ has a larger real part 
than that of the leading singularity (one in the $j$-plane with the largest real part),
 $j = \alpha_{\P 1}(t)$. 
It is easy to see this in $\lambda_{\rm eff.}$ (which we have 
already discussed in section \ref{sssec:small-(-t)}).
\begin{align}
 \lambda_\text{eff}(x,t,q^2)&=
\begin{cases}
\alpha_{\P, 1}(t) -1 & q<q_c(x,t),
\\
j^{\ast} - 1  & q>q_c(x,t),
\end{cases}
\label{eq:phase-transition-in-lambda-eff}
\end{align}
where the transition is induced (assuming that $\gamma(j)$ is a decreasing
function of $j$ along the real axis) at 
\begin{equation}
 \left. \frac{\partial \gamma(j)}{\partial j} 
 \right|_{j = \alpha_{\P, 1}(t)} = 
 \left. \frac{\partial \gamma(j)}{\partial j} \right|_{j = j^*(q=q_c)}
  =  \frac{\ln (1/\sqrt{\lambda}x )}{\ln (q_c /\Lambda)}.
\end{equation}
Schematically, it behaves as in Figure~\ref{fig:transition}~(a).
In the following, we will refer to the two phases\footnote{The KK-sum
Spin-sum phase (saddle point phase) is divided into two phases, when 
the real part of the scattering amplitude is studied. One of them 
is still called KK-sum Spin-sum phase (saddle point phase), and the
other as KK-sum low-spin phase (spin-2 phase). See \cite{Hatta:2007} 
and section \ref{ssec:real-part} of this article.} as 
\begin{itemize}
 \item {\bf Low-KK Spin-sum phase} or {\bf Leading pole phase} (leading
       singularity phase), where the leading singularity has a larger real part 
       than the saddle point, and 
 \item {\bf KK-sum Spin-sum phase} or {\bf Saddle point phase},
       where the saddle point value is larger than the real part of the 
leading singularity in the $j$-plane. 
\end{itemize}
It is intuitively obvious in gravity dual descriptions 
that the saddle point phase 
is realized (and such observables as $\gamma_{\rm eff.}$ and $\lambda_{\rm eff.}$ are controlled by 
the saddle point value $j^*$) for larger $q^2$, not in the other 
way around; the holographic wavefunction of photon with larger virtuality 
makes the photon-[spin-$j$ string] couplings weaker, and 
the couplings for larger spin states are affected more severely than 
those for smaller spin states, because of the stronger power-law 
behavior of the holographic wavefunctions of the higher spin states. 
See (\ref{eq:C(i)-def}) and definition of 
$\gamma_{\gamma^* \gamma^* \P n}(j)$.
The factor $(\Lambda/q)^{2j + \gamma(j)}$ comes from there, and 
consequently, the saddle point value $j^*$ shifts to the right in the
$j$-plane to enter into the saddle point phase. 
See Figure~\ref{fig:phasediagramSC}.

In fact, the transition between the two phases is not a singular phase 
transition but a crossover, unless we literally take a small $x$ limit. 
As long as $\ln(1/x)/\sqrt{\lambda}$ remains finite, the saddle point 
approximation is never exact, and the notion of the saddle point itself 
should be accompanied by a width proportional to 
$[\ln (1/x) /\sqrt{\lambda}]^{-1/2}$. Although we defined the two phases 
by simply comparing the real part of the leading singularity
$j = \alpha_{\P, n=1}(t)$ and the saddle point $j^*$, we should also keep 
in mind that the $n = 2$ and higher Kaluza--Klein contributions give
rise to finite corrections (for finite $\ln (1/x)/\sqrt{\lambda}$) 
to the leading pole contribution in the leading pole phase; 
the corrections (and cancellation) may be sizable for negative $t$ 
and finite $\ln(1/x)/\sqrt{\lambda}$. For those reasons, the transition 
between the two phases can be a singular genuine phase transition only 
in the $\ln(1/x)/\sqrt{\lambda} \rightarrow \infty$ limit.

Phases are determined by kinematical variables $(x,t,q^2)$, and 
kinematical variable dependence of all the observables, not just 
$\lambda_{\rm eff.}$, will be different for different phases. 
Real part to imaginary part ratio of the scattering amplitude, 
which we study in section \ref{ssec:real-part}, is an example. 
Note also that the saddle point value $j^*$ depends on momentum 
transfer $t$, not just on $(x,q^2)$, and the leading Pomeron pole 
also depends on $t$. Thus, the crossover should also be 
induced by the kinematical variable $t$ (at least somewhere in the $t$-plane).
In section \ref{sssec:slope}, we study another observable, $t$-slope 
parameter $B(x,\eta = 0, t,q^2)$, and discuss its crossover behavior 
(like in Figure~\ref{fig:transition}~(b)).

\subsubsection{Slope parameter of the forward peak}
\label{sssec:slope}

We have so far focused on various features of the DDVCS amplitude 
that are robust and do not depend on detailed difference of 
holographic models. Individual holographic models, however, 
have full control over non-perturbative aspects of hadron physics, 
and more (possibly model dependent) information can be extracted. 
The expression (\ref{eq:Ii-saddle-point}), for example, tells us 
how to calculate full $t$-dependence of the (imaginary) part of 
the DDVCS amplitude, not just for $t=0$ and the $\Lambda^2 \ll -t$
asymptotic region. 

The slope parameter of forward peak (t-slope parameter) in the elastic scattering of 
two hadrons, 
\begin{equation}
 B(s,t) \equiv \frac{\partial}{\partial t}
   \ln \left[ \frac{d \sigma_{\rm el}}{dt} (s,t) \right], 
\label{eq:introduction-SP}
\end{equation}
is an observable that characterizes the transition from $t \approx 0$
region to $\Lambda^2 \ll -t$ region, and has been measured for 
$p+p$ and $p + \bar{p}$ scattering at various energy scales 
(e.g., \cite{BaronePredazzi}).
The slope parameter has also been measured in HERA experiment for 
$\gamma^* + p \rightarrow \gamma + p$ scattering 
(DVCS) \cite{Aktas:2005ty, Aaron:2007cz, Aaron2009}. In this 
section \ref{sssec:slope}, we will first use the hard wall model to derive 
an explicit prediction of the slope parameter in the photon--hadron 
scattering at vanishing skewedness $\eta = 0$ (DDVCS process) for simplicity. 
Later on, we will discuss how much predictions on the slope parameter 
could be different for different holographic models, and discuss the
possible crossover to be seen in the slope parameter (for 
some holographic models). 

We define the slope parameter in non-skewed DDVCS by\footnote{
We could define the slope parameter 
$B$ by the absolute value $|I_i|$,
 not by the imaginary part ${\rm Im}\: I_i$ as in (\ref{eq:def-slope-parameter}).
This choice, however, makes no difference as long as $(-t)\lesssim \Lambda^2$,
 and $\gamma_\text{eff.}<0$ (the condition for the real part to be outside spin-2 phase); see section 4.3.1.
This is because in this region the real-to-imaginary ratio of $I_i$ is not dependent on $t$, (\ref{eq:phase-transition-of-ratio}).
It is very likely that the most of the kinematical reach of DVCS measurement in HERA
 \cite{Aktas:2005ty, Aaron:2007cz, Aaron2009} is also in this region.
}
\begin{align}\label{eq:def-slope-parameter}
B_i(x, \eta=0, t, q^2)= 2 \;  \frac{\partial}{\partial t}
  \ln {\rm Im} \; I_i(x,\eta=0,t,q^2).
\end{align}
Let us first work out the prediction of the slope parameter in the 
saddle point phase, and then present the result in the leading pole
phase later. 

In the saddle point phase (KK-sum spin-sum phase), 
the factor $(1/x)^{j^*} \times (\Lambda/q)^{\gamma(j^*)}$ 
in (\ref{eq:Ii-saddle-point}) does not contribute to the slope parameter, 
because the saddle point value $j^* = j_{\nu^*}$ does not depend on $t$ for 
$0 \lesssim (-t) \lesssim \Lambda^2$. The slope parameter comes entirely 
from Pomeron--hadron form factor 
$\left[\Gamma_{hh \P^*}(j^*, t)\right]_{1/\epsilon} \propto 
g^h_{i\nu^*}(\sqrt{-t}/\Lambda)$, which is regarded as 
a ``spin $j = j^*$ form factor''.
The Pomeron--hadron--hadron coupling behaving like a form factor 
is similar to the idea advocated in \cite{Landshoff:1971gm,
Landshoff:1971pw}, but the form factor turns out not to be precisely 
the same as the electromagnetic (spin 1) or gravitational (spin 2) one. 
There exists a notion of form factor $\Gamma_{hh \P^*}(j, t)$ that 
is holomorphic in spin $j$, and the one with the saddle point value of 
$j = j^*$ is relevant for the DDVCS amplitude; the saddle point value 
$j^* = j_{\nu^*}$ is determined by $(x,t,q^2)$ as we have already seen in 
sections \ref{sssec:small-(-t)}--\ref{sssec:large-(-t)}. 

\begin{figure}[tbp]
  \begin{center}
  \includegraphics[scale=0.7]{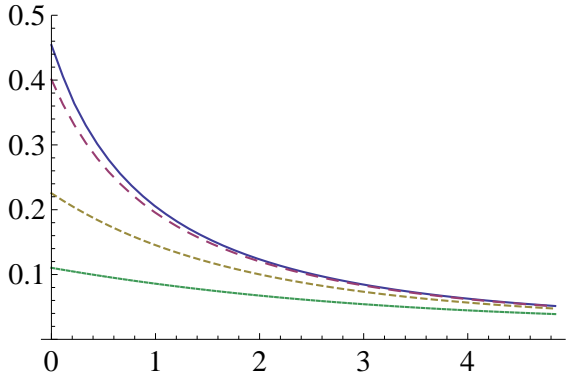} 
\caption{\label{fig:slope} (color online) 
Slope $B$ of the forward peak in DDVCS. 
The dimensionless value $B \times \Lambda^2$ is shown as a function 
of $i\nu^*$ (\ref{eq:nu ast for small -t}); from top to bottom, blue 
(solid line) curve is for $\sqrt{-t}/\Lambda \simeq 0.01\mbox{--}0.1$, 
red (long dashed) one for $\sqrt{-t}/\Lambda = 1.$, yellow (dashed) one 
for $\sqrt{-t}/\Lambda = 3.$, and green (short dashed) one for 
$\sqrt{-t}/\Lambda = 6$. We used the wavefunction of the target hadron 
(\ref{eq:normalizable mode}) for the first excited ($n=1$) mode 
with conformal dimension $\Delta = 5$, and the 't Hooft coupling was 
set to $\sqrt{\lambda} = 10$ in this calculation. 
$t$-dependence is very weak for $0 \leq -t \lesssim \Lambda^2$.}
  \end{center}
\end{figure}

It is straightforward to calculate the slope parameter in the saddle
point phase, 
\begin{equation}
 B(x,\eta = 0, t, q^2) \simeq 2\frac{\partial}{\partial t} 
   \ln \left[ \Gamma_{hh \P^*}(j^*, t) \right],
\label{eq:slope-via-saddlepoint}
\end{equation}
by using the explicit expression of the form factor\footnote{
A grossly incomplete list of literatures on fixed spin form 
factor in holographic methods will include \cite{Hong:2003jm, Hong:2005np, 
Hong:2007ay, Hashimoto:2008zw, deTeramond:2006xb, Brodsky:2007hb, 
Grigoryan:2007vg, RodriguezGomez:2008zp, Abidin:2008ku, Abidin:2008hn, 
Brodsky:2008pf, Carlson:2008ha}.}  (\ref{eq:spin-j-form-factor}).
The result is shown in Figure~\ref{fig:slope}.
The larger the spin $j^*$ (and hence $i\nu^*$), the smaller the slope. 
At $t = 0$, the slope parameter is now the same as the ``charge radius
square'' of the hadron under a ``spin-$j^*$ probe''. 

When the slope parameter is seen as a function of $q^2$ and $W^2$, 
it changes only through the change in the saddle point value 
$j^* = j_{\nu^*}$. Because 
\begin{equation}
 \frac{\partial [i\nu^*(q^2, W^2)]}{\partial \ln(q^2/\Lambda^2)} 
 \simeq \frac{i\nu^*}{\ln(q^2/\Lambda^2)} > 0, \qquad 
 \frac{\partial [i\nu^*(q^2, W^2)]}{\partial \ln (W^2/\Lambda^2)}
 \simeq - \frac{i\nu^*}{\ln (W^2/\Lambda^2)} < 0,   
\label{eq:saddle-pt-on-QandW}
\end{equation}
the slope parameter decreases for larger $\ln (q^2/\Lambda^2)$, 
and increases for larger $\ln (W^2/\Lambda^2)$ in this 
saddle point phase prediction. It must be more sensitive to 
$\ln(q^2/\Lambda^2)$ than to $\ln (W^2/\Lambda^2)$ for small $x$
[$\ln(1/x) \gg \ln(q^2/\Lambda^2)$], because of the denominators 
in (\ref{eq:saddle-pt-on-QandW}).
That is, we find the following qualitative prediction of the 
saddle point phase:\footnote{In this article, we assume the 
generalized Bjorken regime, 
(\ref{eq:double deeply virtual}, \ref{eq:Bjorken-2}), 
exponentially small $x$ (\ref{eq:exp-small-x}), and 
large 't Hooft coupling, with an extra constraint 
$j^* \lesssim {\cal O}(1)$ (i.e., $i\nu^* \lesssim \lambda^{1/4}$).} 
\begin{align}
 \frac{\partial B}{\partial \ln (q/\Lambda)} &<0,
&
\frac{\partial B}{\partial \ln (W/\Lambda)} &>0,
&
\left|\frac{\partial B}{\partial \ln (W/\Lambda)}\right| 
&\ll
\left|\frac{\partial B}{\partial \ln (q/\Lambda)}\right|.
\label{eq:3properties-of-slope-parameter}
\end{align} 

There are three remarks here, before we move on to discuss predictions 
on the slope parameter from the leading pole phase.
The first remark is on the size of the slope parameter 
of the {\it forward peak}. 
It is of order $1/\Lambda^2$ in this prediction of the saddle point
phase, and is not tied to the slope of {\it Pomeron trajectories};
the asymptotic slope of the trajectories is of order
$(1/\Lambda^2) \times (1/\sqrt{\lambda})$ 
(see (\ref{eq:FT-of-linear-traj})). 

Secondly, let us discuss the momentum transfer dependence 
of the slope parameter within the saddle point phase.
All the physical kinematical region $t \leq 0$ is in the 
saddle point phase in the hard wall model, and the phase 
appears at least for sufficiently negative $t$ ($t < t_{c,1}$) for 
any other holographic models.
Because of the power-law behavior of the form factor in UV conformal 
theories, the slope parameter behaves as $\propto 1/(-t)$ 
for large $(-t)$. The slope parameter does not diverge toward 
$t \rightarrow 0^{-}$ (if all the range of $t \leq 0$ is in the 
saddle point phase), however, and it approaches a finite
plateau value of order $1/\Lambda^2$ instead. 
Whether there is a range of $t$ (in small $|t|$) where the amplitude 
shows exponential fall-off in $t$ is about the stability of the plateau 
value of $B$ for a finite range of $t$, and about the plateau 
range and plateau value. This is a quantitative question whose answer 
depends on details of holographic models, and we do not discuss more 
about this question in this article.  

Finally, it is important to note that the argument above 
on the $W^2$ and $q^2$ dependence of the slope parameter $B$
 (\ref{eq:3properties-of-slope-parameter}) can 
be applied to other observables of the DDVCS amplitudes, as long 
as they depend on $q^2$ and $W^2$ only through the saddle point value
$j^*$. Their $\ln(q^2/\Lambda^2)$ and $\ln (W^2/\Lambda^2)$ dependence 
are always opposite, and the $\ln (q^2/\Lambda^2)$ dependence must 
be stronger at sufficiently small $x$,
because the saddle point $j^\ast$ has three properties:
\begin{align}
  \frac{\partial j^\ast}{\partial \ln (q/\Lambda)} &>0,
&
\frac{\partial j^\ast}{\partial \ln (W/\Lambda)} &<0,
&
\left|\frac{\partial j^\ast}{\partial \ln (W/\Lambda)}\right| 
&\ll
\left|\frac{\partial j^\ast}{\partial \ln (q/\Lambda)}\right|.
\label{eq:3properties-for-saddle-point}
\end{align}
This argument can be applied, for example, to $\lambda_{\rm eff.}$ and $\gamma_{\rm eff.}$.

Let us now move on to the leading pole phase. 
Although all the physical kinematics $t \leq 0$ is in the saddle point 
phase in the hard wall model,
 there may also be a range of the leading pole phase within
the physical region in some holographic models, such as 
asymptotic conformal models with negative $t_{c,1}$ and asymptotic
free models, as we have seen in section \ref{sssec:regge}. 
In such holographic models, the crossover between the two phases 
may be observed in the slope parameter within the physical kinematic 
range $t \leq 0$.
The slope parameter in the leading pole phase comes mainly from 
the factor
\begin{equation}
 \left.
 \left[
 \left(\frac{W^2}{\Lambda^2}\right)^j \gamma_{\gamma^* \gamma^* \P_1}(j)
 \right]
 \right|_{j=\alpha_{\P, 1}(t)} \sim 
 \left.
 \left[\left(\frac{1}{x}\right)^{j}  
       \left(\frac{\Lambda}{q}\right)^{\gamma(j)}
 \right]
 \right|_{j=\alpha_{\P, 1}(t)}
\end{equation}
in the small $x$ and large $q^2$ region; the $t$-dependence 
in the Pomeron coupling $\gamma_{\gamma^\ast \gamma^\ast \P_1}(\alpha_{\P,1}(t))$ is now ignored. Thus, we have 
\begin{equation}
\label{eq:slope-paramter-in-pomeron-pole-phase}
 B(x, \eta = 0, t, q^2) \simeq 
  2 \frac{\partial \alpha_{\P, 1}(t)}{\partial t}
\left[
 \ln (1/x) + 
\left.\frac{\partial \gamma}{\partial j}\right|_{j=\alpha_{\P,1}(t)}
\ln (\Lambda/q)
\right].
\end{equation}
For sufficiently small $x$, this is of the order of 
\begin{equation}
B \simeq \frac{1}{\Lambda^2} \frac{\ln (\frac{1}{x})}{\sqrt{\lambda}}.  
\end{equation}
This is larger than the slope parameter in the saddle point 
phase, $B \sim {\cal O}(1/\Lambda^2)$, in the small $x$ 
regime (\ref{eq:exp-small-x}) we have been studying in this article. 
Thus, the crossover between the two different phases are induced
schematically as in Figure~\ref{fig:transition}~(b), where the 
large slope fall-off of the scattering amplitude in larger (more positive) $t$ breaks into 
smaller slope behavior in smaller (more negative) $t$, when 
$[\partial \gamma/\partial j]|_{j=\alpha_{\P, 1}(t)}$ comes close to 
$[\partial \gamma/\partial j]|_{j=j^*}$.

Even in asymptotic conformal theories with negative $t_{c,1}$ and in 
asymptotic free theories, where the leading pole phase may exist 
in the physical kinematical region $t \leq 0$, one always enters into 
the saddle point phase for sufficiently high $q^2$; this behavior is
understandable just like in the case of $\lambda_{\rm eff.}$.

It is an interesting question, at least from a theoretical 
perspective,\footnote{In reality, unitarity limit is reached and 
$1/N_c$-suppressed contributions (that we ignored throughout in this
article) also become important at sufficiently high energy.
Such high energy region is described as phase IV in figure \ref{fig:phasediagramSC}.
} and also from the context of fitting 
experimental data, 
whether such a crossover should be expected within the physical 
kinematical range $t \leq 0$.
The answer is ``no'' in the hard wall model. But, the answer 
depends on infrared geometry of holographic models, 
and a robust conclusion cannot be drawn only from the experience 
in the hard wall model. It is interesting to know the answer to 
this question in holographic models whose infrared geometry is 
fully faithful to the equation of motions of Type IIB string theory 
\cite{future}.

\subsection{Impact Parameter Dependence of the Imaginary Part}
\label{ssec:b-space}

One can take a Fourier transform of ${\rm Im} \; V_i(x, \eta, t, q^2)$'s 
with $\eta = 0$, to obtain distributions in the impact parameter space. 
Such distributions are interesting on their own, because they show 
transverse spacial distribution of longitudinal momenta in the 
target hadron \cite{Burkardt:2002hr}.
Also, unitarization of high energy elastic scattering amplitude and DIS 
cross section needs to be discussed in the impact parameter space, 
because all the partial wave amplitudes, i.e., for any values of 
impact parameter, should be unitary (e.g. \cite{BaronePredazzi}).
Although the impact parameter dependent amplitude of hadron--virtual 
photon elastic scattering has already been formulated in 
\cite{BrowerJHEP0712:0052007, Brower2007a, Hatta:2007, Brower2007, 
Brower:2010wf} in holographic calculations, we elaborate more on this 
in the following, and find a result
(\ref{eq:trans-profile-Lq}--\ref{eq:trans-profile-Sq-Sb}) and 
the phase diagram in Figure~\ref{fig:phase}, which certainly refines 
the results that are already found in the literature.  

The impact parameter dependence, and hence the momentum transfer 
dependence, comes mainly from the Pomeron kernel.
The profile from the Pomeron kernel 
is extended over the size of order $b\Lambda\sim [\ln(1/x)/\sqrt{\lambda}]^{1/2}\gg 1$,
as we will also see explicitly later.
Although the virtual photon 
wavefunction also have $t = - \Delta^2$ dependence through 
$q_{1,2}^2 = q^2 + \Delta^2/4$, this dependence is relevant to the 
transverse profile only in the short distance of order $b \sim 1/q$,
so we neglect this contribution. 
Thus, the impact parameter dependence purely comes from the Pomeron 
kernel. Fourier transform of the kernel is 
\begin{align}\label{def:kernel in b space}
 {\cal K}(s,b,z,z')=\int \frac{d^2 \vec{\Delta}}{(2\pi)^2}
    e^{-i\vec{\Delta}\cdot \vec{b}} {\cal K}(s,t=-\Delta^2,z,z')
  = \frac{1}{4\pi}\int_0^\infty d \Delta^2 J_0(b\Delta) {\cal K}(s,t,z,z').
\end{align}

The confinement effects (finite $\Lambda$ effects)
 are crucial for the impact parameter profile at long distance, $b\Lambda \gg 1$.
The profile in hard wall model shows quite different behaviors
from one in conformal theory (corresponding to $\Lambda\rightarrow 0$) \cite{Brower2007}.
Because we are interested in hadrons in confinement,
we examine the explicit form of the Pomeron kernel 
including the second term of (\ref{eq:kernel in hw2}), in detail.

\begin{figure}[tbp]
\begin{center}
\begin{tabular}{ccc}
  \includegraphics[scale=0.6]{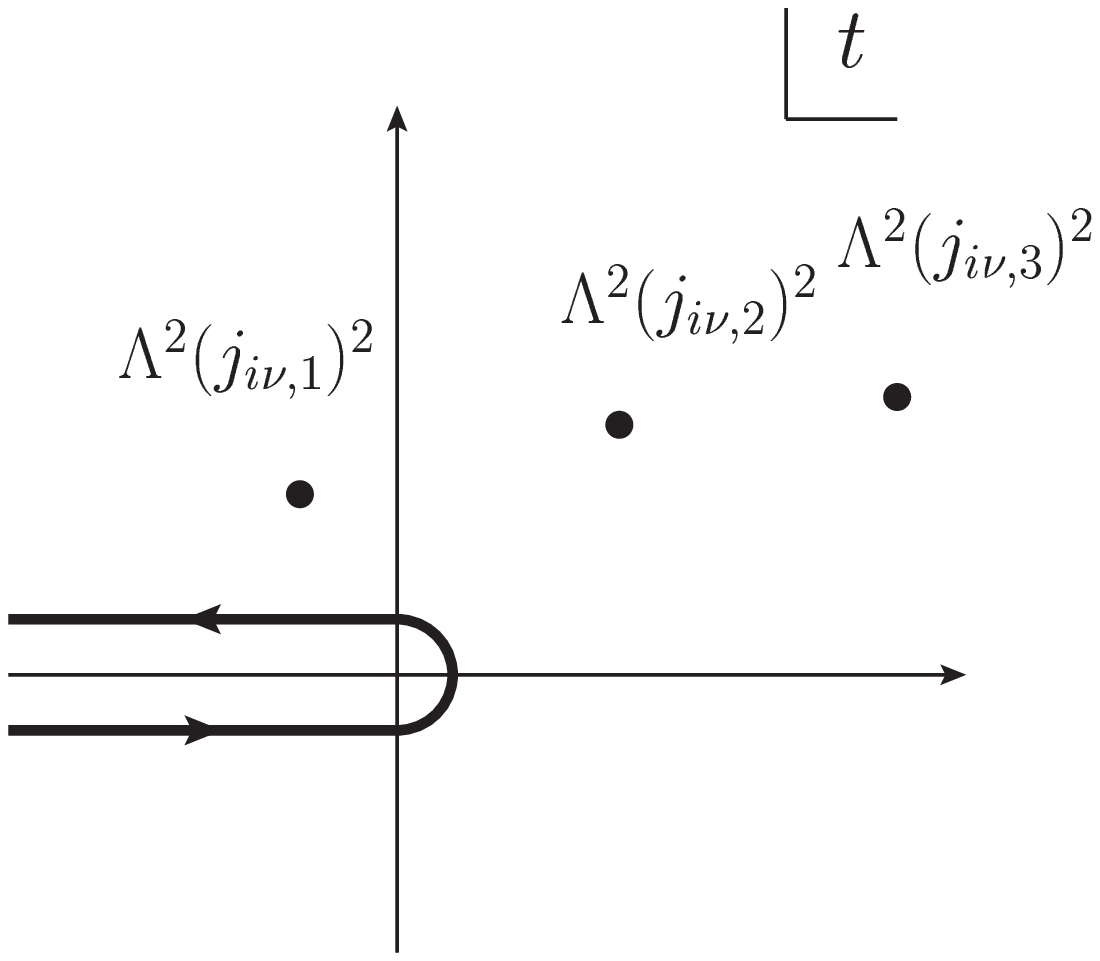} & &
  \includegraphics[scale=0.6]{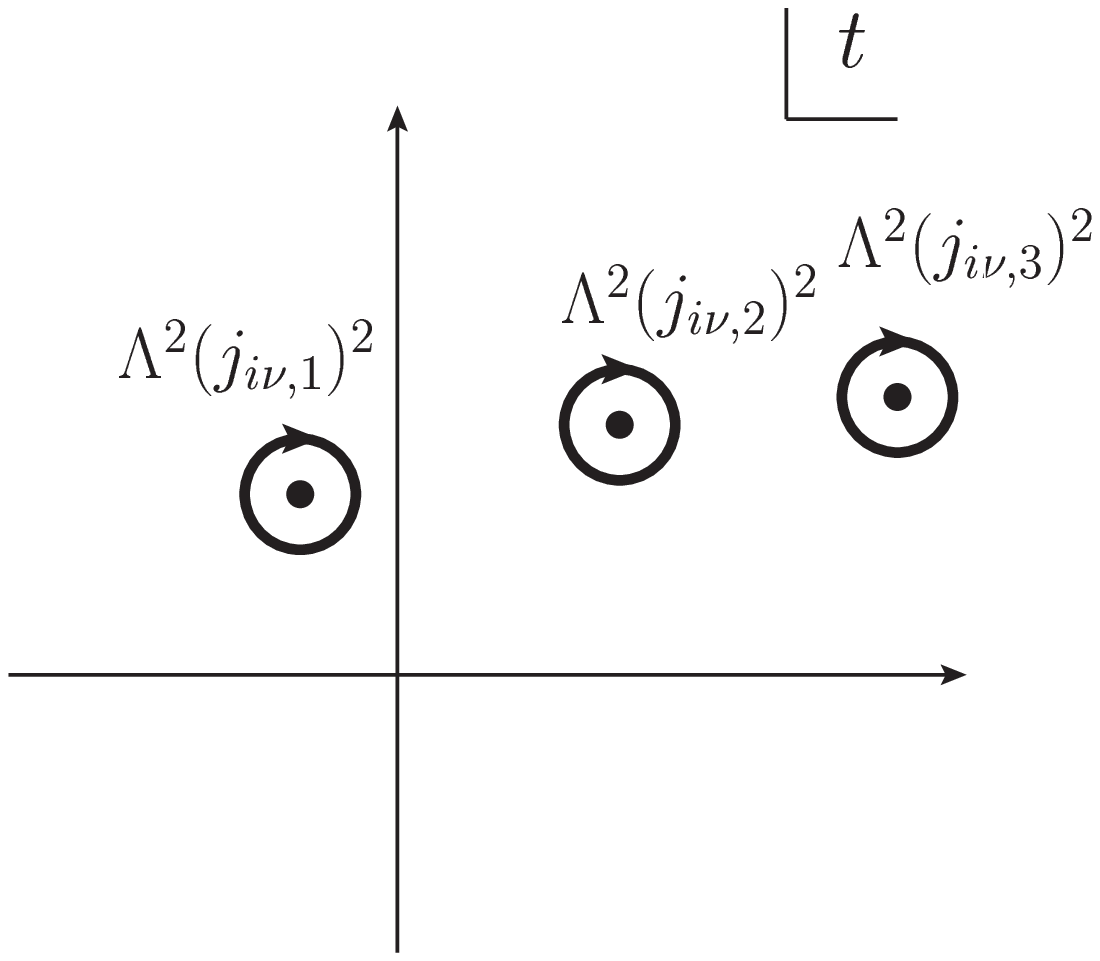} \\
  (a) & & (b)
\end{tabular}
\caption{\label{fig:t-space-contour} (a) contour in the $t$-plane before 
deformation, along with location of singularities (for some fixed $\nu$),  
and (b) the contour after deformation. }
\end{center}
\end{figure}

As we have already mentioned in section 3 and section 4.1, 
the integrand of the Pomeron kernel
(\ref{eq:pomeron kernel}, \ref{eq:kernel in hw2})\footnote{The
``imaginary'' part means the imaginary part when both $s$ and $t$ 
are real. The ``imaginary part'' and ``real part'' of the kernel
separately become holomorphic functions of complex valued $t$.} 
is holomorphic (except some singularities) not just in spin $j$ and 
anomalous dimension $(i\nu - j)$, but also in momentum transfer $t$,
except at poles $t = (\Lambda j_{i\nu, n})^2 = (m^{(\nu)}_n)^2$ in 
(\ref{eq:poles-of-integrand-of-kernel}).
Thus, one can rewrite the kernel for $t = - \Delta^2 \leq 0$ on the real negative axis  as
\begin{equation}
   {\cal K}(s,t,z,z') = \frac{1}{2 \pi i}
      \int_C dt' \frac{ 
                       {\cal K}(s,t',z,z')}{t'-t},
\end{equation}
using the function ${\cal K}(s,t',z,z')$ holomorphic in $t'$ (except 
the poles) and an integration contour $C$ which goes around the 
non-positive real axis counterclockwise 
(Figure~\ref{fig:t-space-contour}~(a)).
Its impact parameter space representation becomes\footnote{
The following relation is used:
\begin{align}
 \int_0^\infty d \Delta^2 \frac{J_0(b\Delta)}{t'+\Delta^2}=
 2K_0(bt'^{1/2}),\;\;\; -\pi<\arg t'<\pi.
\end{align}
} 
\begin{equation}
 {\cal K}(s,b,z,z') = \frac{1}{2 \pi i} \int_C dt'
  \frac{K_0(b (t')^{1/2})}{2\pi} 
  {\cal K}(s,t',z,z'). 
\end{equation}

Because of the holomorphicity in the complex $t'$-plane, the contour 
$C$ can be deformed as in Figure~\ref{fig:t-space-contour}~(b), picking 
residues at the $\nu$-dependent poles in the $t'$-plane; we can do this, 
because the integrand (for given $\nu$) vanishes exponentially at
$|t'| \rightarrow \infty$.  
One can show that 
\begin{align}
{\cal K}(s,b,z,z') & = - \frac{8 \sqrt{\lambda} \Lambda^2}{\pi^2} 
   \int_{-\infty}^{+\infty} d\nu \; i \nu  
      \left[\frac{1+e^{- \pi i j_\nu}}{\sin (\pi j_\nu)}\right]
      \frac{1}{\Gamma^2(j_\nu/2)}
      \left( \frac{\alpha' \tilde s}{4} \right)^{j_\nu} \notag \\
   & \qquad \qquad \qquad \sum_{n=1}^{\infty} 
        e^{- 2A(z)}\frac{J_{i \nu}(m^{(\nu)}_n z)}{J'_{i\nu}(j_{i\nu,n})}
        \frac{J_{i \nu}(m^{(\nu)}_n z')}{J'_{i\nu}(j_{i\nu,n})} e^{-2A(z')} \; 
        K_0(m^{(\nu)}_n b).
\label{eq:kernel-in-b-space}
\end{align}
The imaginary part [resp. real part] of ${\cal K}(s,b,z,z')$ is obtained 
by taking imaginary part [resp. real part] of $[1+ e^{- \pi i j_\nu}]$, as the 
Fourier transform works separately for the imaginary part and real part 
in (\ref{def:kernel in b space}). 
The $n$-th term corresponds to the $n$-th term 
of (\ref{eq:Regge-Pole-contribution-to-GG-amplitude-2}). 

Let us examine the profile of the imaginary parts of $I_{0,1}$ in the 
impact parameter space. We will focus on a region 
\begin{align}
 b \gg \Lambda^{-1},
\label{eq:b>Lambda^-1}
\end{align}
where the series of contributions from Kaluza--Klein tower of 
Pomeron trajectories (\ref{eq:kernel-in-b-space}) converges quickly;
since the real part of $m^{(\nu)}_n$ is of the same order or larger than 
the hadronic scale $\Lambda$ at least for moderate value of $\nu \in \R$ 
and $n \in \N$, we can approximate
\begin{align}
 K_0(m^{(\nu)}_n b)\simeq 
   \sqrt{\frac{\pi}{2m^{(\nu)}_n b}}e^{- m^{(\nu)}_n b}. 
\label{eq:assym-of-K_0}
\end{align}
Therefore, in this region, (\ref{eq:b>Lambda^-1}), the amplitudes 
(\ref{eq:kernel-in-b-space}) are dominated by the leading 
(lowest Kaluza--Klein) Pomeron trajectory, $n=1$, 
and the contributions from the higher Kaluza--Klein trajectories
are suppressed by of order $e^{- (b \Lambda )(\pi/2) n}$.

At small $x$, (\ref{eq:high-e-scatter}, \ref{eq:exp-small-x}), 
the impact-parameter dependent Pomeron kernel can also be evaluated 
by the saddle point method.
 In addition to the 
$(\alpha' \tilde s)^{j_\nu}$ and $K_0(b m^{(\nu)}_1)$ factors, another 
factor $J_{i \nu}(m^{(\nu)}_1 z)$ may also give rise to a large dependence 
in $\nu$; dominant contribution to scattering amplitudes 
come only from small $z \sim 1/q$ region in DDVCS as well as in DIS, 
and hence $J_{i \nu}(m^{(\nu)}_1 z) \sim (m^{(\nu)}_1 z)^{i \nu}$.
The saddle point $\nu^\ast(q/\Lambda, x; b)$ is, therefore, determined by
\begin{align}\label{eq:the equation determining saddle point in b}
i\nu^\ast (q/\Lambda, x; b) =
\frac{
\ln\sfrac{q}{\Lambda}+b\Lambda
 \left.\frac{\partial j_{i\nu,1}}{\partial i\nu}\right|_{i\nu=i\nu^\ast}
}
{
\frac{1}{\sqrt{\lambda}}\ln\sfrac{q/\Lambda}{\sqrt{\lambda}x}
}.
\end{align}
Therefore, we find that 
\begin{equation}
 \frac{{\rm Im} \; I_i(x, \eta = 0, \vec{b}, q^2)}
      {{\rm Im} \; I_i(x, \eta = 0, t = 0, q^2)} \sim 
 \frac{ \left(\frac{1}{\sqrt{\lambda}x}\right)^{j_{\nu^*(b)}} 
        \left(\frac{\Lambda}{q}\right)^{\gamma_{\nu^*(b)}}
        \Lambda^2 e^{-m^{(\nu^*(b))}_1 b} 
      }
      { \left(\frac{1}{\sqrt{\lambda}x}\right)^{j_{\nu^*(t=0)}} 
        \left(\frac{\Lambda}{q}\right)^{\gamma_{\nu^*(t=0)}}
       }.
\end{equation}
Note that the saddle point value $i\nu^*(b)$ in the numerator 
is determined by (\ref{eq:the equation determining saddle point in b}),
while $i\nu^*(t=0)$ in the denominator by (\ref{eq:nu ast for small -t}).
The $x$-dependent transverse profile of parton distribution\footnote{
The impact parameter dependent Pomeron kernel can be used 
also to determine the phase shift of the scattering amplitude in the
impact parameter space, $\chi(s,\vec{b},z,z')$ \cite{Brower2007a}. 
The relation between them is $I_i(x,\eta=0,\vec{b},q^2) \sim 
(s N_c^2) \; \chi(s, \vec{b}, z, z')|_{z\sim 1/q, z' \sim 1/\Lambda}$.
In section \ref{ssec:real-part}, where we discuss the real part of the 
scattering amplitude, we will study the $b$-dependent phase shift,
rather than $b$-dependent $I_i$.} 
is given by ${\rm Im} \; I_i (x, \eta = 0, \vec{b},q^2) \; dx d^2 \vec{b}$.

The transverse profile above is not a simple function, especially
because $m^{(\nu^*(b))}_1$ depends on $x$, $q^2$ and $b$
through (\ref{eq:the equation determining saddle point in b}).
In order to extract the qualitative feature of the profile, let us 
first consider situations where the saddle point value $i\nu^*(b)$ is 
much smaller or much larger than unity. Then the following expansion 
of the zeros of Bessel functions of order $i\nu$, $j_{i\nu, n}$, can be 
exploited in order to understand how the saddle point value
$i\nu^*(q/\Lambda, x; b)$ is determined by $x$, $q^2$ and $b$ 
through (\ref{eq:the equation determining saddle point in b}): 
\begin{align}\label{eq:besselzero at small nu}
 j_{i\nu,n} & \simeq j_{0,n}+c_n i\nu +{\cal O}((i\nu)^2), \\
 \label{eq:besselzero at large nu}
 j_{i\nu,n} &\simeq i\nu + d_n (i\nu)^{1/3}+{\cal O}((i\nu)^{-1/3})
\end{align}
with positive numbers $c_n$ and $d_n$ of order 
unity \cite{Math-Handbook}.\footnote{$c_n \rightarrow \pi/2$ for large $n$.}
The saddle point stays within $|\nu^\ast| \ll 1$, when 
\begin{align}\label{eq:small nu cond}
 \frac{\ln(q/\Lambda)}{\ln(1/x\sqrt{\lambda})/\sqrt{\lambda}}\ll 1 
{\text{ and\; }} \frac{b\Lambda}{\ln(1/x\sqrt{\lambda})/\sqrt{\lambda}}\ll 1
\end{align}
and the saddle point is 
\begin{align}
 i\nu^\ast (q/\Lambda, x; b) \simeq  \frac{\ln(q/\Lambda) + c_1 b \Lambda}
                               {\frac{1}{\sqrt{\lambda}}\ln\sfrac{q/\Lambda}{\sqrt{\lambda}x}}.
\label{eq:nu-ast-small-bspace}
\end{align}
On the other hand,
when either of the left hand sides of (\ref{eq:small nu cond}) is much larger than unity,
 the saddle point becomes $|\nu^\ast|\gg 1$ and
\begin{equation}
 i\nu^\ast (q/\Lambda, x; b) \simeq 
\frac{\ln(q/\Lambda)+b\Lambda}{\frac{1}{\sqrt{\lambda}}\ln\sfrac{q/\Lambda}{\sqrt{\lambda}x}}.
\label{eq:nu-ast-large-bspace}
\end{equation}

Now it is easy to see the transverse profile for large $q^2$ satisfying 
$\ln\left(\frac{q}{\Lambda}\right) \gg \ln\left[\frac{q/\Lambda}{\sqrt{\lambda}x}\right]/\sqrt{\lambda}$;
this situation corresponds to $(x, q^2)$ where
$\gamma_{\rm eff.}$ is positive, that is, PDF decreases in DGLAP
evolution, as we saw in section \ref{sssec:small-(-t)}.
\begin{equation}
 \frac{{\rm Im} \; I_i(x, \eta = 0, \vec{b}, q^2)}
      {{\rm Im} \; I_i(x, \eta = 0, t = 0, q^2)} \sim 
 \Lambda^2 \exp \left[ - 
     \frac{ b\Lambda \ln\left(\frac{q}{\Lambda}\right) + \frac{1}{2}(b\Lambda)^2 }
          { \frac{1}{\sqrt{\lambda}} 
            \ln \left(\frac{q/\Lambda}{\sqrt{\lambda}x}\right)
          }
                \right] \sim 
  \left\{
\begin{array}{ll}
 e^{- \frac{ \sqrt{\lambda} \ln\left(\frac{q}{\Lambda}\right) }
           { \ln\left(\frac{q/\Lambda}{\sqrt{\lambda}x}\right) }
      b\Lambda }
 &  \left[ b\Lambda \ll \ln \left(\frac{q}{\Lambda} \right) \right], \\
 e^{- \frac{(b\Lambda)^2 }
           { \frac{2}{\sqrt{\lambda}}
             \ln\left(\frac{q/\Lambda}{\sqrt{\lambda}x}\right)
            }          
   }
 &  \left[ \ln \left(\frac{q}{\Lambda}\right) \ll (b\Lambda) \right]. 
\end{array}
 \right.
\label{eq:trans-profile-Lq}
\end{equation}
The profile remains linear exponential in the impact parameter $b$ for
smaller $b$, until it turns into Gaussian for larger $b$.

Let us now study the transverse profile for smaller $q^2$ satisfying 
$\ln\left(\frac{q}{\Lambda}\right) \ll \ln\left[\frac{q/\Lambda}{\sqrt{\lambda}x}\right]/\sqrt{\lambda}$
instead. This condition on $(x, q^2)$ corresponds to 
$i\nu^*(t = 0) \ll 1$, when the PDF still increases under the DGLAP
evolution, $\gamma_{\rm eff.} < 0$. The saddle point value 
$i\nu^*(q/\Lambda, x; b)$ becomes large for large $b$ which violates 
the second condition of (\ref{eq:small nu cond}). In this case, 
by using (\ref{eq:nu-ast-large-bspace}), the transverse profile turns 
out to be Gaussian approximately:
\begin{equation}
 \frac{{\rm Im} \; I_i(x, \eta = 0, \vec{b}, q^2)}
      {{\rm Im} \; I_i(x, \eta = 0, t = 0, q^2)} \sim 
 \Lambda^2
 e^{- \frac{(b\Lambda)^2 }
           { \frac{2}{\sqrt{\lambda}}
             \ln\left(\frac{q/\Lambda}{\sqrt{\lambda}x}\right)
            }          
   } \qquad 
\left[{\rm for~} \frac{1}{\sqrt{\lambda}}
         \ln\left[\frac{q/\Lambda}{\sqrt{\lambda}x}\right] \ll (b\Lambda)   
\right]. 
\label{eq:trans-profile-Sq-Lb}
\end{equation}
In a range of smaller impact parameter where $i\nu^*(b) \lesssim 1$, 
however, the expression cannot be simpler than 
\begin{equation}
 \frac{{\rm Im} \; I_i(x, \eta = 0, \vec{b}, q^2)}
      {{\rm Im} \; I_i(x, \eta = 0, t = 0, q^2)} \sim 
 \Lambda^2 e^{- b m^{(\nu^*(q/\Lambda,x; b))} } e^{+
 \frac{ \left( \left. \frac{\partial j_{\mu,1}}{\partial \mu}
               \right|_{\mu=i\nu^*} \!\!\!\!\! b\Lambda \right)^2 
      }{
        \frac{2}{\sqrt{\lambda}}
        \ln \left( \frac{q/\Lambda}{\sqrt{\lambda}x}\right)
      }                                           } \quad  
\left[{\rm for~}  (b\Lambda) \ll \frac{1}{\sqrt{\lambda}}
         \ln\left[\frac{q/\Lambda}{\sqrt{\lambda}x}\right]   
\right]. 
\label{eq:trans-profile-Sq-Sb}
\end{equation}
The transverse profile in this range of impact parameter $b$ is
approximately linear exponential in $b$ (c.f. \cite{Brower:2010wf}), 
with the mass parameter $m^{(\nu^*(q/\Lambda ,x; b))}_1$ changing slowly 
in $(q/\Lambda, x, b)$. This approximate linear exponential profile 
(\ref{eq:trans-profile-Sq-Sb}) smoothly turns into the Gaussian profile 
(\ref{eq:trans-profile-Sq-Lb}), because of the $b$ dependence of the 
saddle point $i\nu^*(q/\Lambda,x; b)$. All of these results\footnote{
Note that for $(b\Lambda) \gg \ln (1/(\sqrt{\lambda}x))$, 
neither (\ref{eq:trans-profile-Lq}) nor (\ref{eq:trans-profile-Sq-Lb}) 
are reliable, as we have used the form of $j_r(\nu)$ that is reliable 
only in $|\nu| \ll \sqrt{\lambda}$. For such a large impact parameter, the 
saddle point value of $\nu$ is not within this range. Similarly, for 
$\ln (q/\Lambda) \gg \ln (1/(\sqrt{\lambda}x))$, 
(\ref{eq:trans-profile-Lq}) cannot be trusted.}  are
summarized in Figure~\ref{fig:phase}~(a).

Regardless of whether $q^2$ is large (\ref{eq:trans-profile-Lq}) or 
small (\ref{eq:trans-profile-Sq-Lb}), the imaginary part of the 
(sphere level) DDVCS amplitude shows Gaussian profile for sufficiently 
large impact parameter $b$. The Gaussian profile in the transverse
direction has been used for phenomenological fit of GPD, and 
the holographic calculation above justifies the Gaussian ansatz at least 
for sufficiently large impact parameter. 

This agreement between the holographic calculation and the conventional 
phenomenological ansatz is not an accident. The phenomenological 
Gaussian ansatz is a direct consequence of a linear trajectory 
(\ref{eq:linear-traj}) in the traditional Regge ansatz 
(e.g. \cite{BaronePredazzi}). Although the Pomeron trajectories
(\ref{eq:j-nu-t}) in holographic QCD are not linear at all, they 
become approximately linear at large $t$ and $j$ 
(but not too large $j$ \cite{Gubser:2002tv}), 
because of (\ref{eq:besselzero at large nu});
\begin{align}
 \alpha_{\P,n}(t)=j_0+\frac{1}{2\sqrt{\lambda}}\sfrac{t}{\Lambda^2}+\frac{1}{2\sqrt{\lambda}}{\cal O}\left(\sfrac{t}{\Lambda^2}^{2/3}\right).
\label{eq:FT-of-linear-traj}
\end{align}
Since the saddle point for large $b$ is determined by the large $i\nu$ 
behavior (and hence by the large $t$ (and large $j$) behavior) of the
Pomeron kernel, the asymptotically linear 
trajectory (\ref{eq:FT-of-linear-traj}) gives rise to the same
large-$b$ behavior (that is, Gaussian profile) as in the
phenomenological ansatz with a linear trajectory. 
The width-square of the Gaussian profile should be given by 
using the asymptotic slope parameter of the Pomeron trajectory 
$j = \alpha_{\P, 1}(t)$ in (\ref{eq:FT-of-linear-traj}): 
\begin{equation}
 2 \alpha'_{\P} \ln \left(\frac{q/\Lambda}{\sqrt{\lambda}x} \right) = 
 \frac{1}{\sqrt{\lambda} \Lambda^2} 
    \ln \left(\frac{q/\Lambda}{\sqrt{\lambda}x} \right),
\end{equation}
which is indeed the case 
in (\ref{eq:trans-profile-Lq}, \ref{eq:trans-profile-Sq-Lb}).

The Pomeron trajectories in holographic QCD are far from being linear,
however, except for the large $j$--$t$ asymptotic region. 
As a result, the transverse profile may well be different from 
the Gaussian profile at smaller $b$ (the region where $b\Lambda\ll \ln(1/x)/\sqrt{\lambda}$). 
Indeed, in this region, the results (\ref{eq:trans-profile-Lq},
\ref{eq:trans-profile-Sq-Sb}) show that the profile is not Gaussian, 
but approximately linear exponential in the hard wall model. 
Interestingly, the mass scale of linear exponential profile 
for small $q^2$ in (\ref{eq:trans-profile-Sq-Sb}), 
$m^{(\nu^*(q/\Lambda,x;b))}_n$, depends on kinematical variables, and 
is different from mass eigenvalues of any 4D hadron states; this 
linear exponential profile is not associated with a $t$-channel exchange 
of a single particle, but emerges after summing up all the stringy
states with different spins in the $n$-th Kaluza--Klein trajectory. 
The mass scale $m^{(\nu=0)}_{n=1}$ appearing in the small $x$ limit 
(for fixed $q^2$ and $b$) is smaller than any one of the mass
eigenvalues of the stringy (and graviton) states in the trajectory, and 
the range is longer than a simple $t$-channel exchange of a glueball.   

Although the expressions 
(\ref{eq:trans-profile-Lq}--\ref{eq:trans-profile-Sq-Sb}) rely 
on the hard wall model, most of its qualitative aspects are expected 
be in common with other holographic models. The transition from
(\ref{eq:trans-profile-Sq-Lb}, \ref{eq:trans-profile-Sq-Sb})  
to (\ref{eq:trans-profile-Lq}) for larger $q^2$, for example, 
will be induced in various models. The saddle point is dragged to 
larger $j$, $i\nu$ and $t$ for a given value of $b$, just like we
discussed in section \ref{sssec:regge}.

\subsection{Real Part of the Amplitudes}
\label{ssec:real-part}

The imaginary parts of the structure functions are related to the 
GPDs, whose Fourier transforms are interpreted as the distributions of 
partons in the transverse spacial directions. Although the real part 
of the (D)DVCS amplitude does not have such an interpretation, it still 
contributes to the DVCS cross section.

The large impact parameter $b$ behavior of the real part has been
discussed in \cite{Brower2007}, and the behavior in the small $b$ region 
(but not as small as $b \Lambda \lesssim 1$) is covered 
by \cite{Brower:2010wf}. 
Overall normalization is found in \cite{Hatta:2007}. 
The following discussion provides a clear description of how the small 
$b$ behavior of (\cite{Brower:2010wf}) smoothly turns into the large 
$b$ behavior of \cite{Brower2007}, as well as careful interpretation of 
the physics behind this phenomenon. 
This subject has, in fact, quite a long history
(e.g., \cite{Gribov-cpx-ang-mom}), but we hope that the following 
presentation using the saddle point value $j^*$ as a key concept 
helps clarify things a little bit. 

\subsubsection{Momentum Transfer Dependence}

It is convenient (and customary) to use a variable
\begin{equation}
 \rho(t,x,\eta = 0,q^2) \equiv 
  \frac{{\rm Re} \; I_i(x, \eta = 0,t,q^2)}
       {{\rm Im} \; I_i(x, \eta = 0,t, q^2)}
\end{equation}
in characterizing the real part of the amplitude, as we already 
know the behavior of the imaginary part. The value of 
$\rho (t)$ at $t = 0$ is often denoted by $\rho$ in elastic 
scattering of hadrons.\footnote{
The real part to imaginary part ratio is often denoted by $\eta(t)$, 
but $\eta$ is reserved for skewedness in this article.}  
The real part of the amplitude is 
obtained by simply taking the real part of $[1 + e^{- \pi i j}]$ 
in the Pomeron kernel (\ref{eq:pomeron kernel}). 
The ratio $\rho(t)$ is simply given by 
\begin{equation}
 \rho(t,x,\eta = 0,q^2) = 
   \tan \left( \frac{\pi}{2}\left(j^*(x,\eta = 0,t,
			     q^2)-1\right)\right), 
\label{eq:eta(t)-saddle}
\end{equation}
using the saddle point value $j^* = j_{\nu^*(x,q^2,t)}$ in the hard wall
model \cite{Hatta:2007} for the entire range of physical momentum 
transfer $t \leq 0$. The $j$--$\nu$ integration in the kernel can be 
evaluated by the saddle point method for small $x$
(\ref{eq:exp-small-x}), unless $q^2$ is too large to satisfy 
\begin{equation}
 i \nu^* < 2 \qquad ({\rm equivalently~} j^* < 2).
\label{eq:i-nu-ast-<-2}
\end{equation}
The expression for $\rho(t)$ above remains valid in other holographic 
models, as long as $(x,t,q^2)$ is in the saddle point phase. 
$\rho(t) \gg 1$ is predicted in gravitational dual in general, because $j^*$ is 
closer to 2 than to 1 for $\lambda \gg 1$ \cite{Hatta:2007, Brower2007}. 

In holographic models other than the hard wall model, 
there may be a leading pole phase in the kinematical variable 
space $(x,t,q^2)$. There, the ratio $\rho(t)$ is given by the 
expression (\ref{eq:eta(t)-saddle}) with the saddle point value $j^*$ 
replaced by the leading Pomeron pole $\alpha_{\P, 1}(t)$ \cite{BaronePredazzi}.

The relation (\ref{eq:eta(t)-saddle}) follows immediately from
derivative analyticity relation \cite{Bronzan:1974jh}, if $j^* -1$ 
is understood as $\lambda_{\rm eff.}$, the effective exponent of 
 $s = W^2 \propto 1/x$ of photon--hadron scattering; 
such things as $j$-plane representation of the scattering
amplitude\footnote{
Dispersion relation 
\begin{equation}
 {\rm Re} \; A^{(+)}(s,t) = \frac{1}{\pi} P\int ds' 
   \frac{{\rm Im} \; A^{(+)} (s',t)}{s'-s}
\end{equation}
in the convolution form becomes a simple product form after Mellin 
transformation, $[{\rm Re} \; A^{(+)}(j,t)] = - \cot(\pi j/2) \times 
 [{\rm Im} \; A^{(+)}(j,t)]$. This is why the ratio is better described in the $j$-plane language.
} 
or its saddle point value do not have to be invoked in deriving the 
ratio (\ref{eq:eta(t)-saddle}). But, we still find the expression interesting, 
in that not only $\rho$ and $\lambda_{\rm eff.}$ but also 
$\gamma_{\rm eff.}$ and $B$ are predicted here to depend on the kinematical 
variables $(x,t,q^2)$ through a single value $j^*$. Furthermore, 
the limit of applicability of (\ref{eq:eta(t)-saddle}), 
the condition (\ref{eq:i-nu-ast-<-2}), is translated into 
$\gamma_{\rm eff.} < 0$, which is to say that the GPD at that 
$(x,t,q^2)$ increases in the DGLAP evolution. 
This observation is based simply on the $j$-plane representation 
of the scattering amplitude and an assumption that the kinematical
variables $(x,t,q^2)$ are in the saddle point phase, and thus, 
does not rely on details of the hard wall model.

The condition (\ref{eq:i-nu-ast-<-2}) is not satisfied, however, for 
sufficiently large $q^2$; that is when $\gamma_{\rm eff.} > 0$, and 
GPD/PDF decreases under the DGLAP evolution. 
\begin{figure}[tbp]
\begin{center}
\begin{tabular}{ccc}
  \includegraphics[scale=0.38]{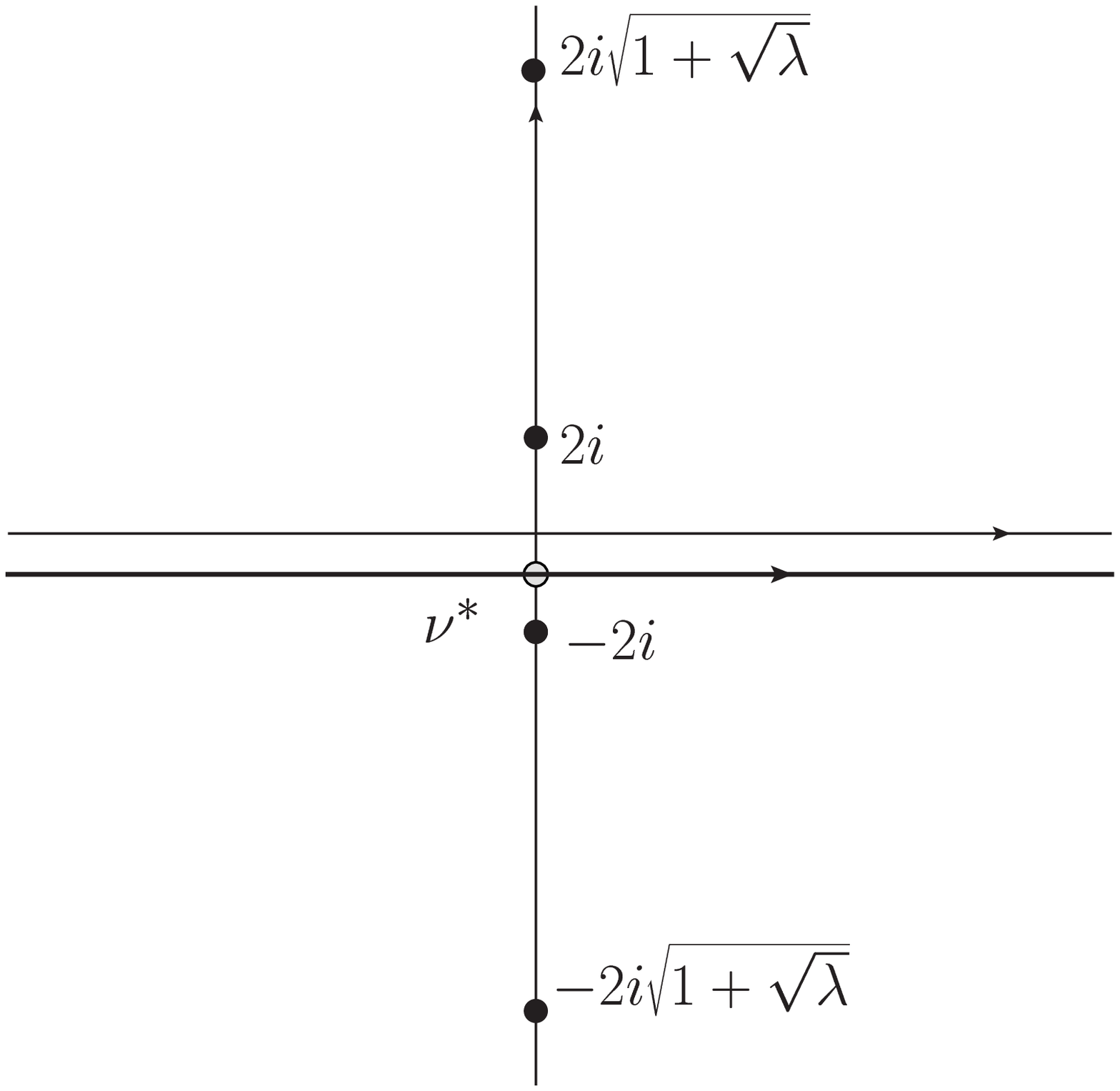} &
& 
  \includegraphics[scale=0.38]{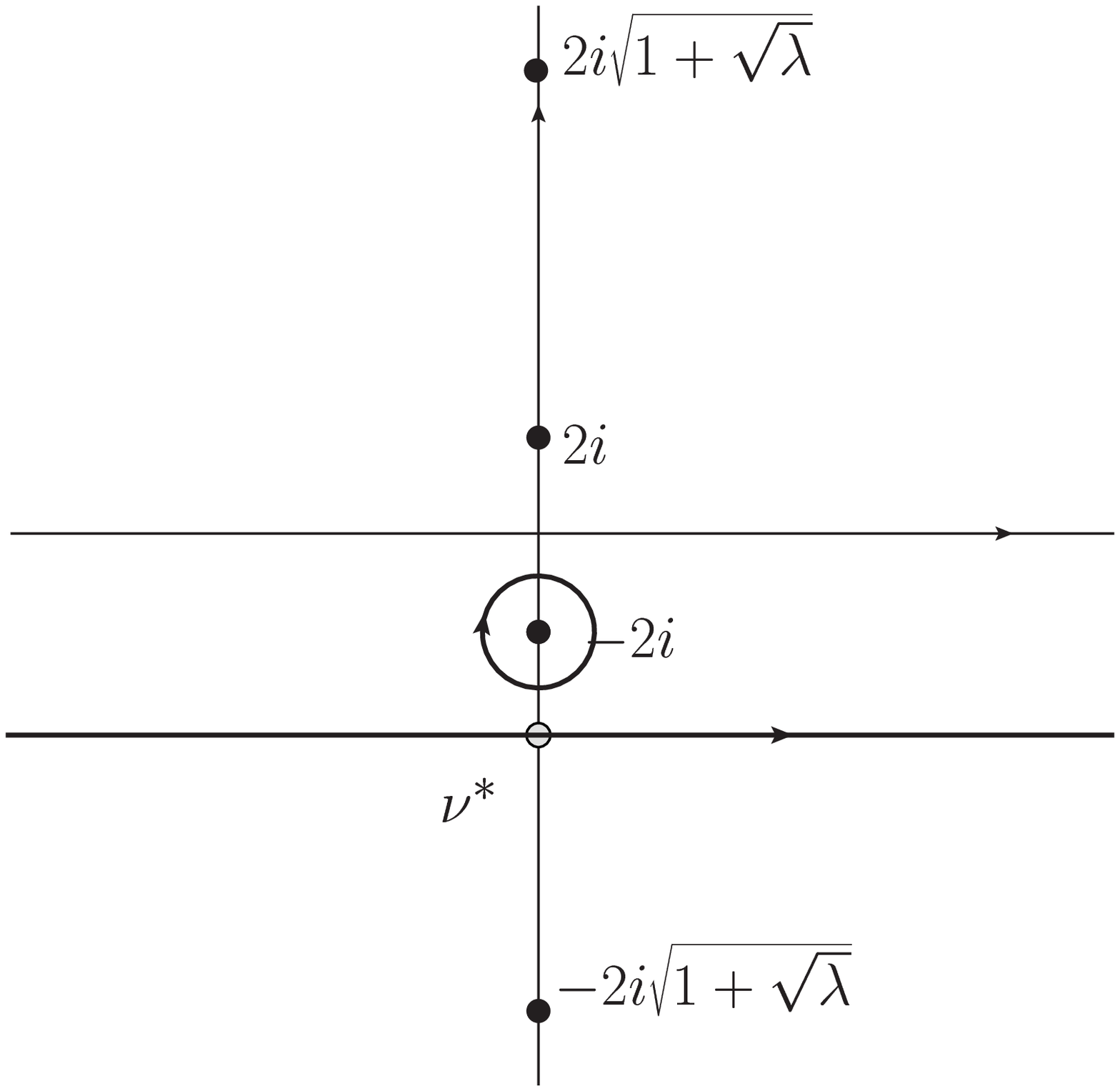} \\
 (a) & & (b)
\end{tabular}
\end{center}
\caption{The integration contour in the $\nu$-plane.
The original contour (for $t \leq 0$) is on the real axis from negative 
infinity to positive infinity.
The contour can be chosen as it is so long as $i\nu^\ast<2$, as 
shown in (a). When the saddle point is at $i\nu^\ast>2$, however, 
one will deform the contour to make it pass the saddle point $\nu =
 \nu^*$; the process of deformation leaves an extra contribution 
coming from the pole of $[1/\sin (\pi j_\nu)]$ at $i\nu = 2$ 
($j_\nu = 2$), as in (b).
The $j_\nu = 4$ pole corresponds to $\nu=\pm 2i\sqrt{1+\sqrt{\lambda}}$ 
in (b); when the saddle point value of $i\nu^*$ is even greater than
that, the whole amplitude consists of contributions from the 
$j_\nu = 2,4$ poles as well as from the contour passing through 
the saddle point \cite{Hatta:2007}.
}
\label{fig:path of nu}
\end{figure}
In this case, the contour of $\nu$ integration should be taken as 
in Figure~\ref{fig:path of nu}~(b), and the structure functions 
are expressed as a sum of contributions from $j = 2$ ($i\nu = 2$) pole 
and a continuous integration whose contour passes through the saddle point.
For even larger $q^2$, the saddle point value $i\nu^*$ may be as large 
as $\lambda^{1/4}$ and $j^\ast$ as large as $4, 6, \cdots$, and the structure 
functions are given by a sum of contributions from some finite number of
poles $j = 2,4,6,\cdots \leq j^*$ and the one from the saddle point 
approximation. Using the Kneser--Sommerfeld expansion of Bessel 
functions, these pole contributions ($j=2,4,\cdots \leq j^\ast$) can be written in the form 
\begin{equation}
 {\rm Re} \; I_i^{(j)} \simeq - c'_s 
  \frac{(s/\Lambda^2)^j}{\Gamma^2(j/2)}\frac{1}{(4\sqrt{\lambda})^{j-2}}
 \sum_{n=1}^{\infty}\frac{\Lambda^2}{t - m_{j,n}^2} 
    \gamma_{hh \P n}(m_{j,n}^2) \gamma_{\gamma^* \gamma^* \P
    n}(m_{j,n}^2).
\label{eq:Re-Ii-pole-KK}
\end{equation}
Therefore, they are interpreted as $t$-channel exchange of the $n$-th
Kaluza--Klein mode of the spin $j < j^*$ state in the graviton 
trajectory \cite{Hatta:2007}. 
The remaining continuous integration should then be regarded\footnote{
In this article, we used the dilaton--graviton scattering as the gravity dual model
of the DDVCS process.
If a D-brane is used as the model of the target hadron (baryon),
however, the scattering amplitude is not a sphere amplitude with four NS-NS vertex operator insertion.
} as 
the $u$-channel exchange of $j < j^*$ states and all the sphere level 
contributions associated with $j^* < j$ states. 

Let us first focus on the $0 \leq -t \lesssim \Lambda^2$ region. 
Then, with a simple argument like we had in section \ref{sssec:regge}, 
one can see that the $j = 2$ ($i \nu = 2$) pole contribution dominates, 
other poles give rise to subleading corrections, and the saddle point 
contribution is even smaller than them \cite{Hatta:2007}.

The real parts of the structure functions are well approximated 
by the saddle point contribution for small enough $q^2$, but 
they are expressed as a sum of $j = 2,4, \cdots < j^*$ pole contributions 
and the saddle point contribution, when $q^2$ becomes large enough 
to satisfy $j^* > 2$, that is, 
\begin{equation}
 \sqrt{\lambda} \frac{\ln (q/\Lambda)}
      {\ln \left( \frac{q/\Lambda}{\sqrt{\lambda}x}\right)} = i\nu^* > 2.
\label{eq:j=2-pole-exist}
\end{equation} 
This transition associated with the change in $q^2$ is understood as 
follows; it is usually better to treat the Pomeron trajectory exchange 
as a whole, not as a sum of exchange of individual spin $j=2,4,\cdots$ 
particles, because the amplitude of a spin $j$ particle exchange has 
an ever-increasing factor $s^j$. Large virtuality of the photon $q^2$
and the photon wavefunction localized near the UV boundary, however, 
introduces $\gamma^*$--$\gamma^*$--[spin $j$ string] coupling suppressed
by powers of $(\Lambda/q)$, and the amplitudes of higher spin $j$
exchange are suppressed significantly. The transition means that,  
for large $q^2$, the power suppression $(\Lambda/q)^{\gamma(j)}$ 
becomes more important than $(1/x)^{j}$ for higher spin $j$ stringy
states propagating in the sphere amplitude, and the contributions from 
smaller $j$, that is, $j=2,4, $ etc. ``stand out'' from all the rest. 
That is the physical meaning of the transition observed in the
rearrangement of the $\nu$ integration contour. Therefore, we refer to 
this phase as 
\begin{itemize}
 \item {\bf KK-sum Low-spin Phase} or {\bf Spin-2 Phase}, where the
       saddle point value is larger than 2 (which also means 
       $\gamma_{\rm eff.} > 0$).
\end{itemize}

The discussion so far can be extended to the region 
$\Lambda^2 \ll (-t) \ll q^2$; 
the discussion in section \ref{sssec:regge} can be applicable almost in its form, 
except that the factor $(\Lambda/q)^{\gamma(j)}$ 
in (\ref{eq:general-j-int}) needs 
to be replaced by $(\sqrt{-t}/q)^{\gamma(j)}$ for $\Lambda^2 \ll -t$; 
the saddle point $j^* = j_{\nu^*}$ is the one 
using (\ref{eq:nu ast for large -t}) in the hard wall model.
The spin $j$ pole contributions can also be evaluated explicitly; they
are 
\begin{equation}
 {\rm Re} \, I_i^{(j)} \sim  \frac{c'_s 8 \lambda}{\Gamma^2(j/2)} 
  \left(\frac{1}{4 \sqrt{\lambda}x}\right)^j 
  \left(\frac{\Lambda}{q}\right)^{\gamma(j)} 
  \left(\frac{\Lambda}{\sqrt{-t}}\right)^{2\Delta - 2 - \gamma(j)}.
\end{equation}
For a given value of $x$ and $q^2$ (so that $j^* > 2$), the real part of the 
structure functions scale as $(1/\sqrt{-t})^{2\Delta - 2}$, because the 
$j=2$ pole contribution dominates. For sufficiently large momentum
transfer (so that $j^* < 2$), however, the real part is better described 
by the saddle point contribution alone, and the real part shows the 
same scaling behavior in $\sqrt{-t}$ as the imaginary part of the 
structure functions. Therefore, the real part to imaginary part ratio 
is given by 
\begin{equation}
\label{eq:phase-transition-of-ratio}
 \rho(t,x,\eta = 0, q^2) = 
\begin{cases}
 \left(\frac{1}{\sqrt{\lambda}x}\right)^{-(j^*-2)}
 \left(\frac{\Lambda}{q}\right)^{-\gamma(j^*)} & 
   {\rm for~} -t \lesssim \Lambda^2 
   {\rm ~if~} (\ref{eq:j=2-pole-exist}) {\rm~holds~true}, \\
 \left(\frac{1}{\sqrt{\lambda}x}\right)^{-(j^*-2)}
 \left(\frac{\sqrt{-t}}{q}\right)^{-\gamma(j^*)} & 
   {\rm for~} \Lambda^2 \ll -t 
   {\rm ~while~} \sqrt{\lambda}\frac{\ln(q/\sqrt{-t})}
                    {\ln \left([q/\sqrt{-t}]/[\sqrt{\lambda}x]\right)} \gg 1, \\
 {\rm given~by~} (\ref{eq:eta(t)-saddle}) & {\rm otherwise}.
\end{cases}
\end{equation}
For a given value of $q^2$ and $t$, the ratio in the small $x$ limit 
is eventually given by (\ref{eq:eta(t)-saddle}) or possibly by the one 
with $j^*$ replaced by $\alpha_{\P, n}(t)$.

\subsubsection{Impact Parameter Dependence}

Although the impact parameter space description of the real part 
does not have such an interpretation as transverse distribution of 
partons, it still is important in discussing how the unitarity limit 
is reached in $\gamma^* + h$ scattering. Since the unitarity limit 
is not achieved in the Bjorken limit $\Lambda^2 \ll q^2$ without 
an extremely small $x$ (and extremely large $W^2 \simeq q^2/x$),  
the following discussion may be only of academic interest, but 
we present the result anyway. 

The phase shift $\chi$ in the impact parameter space $(\vec{b},z,z')$ 
is given by \cite{Brower2007a}
\begin{equation}
 \chi^\text{sphere}(s, \vec{b},z,z') \sim \frac{\kappa_5^2}{R^3}
  \frac{e^{2A(z)} e^{2A(z')}}{s}
  {\cal K}(s,\vec{b},z,z')
\end{equation}
at the leading order in $1/N_c$ expansion (sphere-level);
the unitarity limit of the photon--hadron scattering in a ``partial wave'' 
$b$ is indicated when $|\chi^\text{sphere}| \sim {\cal O}(1)$ for $z \sim 1/q$ and 
$z' \sim 1/\Lambda$ \cite{Brower2007a, Hatta:2007, Brower2007}. 
The saddle point in $\nu$ integration in the Pomeron kernel is
determined regardless of whether we are studying impact-parameter
dependent structure functions or phase shift; for $1 \ll
(b\Lambda)$, where only the 
Pomeron trajectory $j=\alpha_{\P,1}(t)$ of the lowest KK 4D hadrons is relevant,
 we find 
\begin{equation}
 \chi^\text{sphere}(s,b,z,z')|_{z \sim 1/q, z' \sim 1/\Lambda} \sim \frac{1}{N_c^2}
  \frac{-[1+e^{\pi i j^*}]}{\sin(\pi j^*)}
  \left(\frac{1}{\sqrt{\lambda}x}\right)^{j^*-1}
  \left(\frac{\Lambda}{q}\right)^{2+\gamma(j^*)} 
 \times e^{-m^{(\nu^*)}_1 b}, 
\label{eq:phase-shift-saddle-phase}
\end{equation}
as long as the saddle point is not too large, $i\nu^* < 2$ and $j^* < 2$.
Here, both the imaginary (absorptive) part and real (diffractive) part
are included. The phase shift above shows expected dependence on 
$N_c$, $x$ and $q^2$, because $1 < j^*$ and 
$2 + \gamma(j^*) \geq 2-j_0 > 0$.

\begin{figure}[tbp]
\begin{center}
\begin{tabular}{ccc}
   \includegraphics[scale=0.75]{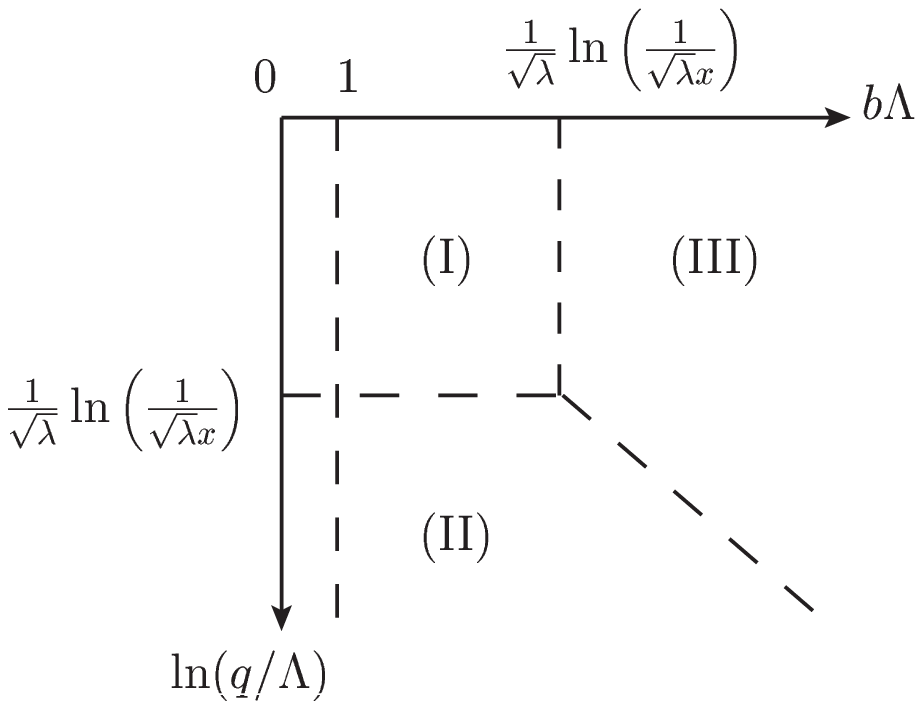} &
 & 
   \includegraphics[scale=0.75]{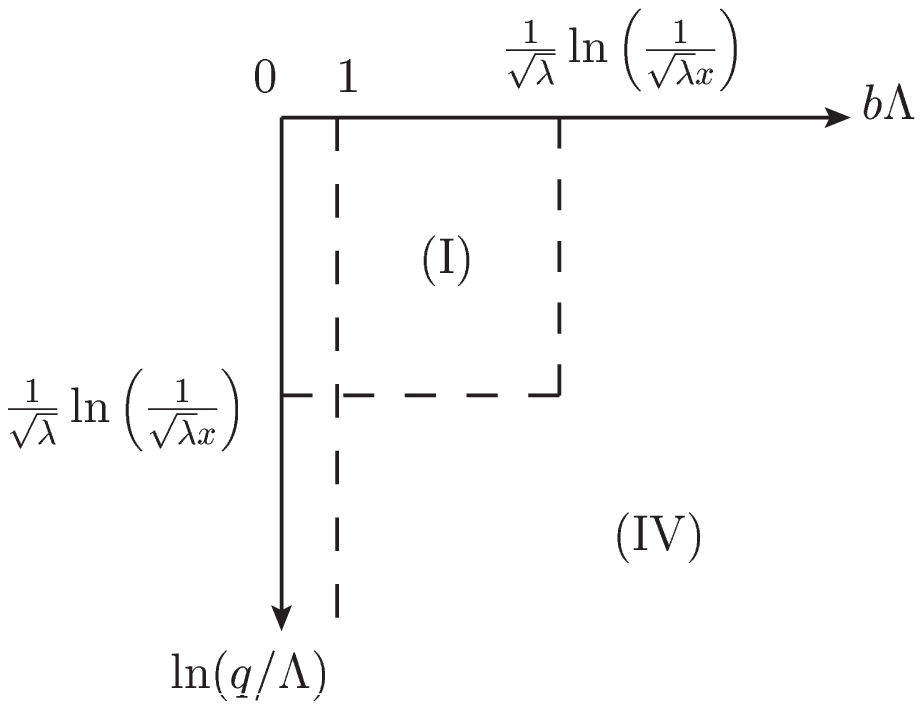} \\
  (a) & & (b)
\end{tabular}
\caption{\label{fig:phase} A schematic picture describing qualitatively
 different behavior of ${\rm Im} \; \chi(b)$ (panel (a)) and 
${\rm Re} \; \chi (b)$ (panel (b)) as functions of the impact parameter $b$. 
Behavior is different for different range of $b$ as well as for
 different value of $q^2$, as shown in this figure. 
 (I): $e^{- m^{(\nu^\ast)}_1 b}$ (see (\ref{eq:trans-profile-Sq-Sb}) for a more precise expression), 
 (II): $e^{- \frac{\sqrt{\lambda} \ln(q/\Lambda)}
                  {\ln ((q/\Lambda)/\sqrt{\lambda}x) }  b\Lambda }$,
 (III):$e^{- \frac{ \sqrt{\lambda}(b\Lambda)^2 }
                  { 2 \ln ((q/\Lambda)/\sqrt{\lambda} x) }  }$, and
 (IV): $e^{- m_{2,1} b}$. 
  }
\end{center}
\end{figure}

When the saddle point becomes large, $i\nu^* > 2$, and hence $j^* > 2$, 
the diffractive (real) part of the phase shift needs to be treated 
separately, as we have already seen in studying the momentum transfer 
dependence. In terms of kinematical variables, $i\nu^* > 2$ for the
lowest Kaluza--Klein mode $(n=1)$ is equivalent to
\begin{align}\label{eq:condition for saddle point < 2}
 \frac{1}{\ln{([q/\Lambda] / [x\sqrt{\lambda}])/\sqrt{\lambda}}} 
   \left(\ln(q/\Lambda)+\left. \frac{\partial j_{\mu,n=1}}{\partial \mu}\right|_{\mu=2}  b\Lambda\right) > 2.
\end{align}
For either large $\ln (q/\Lambda)$ or large $(b \Lambda)$, the
diffractive part is written as a sum of $j = 2$ pole contribution 
and the saddle point contribution, possibly with some other finite 
number of $j = 4,6,\cdots$ pole contributions. The $j=2$ pole
contribution is always the dominant one as long as $i \nu^* > 2$, 
and behaves as 
\begin{equation}
 {\rm Re} \; \chi^\text{sphere}(W^2,b,z,z')|_{z \sim 1/q, z' \sim 1/\Lambda}
 \sim \frac{1}{N_c^2}\left(\frac{1}{x}\right)
 \left(\frac{\Lambda^2}{q^2}\right) \times 
  e^{- m_{2,1}b},
\label{eq:phase-shift-Re-j2-phase}
\end{equation}
where $m_{2,1}$ is the mass eigenvalue of the lowest Kaluza--Klein mode 
of graviton in the warped throat. This impact parameter dependent profile
(\ref{eq:phase-shift-Re-j2-phase}) is also a $j^* \rightarrow 2$ (and
$i\nu^* \rightarrow 2$) limit of the real part 
of (\ref{eq:phase-shift-saddle-phase}). 
All these results are shown schematically in the phase diagram 
in Figure~\ref{fig:phase}.

The total cross section $\sigma^{\gamma^\ast h}$ can be estimated by using the 
 the sphere-level phase shift $\chi^\text{sphere}$ 
at high energy, 
provided that the absolute value of phase shift $|\chi|$ is of order unity or larger 
within the critical radius $b_{c}$ characterized by 
$|\chi^\text{sphere}(b_c)| \sim {\cal O}(1)$.
Thus, the asymptotic form of the total cross section becomes 
\begin{align}
  \sigma^{\gamma^{*}h}_{\rm tot}(W,q^2) \sim 
\frac{1}{(m_{j_{\nu^\ast(b_c)},1})^2}    
\ln^2 \left[\frac{1}{N_c^2}\sfrac{W^2}{\Lambda^2}^{j_{\nu^\ast(b_c)-1}}
\sfrac{\Lambda}{q}^{\gamma(j_{\nu^\ast(b_c)})+2j_{\nu^\ast(b_c)}} \right]
\label{eq:tot-DDVCS-saddle}
\end{align}
as long as $i\nu^\ast(b_c)<2$.
For even smaller $x$ (larger $W$) with a given $q^2$,
however, $b_c$ and $i\nu^\ast(b_c)$ become larger,
and the condition $i\nu^\ast(b_c)<2$ will eventually be violated.
The asymptotic behavior of $\sigma^{\gamma^\ast h}_\text{tot}(W,q^2)$ then turns into
\begin{equation}
 \sigma^{\gamma^{*}h}_{\rm tot}(W,q^2) \sim \frac{1}{m_{2,1}^2} 
   \ln^2 \left[\frac{1}{N_c^2}\frac{W^2 \Lambda^2}{q^4} \right].
\label{eq:tot-DDVCS}
\end{equation}
The total cross section of two light hadrons $\sigma^{hh'}(s)$
can also be discussed in a similar way; it can be calculated by replacing  $q^2$ with  $\Lambda^2$.
Then, the asymptotic behavior of $\sigma^{hh'}(s)$ 
becomes
\begin{align}
   \sigma^{hh'}_{\rm tot}(s) \sim 
\frac{1}{(m_{j_{\nu^\ast(b_c)},1})^2}    
\ln^2 \left[\frac{1}{N_c^2}\sfrac{s}{\Lambda^2}^{j_{\nu^\ast(b_c)-1}}
\right]
\end{align}
for $i\nu^\ast(b_c)<2$.
In the very large $s$ region ($i\nu^\ast(b_c)>2$), 
it becomes the form which is already shown in \cite{Brower2007};
\begin{equation}
 \sigma^{h h'}_{\rm tot}(s) \sim \frac{1}{m_{2,1}^2}
  \ln^2 \left[ \frac{1}{N_c^2} \frac{s}{\Lambda^2}\right].
\label{eq:tot-HH}
\end{equation}
%

\subsection{Structure Functions and GPD}
\label{ssec:GPD}

We have so far used holographic QCD to calculate the DDVCS {\it amplitude} 
and its gauge-invariant {\it structure functions}. While the scattering 
amplitude is all the necessary information in describing experimental 
data, a little more theoretical object {\it GPD}
represents internal structure of hadrons more directly.\footnote{At
least in weak coupling regime, GPD is also considered important, not
just scattering amplitudes, also because GPD can be used to describe 
amplitudes of multiple different scattering processes. } In this subsection, we 
will argue that GPD is a well-motivated notion separately from the 
scattering amplitude or structure functions even in strong coupling 
regime (in holographic QCD), and show how to extract GPD 
from holographic calculation. 

In the real world QCD, where only fermion partons (quarks and
anti-quarks) are charged under the external probe (photon) 
of DDVCS, quark GPD in a scalar hadron $h$ is defined by 
\begin{equation}
 \frac{1}{2}
 \int_{-\infty}^{+\infty} \frac{d\kappa}{2\pi} 
   e^{i \kappa x}
   \bra{h(p_2)}
       \bar{\psi}\left( - \frac{\kappa}{2} \bar{n} \right)
       \bar{n} \cdot \gamma \; 
       \psi \left( +\frac{\kappa}{2} \bar{n} \right)
   \ket{h(p_1)} 
 = H_q(x, \eta, t; q^2), \qquad \bar{n}^\mu = \frac{- q^\mu}{(p \cdot q)}. 
\end{equation}
Its Mellin moment for an even integer $j \in 2\N$ is 
\begin{align}
 & \int_0^{\infty} dx \; x^{j-1} 
   \left[ H_q(x,\eta, t; q^2) + H_{\bar{q}}(x,\eta, t; q^2) \right]
 \notag \\
 = & \int_{-\infty}^{+\infty} dx \; x^{j-1} H_q(x,\eta, t; q^2) 
 =  \sum_{k=0}^{j} A_{j,k}(t) \left(\begin{array}{c}
			     j \\ k
				  \end{array}\right) (-2 \eta)^k
  \equiv A_j(\eta, t),
\end{align}
where $A_{j,k}(t)$ is the matrix element of the spin $j$ (symmetrized
traceless) twist-2 operator,  
\begin{align}
&   \bra{h(p_2)} \left[\bar{\psi} \gamma^{\left\{ \mu_1 \right. } 
     i \mathop{D}^{\leftrightarrow}{}^{\mu_2}
     \cdots 
     i \mathop{D}^{\leftrightarrow}{}^{\left. \mu_j \right\}} 
\psi \right](0) \ket{h(p_1)} \notag \\
 = & \sum_{k=0}^j \frac{A_{j,k}(t)}{k! (j-k)!} \sum_{\sigma \in S_j}
   \Delta^{\mu_{\sigma(1)}} \cdots \Delta^{\mu_{\sigma(k)}}
   p^{\mu_{\sigma(k+1)}} \cdots p^{\mu_{\sigma(j)}}.
\end{align}
Here, $\mathop{}_{D}^{\leftrightarrow} \equiv 
 (\mathop{}_{D}^{\rightarrow}-\mathop{}_{D}^{\leftarrow})/2$.
The quark--anti-quark GPD, 
$H_q(x, \eta, t) + H_{\bar{q}}(x, \eta, t)$, is regarded as the inverse
Mellin transform of $A_j^{(+)}(\eta, t)$, a holomorphic function of 
$j$ which becomes $A_j(\eta, t)$ for $j \in 2 \N$.

Since matrix elements of gauge-invariant local operators are
well-defined notion even in gravitational dual descriptions, we  
characterize GPD in holographic calculation as the inverse Mellin 
transform of a holomorphic function of $j$, $A^{(+)}_j(\eta, t)$, which 
becomes matrix element of ``twist-2'' spin $j$ operator for $j \in 2 \N$.
We have already seen in section \ref{sssec:regge} that 
$\left[ \Gamma_{hh \P^*} (j, t)\right]_{1/\epsilon}$ 
in (\ref{eq:spin-j-form-factor}) is a holomorphic function of $j$ which 
becomes $A_{j, k=0}$ of some ``twist-2'' spin $j$ operator for $j \in 2\N$.
Thus, we can define a GPD renormalized at $\mu = 1/\epsilon$ as the 
inverse Mellin transform of $\left[\Gamma (j, t)\right]_\mu$; 
$(x, t)$ dependence of GPD\footnote{Discussion in this article 
needs to be refined in order to study skewedness dependence.} 
can be calculated by using holographic QCD:
\begin{align}
H(x,\eta=0,t;\mu^2)=\int \frac{dj}{2\pi i} x^{-j}[\Gamma(j,t)]_{\mu}.
\label{eq:GPD-and-Gamma}
\end{align}
The GPD obtained in this way satisfy the expected DGLAP evolution, 
as one can easily see from the $\epsilon$ dependence of 
$\left[\Gamma(j,t)\right]_{\mu= 1/\epsilon}$ in (\ref{eq:spin-j-form-factor}).

Obviously GPD can be estimated in completely the same method  
as we have done in earlier sections for the structure functions. 
For the saddle point phase,
\begin{align}
\label{eq:saddle-phase-GPD}
 H(x,\eta=0,t;\mu^2)\sim
 \sfrac{1}{x}^{j^\ast}
\sfrac{\Lambda}{\mu}^{\gamma(j^\ast)}[\Gamma(j^\ast,t)]_{\Lambda}.
\end{align}
Note that the saddle point value now depends on $\mu$ instead of $q$. 
The form factor $F$ for GPD defined in (\ref{eq:ratio-GPD}) is the same 
as (\ref{eq:GPD-form-factor}) obtained by simply taking the ratio 
of structure functions at arbitrary $t$ to the ones at $t=0$. 
Therefore, all the discussion on the form factor at the end of 
section \ref{sssec:large-(-t)} can be read as statements on the form 
factor of GPD in the saddle point phase of strong coupling regime. 
Note that the $(x, t, \mu)$ dependent form factor
(\ref{eq:GPD-form-factor}) and the effective mass scale
$\Lambda^2_{\rm eff.}$ are compatible with the renormalization group 
evolution.\footnote{Such models of GPD as 
$\Lambda_{\rm eff.}^2 = m_{2g}^2$ in \cite{Frankfurt:2002ka} and 
$\Lambda_{\rm eff.}^{-2} = \left[\alpha' (1-x)^2 \ln (1/x) + B (1-x)^2
 + A x(1-x) \right]$ in \cite{Diehl:2005wq} 
introduce ansatz of $(x,t)$ dependent 
GPD profile at a given renormalization scale. The profile at other 
renormalization scale needs to be determined numerically by following 
the DGLAP evolution. }  For the leading pole phase,
\begin{align}
\label{eq:pole-phase-GPD}
 H(x,\eta=0,t; \mu^2)\sim
\sfrac{1}{x}^{\alpha_{\P,1}(t)}\sfrac{\Lambda}{\mu}^{\gamma(\alpha_{\P,1}(t))}.
\end{align}
This is almost the same as the structure functions at the leading pole phase
 (\ref{eq:Regge-Pole-contribution-to-GG-amplitude-3}).

Certainly one can calculate a ``GPD'' in holographic QCD, but it is not
clear at this moment GPD of {\it which parton} it is. We adopted an
asymptotically conformal ($AdS_5$) geometry for holographic 
calculation, which means that the theory has all the particle contents 
of some conformal gauge theory. The reduced matrix element 
$\Gamma(j,t)$ obtained in holographic calculation must be that of 
some linear combination of ``twist-2'' spin-$j$ operators of gluons, 
fermions and scalars of the gauge theory. 
It must be possible, at least, to specify the linear combination that 
has the smallest anomalous dimension. 
Purely $AdS_5$ background may also be replaced by a more 
realistic background. One could think of a couple of ways to refine 
the physical meaning of GPD obtained in this way, and possibly to obtain more
results, but all of such improvements are beyond the scope of this article. 

Relation between the structure functions of the 
Compton tensor $T^{\mu\nu}$ and GPD 
in the strongly coupled regime (holographic calculation) remains quite similar 
to the one in the weakly coupled regime. To see this, let us first remind
ourselves of relation between them in the weak coupling regime.
Contributions from matrix elements of quark / gaugino twist-2 
operators to the Compton tensor $T^{\mu\nu}$ are formally written 
in the form of operator product expansion as 
\begin{equation}
 T^{\mu\nu} \simeq \sum_{j=1}^{\infty} \frac{1+(-)^j}{2} \sum_{m=0}^{\infty}
 (C_{j, m})^{\mu\nu}_{\rho_1 \cdots \rho_j \sigma_1 \cdots \sigma_m} 
 \bra{h(p_2)} (\partial^m)^{\sigma_1 \cdots \sigma_m} \left[
    \bar{\psi} \gamma (i \mathop{D}^{\leftrightarrow})^{j-1} \psi
		      \right]^{\rho_1 \cdots \rho_j}
 \ket{h(p_1)}; 
\end{equation}
in perturbative QCD, the OPE coefficient 
$(C_{j, m})^{\mu\nu}_{\rho_1 \cdots \rho_j \sigma_1 \cdots \sigma_m} $
can be calculated order by order in QCD coupling $\alpha_s$.
Gauge-invariant structure functions $V_{1,\cdots,5}$ 
can be extracted from the expression above. At tree-level in $\alpha_s$ 
expansion, for example, the quark--anti-quark contributions to (some of) 
the structure functions at $\eta = 0$ are given by 
\begin{eqnarray}
 V_1  \simeq  \sum_{j=2}^\infty \frac{1+(-)^j}{2} \frac{1}{x^j}
  A_{j,0}(t), &&  
 V_3  \simeq  \sum_{j=2}^\infty \frac{1+(-)^j}{2} \frac{-1}{x^j} 
   \frac{1}{q^2} \left[ \frac{j}{j+2} A_{j,0}(t) + 4 x^2 A_{j,2}(t) \right]. 
\end{eqnarray}
Thus, the structure functions at tree level at $\eta = 0$ are
expressed in terms of $j$-plane integrals (inverse Mellin transforms) as
in  
\begin{eqnarray}
 V_1(x,\eta = 0,t) & \simeq &
  \int \frac{dj}{4i} \frac{-[1+e^{- \pi i j}]}{\sin (\pi j)}
  \frac{1}{x^j} A^{(+)}_j(\eta = 0, t), \label{eq:V1-tree} \\
  & = & \int_0^{\infty}\frac{d\xi}{\xi} \frac{-1}{2}
   \left[\frac{1}{1-x/\xi+i\epsilon} + \frac{1}{1+x/\xi - i\epsilon}\right]
   \left(H_q + H_{\bar{q}}\right)(\xi,\eta = 0,t), \nonumber
\end{eqnarray}
\begin{eqnarray}
 q^2 V_3 (x,\eta = 0, t) & \simeq &
  \int \frac{dj}{4i} \frac{-[1+e^{- \pi i j}]}{\sin (\pi j)}
  \frac{-1}{x^j} 
  \left[\frac{j}{j+2} A^{(+)}_{j} 
   + \frac{4x^2}{j(j-1)} \frac{\partial^2}{\partial (2 \eta)^2} A^{(+)}_{j}
  \right]_{\eta = 0},  \nonumber \\
  & = & \int_0^\infty \frac{d\xi_1}{\xi_1}
        \int^{+\infty}_{0} \frac{d\xi_2}{\xi_2}
   \frac{-1}{2}
   \left[\frac{1}{1-x/\xi_2+i\epsilon} + \frac{1}{1+x/\xi_2 -
    i\epsilon}\right]  \label{eq:V3-tree}\\
  & & \qquad 
   \left[ \delta \left(1-\frac{\xi_2}{\xi_1} \right)
         - 2 \frac{\xi_2^2}{\xi_1^2}
           \Theta\left(1-\frac{\xi_2}{\xi_1}\right)
   \right]
  \left(H_q + H_{\bar{q}}\right)(\xi_1,\eta = 0, t) + \cdots . \nonumber 
\end{eqnarray}
Since the non-skewed structure functions in the complex $j$-plane
representation are given by products of the signature factor 
$-[1+e^{-\pi i j}]/\sin(\pi j)$, operator reduced matrix element $A_j$ 
(or its $\eta$-derivative) and its OPE coefficient $C_{j}$, 
the structure functions become convolution of 
the inverse Mellin transforms of those factors. The factors 
$[(\xi-x+i \epsilon)^{-1} + (\xi + x - i \epsilon)^{-1}]$ 
in (\ref{eq:V1-tree}) and (\ref{eq:V3-tree}) originate from 
the inverse Mellin transform of the signature factor \cite{bateman} here, 
but it also arises from the light-cone singularity of the fermion 
parton in direct calculation of the Compton tensor using Feynman 
diagrams \cite{Ji:1996nm}. 

In the holographic calculation of the structure functions of the DDVCS
amplitude, we have seen that all the structure functions $V_{1,\cdots,
5}$ are given by $I_{0,1}$, and $I_{0,1}$ are given by $\nu$ integration
(\ref{eq:Im-Ii-cm}).
The $\nu$-integration becomes $j$-plane integration, when the
integration variable is changed through $j = j_\nu$. The integrand 
is a product of the signature factor $-[1+e^{-\pi i j}]/\sin(\pi j)$, 
reduced matrix element $\left[\Gamma(j,t)\right]_\mu$ of some 
``twist-2'' spin $j$ operator and a remaining $j$-dependent factor, which 
is to be interpreted as the OPE coefficient and 
$(q \cdot p)^j \propto x^{-j}$. Therefore, the structure functions 
of DDVCS amplitude are given even in the strong coupling regime 
by a convolution of a ``propagator'' 
$[(\xi-x+i\epsilon)^{-1} + (\xi+x-i\epsilon)^{-1}]$, GPD (inverse Mellin
transform of $[\Gamma(j,t)]_\mu$) and inverse Mellin transform of 
OPE coefficient, just like in the weak coupling regime. The OPE coefficients 
in the strongly coupled regime, however, are calculated through the 
vertex operator OPE on world sheet; the OPE coefficients may be 
improved order by order in $1/\sqrt{\lambda}$ expansion, rather than
in $\alpha_s$ expansion in perturbative QCD. 

\section{Implication for the Real World QCD}
\label{sec:real-world}

We have so far studied DDVCS and GPD by using holographic QCD.
In this section, we discuss how much we can apply our results 
in gravity dual to the real world QCD.
We should keep in mind that the hard wall model or other UV conformal 
holographic models are not equivalent to the real world QCD.
Dual gauge theories of UV conformal gravity descriptions remain 
strongly coupled, even in UV scale, if the gravity descriptions are 
to be reliable. If one considers an asymptotic free holographic model, 
so that it matches on to the asymptotic free QCD of the real world, 
then such a gravity description will not be a useful framework in the 
UV region of the holographic radius. 
Although there is such a difference, and a big gap that is very
difficult to fill, it is also worthwhile to keep in mind that 
there are several features in GPD and structure functions that are 
shared by gravity dual descriptions and the real world
QCD / perturbative QCD.

Factorization theorem holds for (D)DVCS and leptoproduction of mesons
in the real world QCD \cite{Brodsky:1994kf, Collins1997}.
The structure functions are given by a convolution of hard kernel and
GPD as in (\ref{eq:V1-tree}, \ref{eq:V3-tree}).
The hard kernel is calculable in the perturbation of the gauge coupling,
whereas the GPD is incalculable, and should be given as non-perturbative 
input.
It would be nice if GPD profile can be determined by experimental 
data alone, but that is known not to be possible.
Even in setting constraints on the profile by using data, 
GPD has to be modeled and parameterized, because observable 
scattering amplitudes involve convolution of GPD, not just pure GPD, 
and furthermore, depend on GPD only in the limited region of kinematical
variables. 
It is thus inevitable to construct physically motivated model of GPD. 

Modeling and parametrization of GPD has already grown into a large
field, as one can see in review articles 
(e.g., \cite{DiehlPhys.Rept.388:41-2772003, Belitsky:2005qn}).
The simplest model imaginable assumes factorized form of GPD into 
PDF $q(x)$ and a $t$-dependent factor \cite{Vanderhaeghen1998, 
Vanderhaeghen1999},
\begin{align}
 H(x,\eta=0,t)=q(x)F(t), 
\label{eq:VGG-model}
\end{align}
and models that mimic small $x$ Regge behavior were also considered 
\cite{Goeke2001, Belitsky2002, Diehl2005, Guidal2005}:
\begin{align}
 H(x,\eta=0,t)=q(x)e^{tg(x)}
\label{eq:Regge-motivated-model}
\end{align}
with some $x$-dependent function $g(x)$.
In order to implement basic theoretical requirements on GPD, such 
as skewedness-polynomiality and proper DGLAP evolution, however, 
it is more convenient to have a framework of parameterizing GPD where 
such requirements are implemented systematically.
There are two such frameworks of systematic parametrization.
The first one is double distribution \cite{Mueller1994, Radyushkin1996a,
Radyushkin1997} along with $D$-term \cite{PolyakovWeiss1999}; 
transformation form this parametrization to GPD (Radon transformation) 
and its inverse is known \cite{Radyushkin1999, Belitsky2001, Teryaev2001}.
GPD models in this parametrization framework are found, for example, 
in \cite{Radyushkin1999, Radyushkin1999a}.
The alternative framework of systematic parametrization is the 
collinear factorization approach, which is also known as dual-parametrization.
In this approach, models of GPD are constructed as an amplitude 
on the conformal moment space \cite{Belitsky1998, Polyakov1999, Polyakov2002, Mueller2006, Kumericki:2007sa};
GPD is given by a transformation \cite{Mueller2006} that becomes inverse Mellin transformation
of $[\Gamma(j,t)]_{\mu}$ for zero-skewedness.

Lessons from gravity dual calculation of DDVCS amplitude and GPD 
can be passed on to the understanding of GPD of the real world 
best in the form of $j$-plane representation, $[\Gamma(j, t)]_{\mu}$, 
that is, in the collinear factorization approach (dual parametrization). 
The transformation between the $j$-plane representation and 
Bjorken $x$ representation
is just a pure mathematical one, and does not 
rely on perturbation theory or even on existence of a calculable 
theoretical framework.
The description of the (D)DVCS amplitude in $j$-plane is also based 
on OPE, which once again does not rely on perturbation 
theory \cite{PolchinskiJHEP0305:0122003}. 
While there is no way calculating hadron matrix element
$[\Gamma(j,t)]_\mu$ in perturbation theory, or in lattice gauge theory
for complex valued $j$, various gravity dual models yield at least some
results (if not truly faithful to the real-world QCD), as we have 
seen in earlier sections. 

The $j$-plane representation of the scattering amplitude and GPD 
also plays an essential role in characterizing the following 
three phases of the behavior of GPD and (D)DVCS amplitude:
\begin{itemize}
 \item $j^* < {\rm Re} \; \alpha_{\P,1} (t)$: leading pole phase (or
       leading singularity phase),
 \item ${\rm Re} \; \alpha_{\P,1} (t) < j^* < 2$: saddle point phase,
       where $\gamma_{\rm eff.} < 0$, 
 \item ${\rm Re} \; \alpha_{\P,1} (t) < 2 < j^*$: spin-2 phase,
       where $0 < \gamma_{\rm eff.}$.  
\end{itemize}
The three phases are also interpreted as
Low-KK Spin-sum phase, KK-sum Spin-sum phase, and KK-sum Low-spin phase, respectively,
from above to below.
The distinction between the last two phases is absent in GPD or in
imaginary part of the scattering amplitude (at leading order in 
$1/N_c$ expansion), and we simply refer to the last two phases as saddle point phase. 
Earlier sections of this article described such a phase structure 
by relying on gravity dual calculation, but the essence is in 
the relative position of the leading singularity, $j=2$ pole and 
the saddle point in the $j$-plane representation, not so much 
on details of gravity dual descriptions.
As we have studied in sections \ref{sssec:small-(-t)} and
\ref{sssec:regge}, the parton dynamics characterized by 
$\gamma_{\rm eff.}$ and $\lambda_{\rm eff.}$ and the phase structure 
 Figure~\ref{fig:phasediagramSC} in $(q^2, x)$ plane (fixed $t$ slice)
in gravity dual 
are surprisingly similar (qualitatively) to the one we know in the real 
world QCD. This similarity is traced back to the fact that the 
saddle point is determined by (\ref{eq:saddle-p-QCD}) in perturbative
QCD as well as in holographic QCD, and that the anomalous dimensions
$\gamma(j)$ in both frameworks are also qualitatively 
similar \cite{BrowerJHEP0712:0052007}; especially the two properties
\begin{align}
 \frac{\partial \gamma(j)}{\partial j}>0,\quad \frac{\partial^2 \gamma(j)}{\partial j^2}<0
\label{eq:common-properties-of-anomalous-dimension}
\end{align}
are shared
for certain range of $j$.

Most of kinematical regions explored in DIS experiments should be 
in the saddle point phase (including the spin-2 phase in
DIS),\footnote{The leading
singularity phase tends to be realized in smaller $x$, lower $q^2$ and 
less negative $t$, and might be found experimentally. It is difficult, 
though, to tell theoretically where the phase boundary is located 
in the $(x,t, q^2)$ variable space in a theory with running coupling 
constant. } 
as is clear from the $(x, q^2)$ dependence of $\gamma_{\rm eff.}$ and 
$\lambda_{\rm eff.}$ in DIS \cite{Breitweg:1998dz}.\footnote{It is not 
very clear in low $q^2$ region, however, whether $\lambda_{\rm eff.}$ 
increases for larger $q^2$ \cite{Adams:1996gu}. 
c.f. Figure~\ref{fig:transition}~(a). 
The leading pole phase may be hidden there.}
Anomalous dimension $\gamma(j)$ in the weak coupling
regime\footnote{There are various approximation schemes in perturbative
QCD, such as DLLA, BFKL and double leading approximation. } should 
be used, because of the value of $\lambda_{\rm eff.}$ (depending weakly
on $q^2$) closer to 0 than to 1 \cite{Breitweg:1998dz}.
The GPD, then, is approximated by the expression (\ref{eq:saddle-phase-GPD}) 
for some form factor $[\Gamma(j, t)]_\Lambda$ evaluated at $j = j^*$.
This suggests (if the skewedness dependence that we did not study in
this article does not screw things up) that various parameters
characterizing the DVCS differential cross section\footnote{This 
parametrization of DVCS cross section by $(\delta, n)$ (conventionally 
adopted in data analysis) corresponds to 
$I_i \sim (W/\Lambda)^{2+\delta/2} (\Lambda^2/q^2)^{n/2}$. }  
\begin{equation}
 \frac{d \sigma_{\rm DVCS}(\gamma^*p \rightarrow \gamma p)}{dt} \sim 
 \frac{\alpha_{QED}^2}{\Lambda^{4}} \times
 \left(\frac{W}{\Lambda}\right)^{\delta}
 \left(\frac{\Lambda^2}{q^2}\right)^n,
\end{equation}
namely, $\delta = 4 (j^* -1)$ and $n = \gamma(j^*) + 2j^*$ above, as
well as the real part to imaginary part ratio $\rho$ and the $t$-slope 
parameter $B$, are all functions of the saddle point value
$j^*$. Because of the way the saddle point value is 
determined, (\ref{eq:saddle-p-QCD}), all the four parameters 
are more sensitive to $\ln(q/\Lambda)$ than to $\ln (1/x)$ or 
$\ln (W/\Lambda)$ in small $x$, for the reason that we have already seen 
in (\ref{eq:3properties-for-saddle-point}). All of 
$\delta = 4 \lambda_{\rm eff.}$, $\gamma(j^*)$, $n$ and $\rho$ are increasing
functions of $j^*$, and hence they increase\footnote{
The HERA measurements give
$\delta=0.44\pm0.19$ for $q^2=2.4$ GeV$^2$,
$\delta=0.52\pm0.09$ for $q^2=3.2$ GeV$^2$,
$\delta=0.75\pm0.17$ for $q^2=6.2$ GeV$^2$,
$\delta=0.84\pm0.18$ for $q^2=9.9$ GeV$^2$, and
$\delta=0.76\pm0.22$ for $q^2=18$  GeV$^2$
in ZEUS \cite{Chekanov:2008vy},
and
$\delta=0.61\pm0.10\pm0.15$ for $q^2=8$ GeV$^2$,
$\delta=0.61\pm0.13\pm0.15$ for $q^2=15.5$ GeV$^2$, and
$\delta=0.90\pm0.36\pm0.27$ for $q^2=25$ GeV$^2$,
in H1 \cite{Aaron2009}.
} for larger 
$\ln (q^2/\Lambda^2)$. They should decrease for larger $\ln (1/x)$ or
larger $\ln(W/\Lambda)$, but only weakly. 
Although it is not immediately clear whether the slope parameter $B$ 
increases or decreases for larger $j^*$ (and hence
for larger $\ln (q/\Lambda)$), yet we found by using the spin $j$ form 
factor in the gravity dual hard wall model (in section
\ref{sssec:slope}) that the slope parameter decreases for large $q^2$, 
which is in nice agreement with the HERA measurements
\cite{Aktas:2005ty, Aaron:2007cz, Aaron2009}.\footnote{In the leading 
pole phase with an anomalous dimension $\gamma(j)$ in the weak coupling 
regime, the slope parameter of the (D)DVCS differential cross section 
{\it increases} for larger $\ln(1/x)$, and depends only weakly on 
$\ln (q/\Lambda)$; $\gamma(j) \propto \alpha_s$ is used 
in (\ref{eq:slope-paramter-in-pomeron-pole-phase}). This prediction 
does not fit very well with the HERA data \cite{Aktas:2005ty, 
Aaron:2007cz, Aaron2009}. In this article, we did not work out 
gravity dual prediction for the $t$-slope parameter $B$ in the spin-2
pole phase, where $\gamma_{\rm eff.} > 0$. 
The contribution from the real part dominated by a tower of spin-2 glueball
exchange should be important there. }

It is also worthwhile to remind ourselves that GPD models in the 
collinear factorization approach (dual parametrization) in the saddle 
point phase are always compatible with the renormalization group 
evolution. In the approximate form (\ref{eq:saddle-phase-GPD}), the 
renormalization scale $\mu$ dependence comes into the form factor 
$[\Gamma(j^*,t)]_\Lambda$ through the saddle point value $j^*$.

All of these phenomenological as well as theoretical successes of 
the GPD models (\ref{eq:saddle-phase-GPD}) come directly from the 
nature of the saddle point phase, as we have already seen above. 
One more necessary ingredient is that the slope parameter 
\begin{equation}
 2 \left.
 \frac{\partial}{\partial t} \ln \left[ \Gamma(j,t) \right]
 \right|_{-t \sim [0 \sim 1] \GEV^2} = B
\label{eq:j-dep-B}
\end{equation}
is a decreasing function of $j$; this property of the hard-wall model 
prediction was used in explaining the $q^2$ dependence of the $t$-slope 
parameter. It is a blessing indeed that the various features of 
experimental data are attributed to such a small number of conditions. 
To our knowledge, the first model of GPD in this category was that 
of \cite{Mueller:2006pm}, which introduces an ansatz 
\begin{equation}
 [\Gamma(j,t)]_{\Lambda} \sim \frac{\beta(t)}{j-\alpha(t)}, \qquad 
 \beta(t)=\left(1-\frac{a t}{\Lambda^2}\right)^{-n};
\label{eq:Mueller-form-factor} 
\end{equation}
$a$ and $n$ are parameters of the model.
All possible holographic models will also have their own predictions of 
the reduced matrix element (spin $j$ form factor) $[\Gamma(j,t)]$, 
and hence successful models of GPD in this category are obtained from 
these models, as long as (\ref{eq:j-dep-B}) is a decreasing function of $j$.

A general lesson from gravity dual calculations of the spin $j$ form
factor $[\Gamma(j, t)]$ is that even the leading trajectory 
(containing graviton) in 5-dimensions (or in 10-dimensions) becomes
a Kaluza--Klein tower of infinite Pomeron trajectories for hadron scattering 
in 4-dimensions.\footnote{On top of this tower of infinitely many Pomeron
trajectories, gravity dual descriptions predict yet another tower
structure of trajectories, because stringy states form a tower of daughter trajectories
already in the 5D (or 10D) description.} The form factor is given in the following form 
\begin{align}
 [\Gamma(j,t)]_{\Lambda}=\sum_n \frac{a_n(j)}{t-m^2_{j,n}},
\label{eq:HQCD-Gamma}
\end{align}
where the $n$-th term corresponds to the $n$-th Kaluza--Klein excited 
Pomeron trajectory; $t = m^2_{j,n}$ is the $({\rm mass})^2$--spin relation.
After summing up all the contributions in a Kaluza--Klein tower 
(\ref{eq:HQCD-Gamma}), however, the form factor has a power-law 
dependence in $(-t)$. This property is understood simply by using 
Green function in warped 5-dimensions; details of the background 
geometry of gravity dual models are irrelevant.
This power-law behavior at large $(-t)$ nicely matches on to the 
power law behavior expected theoretically from naive quark power
counting \cite{Brodsky:1973kr, Matveev:1973ra, Brodsky:1974vy, 
Sivers:1975dg, Donnachie:1979yu}, albeit only at qualitative level.

It should be kept in mind, however, that gravity dual descriptions 
are not an ideal and flawless framework in calculating the form factor. 
Although a gravity dual model with a carefully chosen background may 
serve as a faithful dual to the QCD of the real world, it should work 
as a reliable framework of calculation only in the IR region of the 
holographic radius. The UV region of the geometry should correspond 
to the real world QCD in the weak coupling regime, and hence the 
curvature of the background should be too large for a reliable calculation. 
It will be nice if it is possible to make even a crude estimate of the 
error in gravity dual calculation of the form factor (e.g., the
power-law behavior from UV conformal gravity dual models is dictated by 
the conformal dimension associated with the target hadron, whereas 
the power is determined by the number of valence partons in the 
(empirically successful) naive power counting rule), and to use 
perturbative QCD in some way or other in controlling or reducing the 
error. 

Such a dream may not be totally unrealistic, because at least 
one can compare singularities in the $j$-plane representation. 
Gravity dual models for asymptotic free running coupling (at least up to
some high energy scale), e.g., \cite{Aharony:1998xz}, can be used to 
calculate the isolated poles of $[\Gamma(j,t)]$. It is just to solve 
a Schr\"{o}dinger equation \cite{BrowerJHEP0712:0052007}. 
Poles in the large  ${\rm Re }\: j$ region are reliable, because the
corresponding wavefunctions are localized in the IR region of the warped spacetime, but 
those with smaller ${\rm Re} \; j$ are not. 
On the other hand, the BFKL theory can calculate a spectrum of its kernel.
The spectrum is discrete, when the running coupling effect is taken 
into account, and the discrete spectrum is mapped into poles in the 
complex $j$-plane \cite{Forshaw:1997dc}. The spectrum of the BFKL theory 
in large ${\rm Re} \; j$ region, however, corresponds to wavefunctions 
that are dominantly in the small $k_{\perp}$ (gluon transverse momentum) region, where the coupling 
$\alpha_s$ is large and perturbation in $\alpha_s$ does not work well.
Thus, the BFKL theory might be used in determining the $j$-plane 
singularities of $[\Gamma(j,t)]$ in a region (smaller ${\rm Re} \; j$) 
where the results from gravity dual cannot be reliable.

In the calculation of the form factor, the spectrum of the Pomeron 
kernel (poles in the $j$-plane) is not the only necessary information. 
We also need corresponding wavefunctions of the spectrum, which are 
to be multiplied by a hadron impact factor, and integrated over the
holographic radius $z$, if we are to use the language of gravity
description. Based on the similarity between the $k_{\perp}$ 
factorization formula of perturbative QCD and the expression 
(\ref{eq:two scalar amp}) of hadron scattering amplitudes in gravity
dual, correspondence between the $k_{\perp}$ coordinate of the BFKL theory 
and the holographic radius $z^{-1}$ has been 
suggested \cite{Brower:2002er, Brodsky:2003px, BrowerJHEP0712:0052007}. 
Thus, the integration (and integrand) over the $z$ coordinate in 
the UV region may be replaced by that of the BFKL theory over the
$k_{\perp}$ coordinate, in a crude attempt at improving the form factor 
$[\Gamma(j,t)]$ calculated entirely in gravity dual models.
All of 
such attempts, however, are beyond the scope of this article.

\section{Discussion}
\label{sec:discussion}

In this article, we used holographic QCD to study double deeply virtual 
Compton scattering (DDVCS) amplitude and generalized parton distribution (GPD). 
Analysis presented in this article, however, is restricted to the scattering 
amplitude with vanishing skewedness; in order to exploit holographic methods 
to study non-perturbative aspects of deeply virtual Compton scattering 
(DVCS), analysis in this article needs to be extended to cover the case 
with non-zero skewedness. This subject will be covered in a future 
publication \cite{future}.
 
\begin{figure}[tbp]
  \begin{center}
\begin{tabular}{cc}
  \includegraphics[scale=0.35]{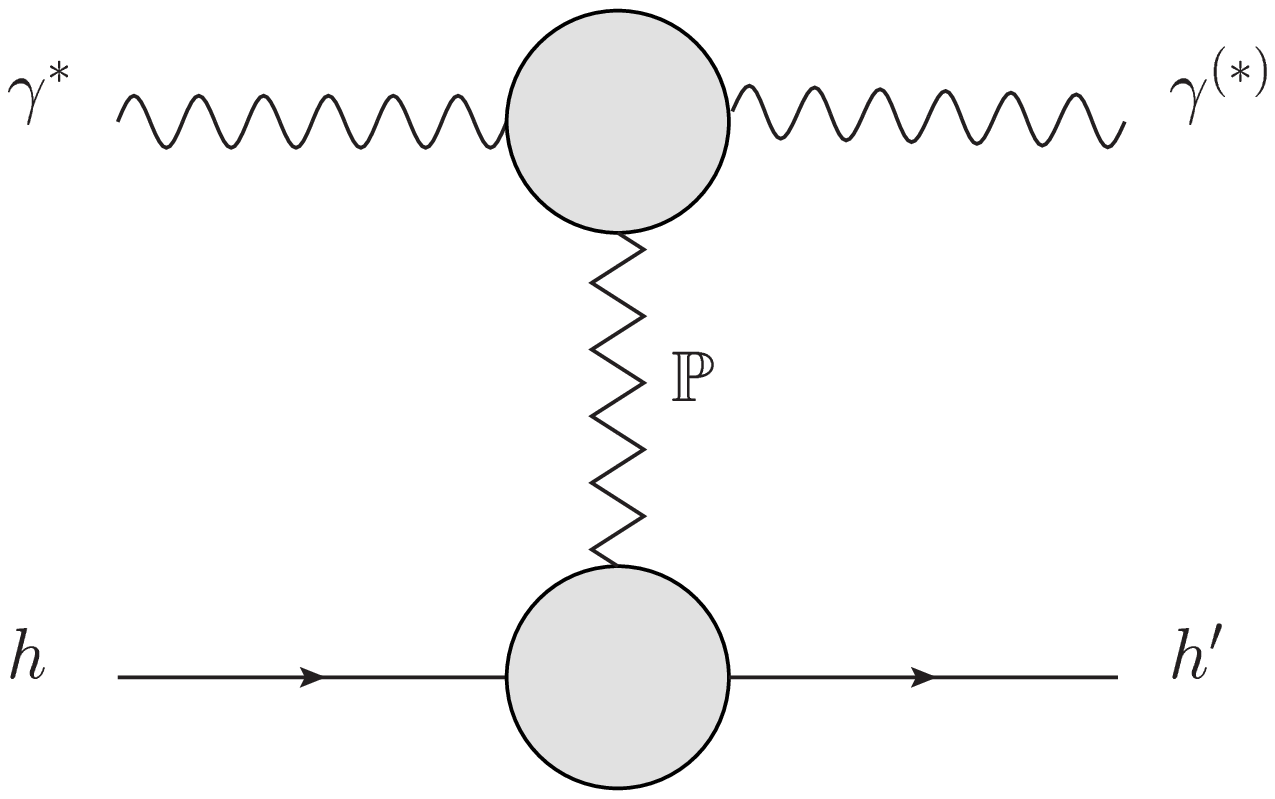} & 
  \includegraphics[scale=0.35]{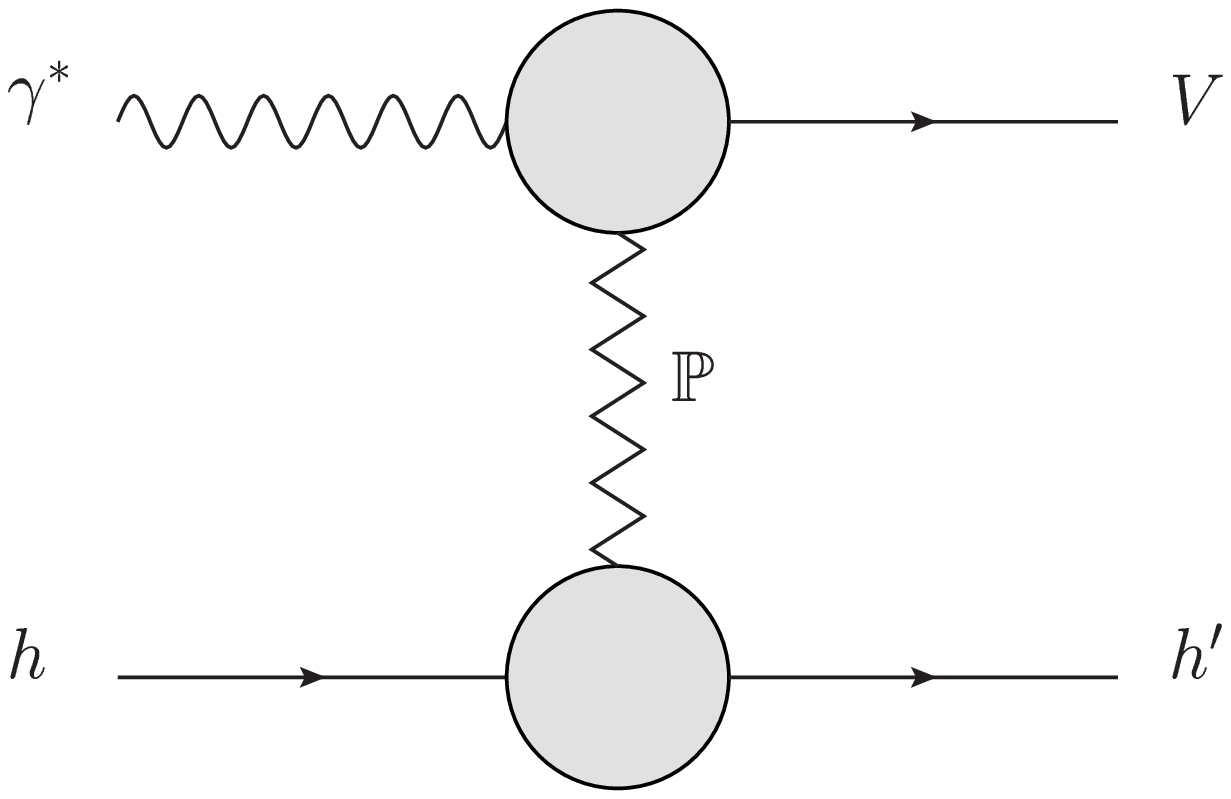} \\ 
  (a) &  (b) 
\end{tabular}
\caption{\label{fig:pomeron-exchanging-diagrams}
One can expect straightforward generalization of various results 
in this article for the following processes.
(a): DVCS/DDVCS or timelike Compton scattering.
The final state hadron $h'$ is allowed here to be different from the initial one $h$.
For example, it can be an excited mode of the initial state hadron.
(b): Vector meson production.
}
  \end{center}
\end{figure}

We have presented in this article an improved conceptual understanding 
of Pomeron couplings and Pomeron form factor, and predicted phase transition 
(crossover, to be more precise) in DDVCS amplitude. Such results, however, 
are not just for DDVCS and DVCS, but are more general in nature;  
some of the statements in this article hold true for some other high energy 
scattering processes only with minor (and almost obvious) modifications. 
An obvious generalization is to relax the constraint that the 
final state hadron $h'$ is the same as the initial state hadron $h$; 
$\gamma^\ast h\rightarrow \gamma^{(\ast)}h'$ 
(Figure~\ref{fig:pomeron-exchanging-diagrams}~(a)).
When the final state hadron $h'$ and the initial state hadron $h$ are different 
Kaluza--Klein modes of the same field in the 5D effective theory, 
Pomerons (with vacuum quantum number) can couple to the transition from 
$h$ to $h'$. One only needs to replace the impact factor $P_{hh}(z')$ by 
$P_{hh'}(z') = c_{\phi} \left[ \Phi_n(z') \Phi_{n'}(z')\right]$ in 
writing down the amplitude. 
Scattering amplitude of vector meson production $\gamma^\ast h \rightarrow Vh'$ 
(Figure~\ref{fig:pomeron-exchanging-diagrams}~(b)) and timelike 
Compton scattering $\gamma^\ast h \rightarrow \gamma^\ast h'$ 
(Figure~\ref{fig:pomeron-exchanging-diagrams}~(a) where the final state 
photon has timelike virtuality) can also be obtained by simply replacing 
the final state photon wavefunction properly. 
Since all of those processes involve an initial state spacelike photon with 
large virtuality, the initial state photon wavefunction localized strongly 
toward UV region gives rise to much the same $q^2_1$ and $W^2$ dependence 
of the Pomeron contribution to those scattering amplitudes. 
All of those processes involve a hadron--hadron--Pomeron vertex 
(lower blobs in Figure~\ref{fig:pomeron-exchanging-diagrams}~(a, b)), and the  
non-perturbative vertex (form factor) in 
those processes should be essentially the same as in DDVCS.
When the saddle point $j=j^*(x,q^2,t)$ of the complex $j$-plane 
representation of the scattering amplitudes has a larger real part than 
all the singularities in the $j$-plane, the form factor 
depends on various kinematical variables (such as $W^2$ and $q^2$) 
only through the saddle point value $j^*$, and the form factor becomes 
that of ``spin $j^*$ current''. 
The $t$-slope parameter can also be 
calculated for these processes, but it will be obvious that the 
our argument in section \ref{sssec:slope} is so generic that its conclusion 
holds true for vector meson production and timelike Compton scattering as well. 
There is sort of universality in the Pomeron--hadron $(h)$--hadron
$(h')$ form factor (coupling), regardless of whether it is used 
for DDVCS, DVCS, time-like Compton scattering or vector meson
production (Figure~\ref{fig:pomeron-exchanging-diagrams}~(a, b)).  

We have also seen in this article that the (Pomeron contribution to the)
DDVCS amplitude shows three qualitatively different behaviors; the three
phases are characterized by the position of the saddle point $j^*$ of 
the $j$-plane representation of the amplitude, relatively to the leading
singularity and $j=2$ pole of the signature factor, as summarized in 
section \ref{sec:real-world}. The essence of this phase classification 
is shared also by DDVCS/DVCS with $h' \neq h$, vector meson production 
and timelike Compton scattering. Since all of these processes involve 
the kinematical variable $q^2_1$, spacelike virtuality of the initial 
state photon, it is not surprising to see a crossover behavior in the 
differential cross sections of these processes.

 \section*{Acknowledgments}  

It is our great pleasure to express cordial gratitude to Kazuki 
Itakura and Yoshitaka Hatta, who gave a series of introductory 
lectures on hadron high-energy scattering at Tsukuba U., '08 and 
at Tokyo U., '10, respectively. We also thank Leonid Frankfurt, Andrei
Mihailov and Shigeki Sugimoto for useful comments, and Simeon Hellerman 
and Tsutomu Yanagida for encouragements. Part of this work was carried 
out during long term programs ``Branes, Strings and Black Holes'' at
YITP, Kyoto, October, 2009 (TW), ``Strings at the LHC and in the 
Early Universe'' at KITP, Santa Barbara, April--May, 2010 (TW), 
``High Energy Strong Interactions 2010'' at YITP, Kyoto, August, 
2010 (RN, TW) and also during a stay at Caltech theory group of TW. 
This work is supported by JSPS Research Fellowships for Young 
Scientists (RN), by WPI Initiative, MEXT, Japan (RN, TW) and 
National Science Foundation under Grant No. PHY05-51164 (TW).

\bibliographystyle{utcaps}
\bibliography{transverse_structure}
\end{document}